\documentclass[format=acmsmall, screen=true, nonacm,pdftex,svgnames]{acmart}

\setcopyright{rightsretained}
\acmPrice{}
\acmDOI{10.1145/3434318}
\acmYear{2021}
\copyrightyear{2021}
\acmSubmissionID{popl21main-p250-p}
\acmJournal{PACMPL}
\acmVolume{5}
\acmNumber{POPL}
\acmArticle{37}
\acmMonth{1}

\usepackage{booktabs}   

\usepackage{xcolor}
\usepackage{colortbl}

\usepackage[utf8]{inputenc}

\usepackage{amsmath}
\usepackage{hyperref}

\usepackage{subcaption}
\usepackage{wrapfig}

\usepackage{xspace}
\usepackage{float}
\usepackage{balance}

\usepackage{cleveref} 

\usepackage{wrapfig}
\usepackage{caption}

\usepackage{longtable}
\usepackage{tabu}

\usepackage{mathtools} 
\usepackage{xfrac} 

\usepackage[qm,braket]{qcircuit}

\newcommand{\sqir}{s\textsc{qir}\xspace}
\newcommand{\qwire}{\ensuremath{\mathcal{Q}\textsc{wire}}\xspace}
\newcommand{\qbricks}{\ensuremath{\mathcal{Q}\textsc{bricks}}\xspace}

\newcommand{\voqc}{\textsc{voqc}\xspace}
\newcommand{\tket}{t$\vert$ket$\rangle$\xspace}

\newcommand{\R}{\ensuremath{\mathbb{R}}\xspace}
\newcommand{\CC}{\ensuremath{\mathbb{C}}\xspace}

\usepackage{tikz} 
\usetikzlibrary{%
  arrows,%
  shapes.misc,
  shapes.geometric, 
  shapes.arrows,%
  shapes.callouts,
  shapes.gates.logic.US,
  chains,%
  matrix,%
  positioning,
  scopes,%
  decorations.pathmorphing,
  decorations.text,
  decorations.pathreplacing, 
  shadows%
}
\tikzset{ machine/.style={
    rectangle,
    minimum width=25mm,
    minimum height=18mm,
    text width=24mm,
    align=center,
    very thick,
    draw=black,
    color=black,
    fill=white,
  }
}

\DeclarePairedDelimiter\abs{\lvert}{\rvert}
\DeclarePairedDelimiter\norm{\lVert}{\rVert}

\makeatletter
\let\oldabs\abs
\def\abs{\@ifstar{\oldabs}{\oldabs*}}
\let\oldnorm\norm
\def\norm{\@ifstar{\oldnorm}{\oldnorm*}}
\makeatother

\makeatletter
\DeclareRobustCommand{\vardivision}{%
  \mathbin{\mathpalette\@vardivision\relax}%
}
\newcommand{\@vardivision}[2]{%
  \reflectbox{$\m@th\smallsetminus$}%
}
\makeatother

\usepackage{booktabs}

\usepackage{alltt}
\usepackage{listings,lstcoq}
\usepackage{MnSymbol}
\definecolor{ltblue}{rgb}{0,0.4,0.4}
\definecolor{dkblue}{rgb}{0,0.1,0.6}
\definecolor{dkgreen}{rgb}{0,0.35,0}
\definecolor{dkviolet}{rgb}{0.3,0,0.5}
\definecolor{dkred}{rgb}{0.5,0,0}
\newcommand{\code}[1]{{\small\texttt{#1}}}

\newcommand{\denote}[1]{\llbracket #1 \rrbracket\xspace}
\newcommand{\pdenote}[1]{\{\hspace{-0.2em}| #1 |\hspace{-0.2em}\}}

\usepackage{newunicodechar}
\let\Alpha=A
\let\Beta=B
\let\Epsilon=E
\let\Zeta=Z
\let\Eta=H
\let\Iota=I
\let\Kappa=K
\let\Mu=M
\let\Nu=N
\let\Omicron=O
\let\omicron=o
\let\Rho=P
\let\Tau=T
\let\Chi=X

\newunicodechar{Α}{\ensuremath{\Alpha}}
\newunicodechar{α}{\ensuremath{\alpha}}
\newunicodechar{Β}{\ensuremath{\Beta}}
\newunicodechar{β}{\ensuremath{\beta}}
\newunicodechar{Γ}{\ensuremath{\Gamma}}
\newunicodechar{γ}{\ensuremath{\gamma}}
\newunicodechar{Δ}{\ensuremath{\Delta}}
\newunicodechar{δ}{\ensuremath{\delta}}
\newunicodechar{Ε}{\ensuremath{\Epsilon}}
\newunicodechar{ε}{\ensuremath{\epsilon}}
\newunicodechar{ϵ}{\ensuremath{\varepsilon}}
\newunicodechar{Ζ}{\ensuremath{\Zeta}}
\newunicodechar{ζ}{\ensuremath{\zeta}}
\newunicodechar{Η}{\ensuremath{\Eta}}
\newunicodechar{η}{\ensuremath{\eta}}
\newunicodechar{Θ}{\ensuremath{\Theta}}
\newunicodechar{θ}{\ensuremath{\theta}}
\newunicodechar{ϑ}{\ensuremath{\vartheta}}
\newunicodechar{Ι}{\ensuremath{\Iota}}
\newunicodechar{ι}{\ensuremath{\iota}}
\newunicodechar{Κ}{\ensuremath{\Kappa}}
\newunicodechar{κ}{\ensuremath{\kappa}}
\newunicodechar{Λ}{\ensuremath{\Lambda}}
\newunicodechar{λ}{\ensuremath{\lambda}}
\newunicodechar{Μ}{\ensuremath{\Mu}}
\newunicodechar{μ}{\ensuremath{\mu}}
\newunicodechar{Ν}{\ensuremath{\Nu}}
\newunicodechar{ν}{\ensuremath{\nu}}
\newunicodechar{Ξ}{\ensuremath{\Xi}}
\newunicodechar{ξ}{\ensuremath{\xi}}
\newunicodechar{Ο}{\ensuremath{\Omicron}}
\newunicodechar{ο}{\ensuremath{\omicron}}
\newunicodechar{Π}{\ensuremath{\Pi}}
\newunicodechar{π}{\ensuremath{\pi}}
\newunicodechar{ϖ}{\ensuremath{\varpi}}
\newunicodechar{Ρ}{\ensuremath{\Rho}}
\newunicodechar{ρ}{\ensuremath{\rho}}
\newunicodechar{ϱ}{\ensuremath{\varrho}}
\newunicodechar{Σ}{\ensuremath{\Sigma}}
\newunicodechar{σ}{\ensuremath{\sigma}}
\newunicodechar{ς}{\ensuremath{\varsigma}}
\newunicodechar{Τ}{\ensuremath{\Tau}}
\newunicodechar{τ}{\ensuremath{\tau}}
\newunicodechar{Υ}{\ensuremath{\Upsilon}}
\newunicodechar{υ}{\ensuremath{\upsilon}}
\newunicodechar{Φ}{\ensuremath{\Phi}}
\newunicodechar{φ}{\ensuremath{\phi}}
\newunicodechar{ϕ}{\ensuremath{\varphi}}
\newunicodechar{Χ}{\ensuremath{\Chi}}
\newunicodechar{χ}{\ensuremath{\chi}}
\newunicodechar{Ψ}{\ensuremath{\Psi}}
\newunicodechar{ψ}{\ensuremath{\psi}}
\newunicodechar{Ω}{\ensuremath{\Omega}}
\newunicodechar{ω}{\ensuremath{\omega}}

\newunicodechar{ℕ}{\ensuremath{\mathbb{N}}}
\newunicodechar{∅}{\ensuremath{\emptyset}}

\newunicodechar{∙}{\ensuremath{\bullet}}
\newunicodechar{≈}{\ensuremath{\approx}}
\newunicodechar{≅}{\ensuremath{\cong}}
\newunicodechar{≡}{\ensuremath{\equiv}}
\newunicodechar{≤}{\ensuremath{\le}}
\newunicodechar{≥}{\ensuremath{\ge}}
\newunicodechar{≠}{\ensuremath{\neq}}
\newunicodechar{∀}{\ensuremath{\forall}}
\newunicodechar{∃}{\ensuremath{\exists}}
\newunicodechar{±}{\ensuremath{\pm}}
\newunicodechar{∓}{\ensuremath{\pm}}
\newunicodechar{·}{\ensuremath{\cdot}}
\newunicodechar{⋯}{\ensuremath{\cdots}}
\newunicodechar{…}{\ensuremath{\ldots}}
\newunicodechar{∷}{~\mathrel{:\!\!\!:}~}
\newunicodechar{×}{\ensuremath{\times}}
\newunicodechar{∞}{\ensuremath{\infty}}
\newunicodechar{→}{\ensuremath{\to}}
\newunicodechar{←}{\ensuremath{\leftarrow}}
\newunicodechar{⇒}{\ensuremath{\Rightarrow}}
\newunicodechar{↦}{\ensuremath{\mapsto}}
\newunicodechar{↝}{\ensuremath{\leadsto}}
\newunicodechar{∨}{\ensuremath{\vee}}
\newunicodechar{∧}{\ensuremath{\wedge}}
\newunicodechar{⊢}{\ensuremath{\vdash}}
\newunicodechar{⊣}{\ensuremath{\dashv}}
\newunicodechar{∣}{\ensuremath{\mid}}
\newunicodechar{∈}{\ensuremath{\in}}
\newunicodechar{⊆}{\ensuremath{\subseteq}}
\newunicodechar{⊂}{\ensuremath{\subset}}
\newunicodechar{∪}{\ensuremath{\cup}}
\newunicodechar{⋓}{\ensuremath{\Cup}}
\newunicodechar{∉}{\ensuremath{\not\in}}
\newunicodechar{√}{\ensuremath{\sqrt}}

\newunicodechar{⊸}{\ensuremath{\multimap}}
\newunicodechar{⊗}{\ensuremath{\otimes}}
\newunicodechar{⨂}{\ensuremath{\bigotimes}}
\newunicodechar{⊕}{\ensuremath{\oplus}}
\newunicodechar{〈}{\ensuremath{\langle}}
\newunicodechar{⟨}{\ensuremath{\langle}}
\newunicodechar{⟩}{\ensuremath{\rangle}}
\newunicodechar{〉}{\ensuremath{\rangle}}
\newunicodechar{¡}{\ensuremath{\upsidedownbang}}
\newunicodechar{∘}{\ensuremath{\circ}}
\newunicodechar{†}{\ensuremath{\dagger}}
\newunicodechar{⊤}{\ensuremath{\top}}
\newunicodechar{⊥}{\ensuremath{\bot}}

\newunicodechar{〚}{\ensuremath{\llbracket}}
\newunicodechar{〛}{\ensuremath{\rrbracket}}

\usepackage{etoolbox} 
\newtoggle{comments}
\toggletrue{comments}

\iftoggle{comments}{
  \newcommand{\fixme}[1]{\textbf{\textcolor{red}{[ Fixme: #1]}}}
  \newcommand{\todo}[1]{\textbf{\textcolor{green}{[ TODO: #1 ]}}}
  \newcommand{\rnr}[1]{\textbf{\textcolor{blue}{[ Robert: #1 ]}}}
  \newcommand{\mwh}[1]{\textbf{\textcolor{olive}{[ Mike: #1 ]}}}
  \newcommand{\khh}[1]{\textbf{\textcolor{orange}{[ Kesha: #1 ]}}}
  \newcommand{\shh}[1]{\textbf{\textcolor{purple}{[ Shih-Han: #1 ]}}}
  \newcommand{\xwu}[1]{\textbf{\textcolor{purple}{[ Xiaodi: #1 ]}}}
  \newcommand{\oth}[2]{\textbf{\textcolor{red}{[ #1: #2 ]}}}
}{
  \newcommand{\fixme}[1]{}
  \newcommand{\todo}[1]{}
  \newcommand{\rnr}[1]{}
  \newcommand{\mwh}[1]{}  
  \newcommand{\khh}[1]{}
  \newcommand{\shh}[1]{}
  \newcommand{\xwu}[1]{}
  \newcommand{\oth}[2]{}
}

\newtoggle{submission}
\togglefalse{submission}

\iftoggle{submission}{
  \newcommand{\aref}[1]{the full version of this paper \cite{VOQC}}
}{
  \newcommand{\aref}[1]{\Cref{#1}}
}

\usepackage{enumitem}
\usepackage[shortcuts]{extdash}
\usepackage{microtype}
\usepackage[utf8]{inputenc}
\usepackage[T1]{fontenc}

\begin{document}

\title{A Verified Optimizer for Quantum Circuits}

\def\titlerunning{A Verified Optimizer for Quantum Circuits}
\def\authorrunning{K. Hietala, R. Rand, S. Hung, X. Wu \& M. Hicks}

\author{Kesha Hietala}
\affiliation{
  \institution{University of Maryland}
  \country{USA}
}
\email{kesha@cs.umd.edu}

\author{Robert Rand}
\affiliation{
 \institution{University of Chicago}
 \country{USA} 
}
\email{rand@uchicago.edu}

\author{Shih-Han Hung}
\affiliation{
 \institution{University of Maryland}
 \country{USA} 
}
\email{shung@cs.umd.edu}

\author{Xiaodi Wu}
\affiliation{
 \institution{University of Maryland}
 \country{USA} 
}
\email{xwu@cs.umd.edu}

\author{Michael Hicks}
\affiliation{
 \institution{University of Maryland}
 \country{USA} 
}
\email{mwh@cs.umd.edu}

\begin{abstract}

We present \voqc, the first fully \emph{verified optimizer for quantum circuits}, written using the Coq proof assistant. Quantum circuits are expressed as programs in a simple, low-level language called \sqir, a \emph{simple quantum intermediate representation}, which is deeply embedded in Coq. Optimizations and other transformations are expressed as Coq functions, which are proved correct with respect to a semantics of \sqir programs. 
\sqir uses a semantics of matrices of complex numbers, which is the standard for quantum computation, but treats matrices symbolically in order to reason about programs that use an arbitrary number of quantum bits. 
\sqir's careful design and our provided automation make it possible to write and verify a broad range of optimizations in \voqc,
including full-circuit transformations from cutting-edge optimizers.

\end{abstract}

\ccsdesc[300]{Hardware~Quantum computation}
\ccsdesc[300]{Hardware~Circuit optimization}
\ccsdesc[500]{Software and its engineering~Formal software verification}

\keywords{Formal Verification, Quantum Computing, Circuit Optimization, Certified Compilation, Programming Languages}

\maketitle

\renewcommand{\shortauthors}{K. Hietala, R. Rand, S. Hung, X. Wu \& M. Hicks}


\section{Introduction}

Programming quantum computers will be challenging, at least in the near term. Qubits will be scarce and gate pipelines will need to be short to prevent decoherence. Fortunately, optimizing compilers can transform a source algorithm to work with fewer resources. Where compilers fall short, programmers can optimize their algorithms by hand. 

Of course, both compiler and by-hand optimizations will inevitably have bugs. 
As evidence of the former, \citet{Kissinger2019} discovered mistakes in the optimized outputs produced by 
the circuit optimizer of \citet{Nam2018}, and \citeauthor{Nam2018} themselves found that the optimization library they compared against (\citet{Amy2013}) sometimes produced incorrect results. 
Likewise, \citet{Amy2018} discovered an optimizer they had recently developed produced buggy results~\cite{Amy2018b}.
Making mistakes when optimizing by hand is also to be expected: as put well by \citet{Zamdzhiev16talk}, quantum computing can be frustratingly unintuitive.

Unfortunately, the very factors that motivate optimizing quantum compilers make it difficult to test their correctness.
Comparing runs of a source program to those of its optimized version is often impractical due to the indeterminacy of typical quantum algorithms and the substantial expense involved in executing or simulating them. Indeed, resources may be too scarce, or the qubit connectivity too constrained, to run the program without optimization!

An appealing solution to this problem is to apply rigorous \emph{formal methods} to prove that an optimization or algorithm always does what it is intended to do. For example, CompCert~\cite{compcert} is a compiler for C programs that is written and proved correct using the Coq proof assistant~\cite{coq}. CompCert includes sophisticated optimizations whose proofs of correctness are verified to be valid by Coq's type checker. 

In this paper, we apply CompCert's approach to the quantum setting.
We present \voqc (pronounced ``vox''), a \emph{verified optimizer for quantum circuits}. \voqc takes as input a quantum program written in a language we call \sqir (``squire''). \sqir is designed to be a \emph{small quantum intermediate representation}, but it is suitable for source-level programming too: it is not very different from languages such as Quil~\cite{Smith2016} or OpenQASM~\cite{Cross2017},
which describe quantum programs as circuits. \sqir is deeply embedded in Coq, similar to how Quil is embedded in Python via PyQuil~\cite{Pyquil}, allowing us to write sophisticated quantum programs. 
\voqc applies a series of optimizations to \sqir programs, ultimately producing a result that is compatible with a specified quantum architecture. For added convenience, \voqc provides translators between \sqir and OpenQASM\@. (\Cref{sec:overview}.)

We designed \sqir to make it as easy as possible to reason about the semantics of quantum programs, which are significantly different from the semantics of classical programs. For example, while in a classical program one can reason about different variables independently, the phenomenon of quantum entanglement requires us to reason about a global quantum state, typically represented as a large vector or matrix of complex numbers. 
This means that reasoning about quantum states involves linear algebra, and often trigonometry and probability too.
As a result, existing approaches to program proofs in Coq tend not to apply in the quantum setting. Indeed, we first attempted to build \voqc using \qwire~\cite{Paykin2017}, which is also embedded in Coq and more closely resembles a classical programming language, but found proofs of even simple optimizations to be non-trivial and hardly scalable. 

To address these challenges, \sqir's design has several key features (\Cref{sec:sqire}). First, it uses natural numbers in place of variables so that we can naturally index into the vector or matrix state. Using variables directly (e.g., with higher-order abstract syntax~\cite{Pfenning1988}, as in \qwire and Quipper~\cite{Green2013}) necessitates a map from variables to indices, which we find confounds proof automation.
Second, \sqir provides two semantics for quantum programs. We express the semantics of a general program as a function between \emph{density matrices}, as is standard (e.g., in QPL~\cite{Selinger2004} and \qwire), since density matrices can represent the \emph{mixed states} that arise when a program applies a measurement operator (\Cref{sec:general-sqire}). However, measurement typically occurs at the end of a computation, rather than within it, so we also provide a simpler \emph{unitary semantics} for (sub\=/)programs that do not measure their inputs. In this case, a program's semantics corresponds to a restricted class of square matrices. These matrices are often much easier to work with, especially when employing automation. 
Other features of \sqir's design, like assigning an ill-typed program the denotation of the zero-matrix, are similarly intended to ease proof.
Pleasantly, unitary \sqir even turns out to be effective for proving quantum programs correct. This paper presents a proof of correctness of \emph{GHZ state preparation}~\cite{Greenberger1989}; in concurrent work~\cite{CPPsub}, we have proved the correctness of implementations of \emph{Quantum Phase Estimation} (a key component of Shor's prime factoring algorithm~\citeyearpar{Shor94}), \emph{Grover's search} algorithm~\citeyearpar{Grover1996}, and \emph{Simon's} algorithm~\citeyearpar{Simon1994}.

At the core of \voqc is a framework for writing transformations of \sqir programs and verifying their correctness. To ensure that the framework is suitably expressive, we have used it to develop verified versions of a variety of optimizations. Many are based on those used in an optimizer developed by \citet{Nam2018}, which is the best performing optimizer we know of in terms of total gate reduction (per experiments described below). We abstract these optimizations into a couple of different classes, and provide library functions, lemmas, and automation to simplify their construction and proof. 
%
We have also verified a circuit mapping routine that transforms \sqir programs to satisfy constraints on how qubits may interact on a specified target architecture. (\Cref{sec:voqc}.)

We evaluated the quality of the optimizations we verified in \voqc, and by extension the quality of our framework, by measuring how well it optimizes a set of benchmark programs, compared to \citeauthor{Nam2018} and several other optimizing compilers. The results are encouraging. 
On a benchmark of 28 circuit programs developed by \citet{Amy2013} we find that \voqc reduces total gate count on average by 17.8\% compared to 10.1\% for IBM's Qiskit compiler~\cite{Qiskit2019}, 10.6\% for CQC's \tket~\cite{tket}, and 24.8\% for the cutting-edge research optimizer by \citet{Nam2018}.
On the same benchmarks, \voqc reduces $T$-gate count (an important measure when considering fault tolerance) on average by 41.4\% compared to 39.7\% by \citet{Amy2013}, 41.4\% by \citeauthor{Nam2018}, and 42.6\% by the PyZX optimizer.
Results on an even larger benchmark suite (detailed in \aref{app:extended-eval}) tell the same story. In sum, \voqc and \sqir are expressive enough to verify a range of useful optimizations, yielding performance competitive with standard compilers. (\Cref{sec:experiments}.) 

\voqc is the first fully verified optimizer for general quantum programs.
\citet{amy18reversible} developed a verified optimizing compiler from source Boolean expressions to reversible circuits and \citet{Fagan2018} verified an optimizer for ZX-diagrams representing Clifford circuits; however, neither of these tools handle general quantum programs. 
In concurrent work, \citet{Shi2019} developed CertiQ, which uses symbolic execution and SMT solving to verify circuit transformations in the Qiskit compiler.
CertiQ is limited to verifying correct application of local equivalences and does not provide a way to describe general quantum states (a key feature of \sqir), which limits the types of optimizations that it can reason about.
This also means that it cannot be used as a tool for verifying general quantum programs.
\citet{Smith2019} presented a compiler with built-in translation validation via QMDD equivalence checking \cite{Miller2006}.
However, QMDDs represent quantum state concretely, which means that the validation time will increase exponentially with the number of qubits in the compiled program.
In contrast to these, \sqir represents matrices \emph{symbolically}, which allows us to reason about arbitrary quantum computation and verify interesting, non-local optimizations, independently of the number of qubits in the optimized program. 
(\Cref{sec:related}.)

Our work on \voqc and \sqir are steps toward a broader goal of developing a full-scale verified compiler toolchain. Next steps include developing certified transformations from higher-level quantum languages to \sqir and implementing optimizations with different objectives, e.g., that aim to reduce the probability that a result is corrupted by quantum noise.
All code we reference in this paper can be found online at \url{https://github.com/inQWIRE/SQIR}.


\section{Overview}
\label{sec:overview}

We begin with a brief background on quantum programs, focusing on the challenges related to formal verification. We then provide an overview of \voqc and \sqir, summarizing how they address these challenges.

\subsection{Preliminaries}
\label{sec:basics}

Quantum programs operate over \emph{quantum states}, which consist of one or more \emph{quantum bits} (a.k.a. \emph{qubits}). A single qubit is represented as a vector of complex numbers
$\langle\alpha, \beta\rangle$ such that $|\alpha|^2 + |\beta|^2 = 1$. The vector $\langle 1,0 \rangle$ represents the state $\ket{0}$ while vector $\langle 0,1 \rangle$ represents the state $\ket{1}$. A state written $\ket{\psi}$ is called a \emph{ket}, following Dirac's notation.
 We say a qubit is in a \emph{superposition} of $\ket{0}$ and $\ket{1}$ when both $\alpha$ and $\beta$ are non-zero. Just as Schrodinger's cat is both dead and alive until the box is opened, a qubit is only in superposition until it is \emph{measured}, at which point the outcome will be $0$ with probability $|\alpha|^2$ and $1$ with probability $|\beta|^2$. Measurement is not passive: it has the effect of collapsing the state to match the measured outcome, i.e., either $\ket{0}$ or $\ket{1}$. As a result, all subsequent measurements return the same answer.

Operators on quantum states are linear mappings. These mappings can be expressed as matrices, and their application to a state expressed as matrix multiplication. For example, the \emph{Hadamard} operator $H$ is expressed as a matrix 
$\frac{1}{\sqrt{2}}\begin{psmallmatrix} 1 & 1 \\ 1 & -1 \end{psmallmatrix}$.
Applying $H$ to state $\ket{0}$ yields state $\langle \frac{1}{\sqrt{2}}, \frac{1}{\sqrt{2}} \rangle$, also written as $\ket{+}$. 
Many quantum operators are not only linear, they are also \emph{unitary}---the conjugate transpose (or adjoint) of their matrix is its own inverse. This ensures that multiplying a qubit by the operator preserves the qubit's sum of norms squared.
Since a Hadamard is its own adjoint, it is also its own inverse: hence $H\ket{+} = \ket{0}$.

\begin{wrapfigure}{R}{.18\textwidth}
$\begin{pmatrix} 
1 & 0 & 0 & 0 \\ 
0 & 1 & 0 & 0 \\ 
0 & 0 & 0 & 1 \\ 
0 & 0 & 1 & 0 
\end{pmatrix}$
\end{wrapfigure}
A quantum state with $n$ qubits is represented as vector of length $2^n$. For example, a 2-qubit state is represented as a vector $\langle\alpha, \beta, \gamma, \delta\rangle$ where each component corresponds to (the square root of) the probability of measuring $\ket{00}$, $\ket{01}$, $\ket{10}$, and $\ket{11}$, respectively. Because of the exponential size of the complex quantum state space, it is not possible to simulate a 100-qubit quantum computer using even the most powerful classical computer!

$n$-qubit operators are represented as $2^n \times 2^n$ matrices. For example, the $\mathit{CNOT}$ operator over two qubits is expressed as the matrix shown at the right.
It expresses a \emph{controlled not} operation---if the first qubit (called the \emph{control}) is $\ket{0}$ then both qubits are mapped to themselves, but if the first qubit is $\ket{1}$ then the second qubit (called the \emph{target}) is negated, e.g., $\mathit{CNOT}\ket{00} = \ket{00}$ while $\mathit{CNOT}\ket{10} = \ket{11}$. 

$n$-qubit operators can be used to create \emph{entanglement}, which is a situation where two qubits cannot be described independently. For example,
while the vector $\langle1, 0, 0, 0\rangle$ can be written as $\langle 1, 0 \rangle \otimes \langle 1, 0 \rangle$ where $\otimes$ is the tensor product, the state $\langle \frac{1}{\sqrt{2}}, 0 , 0, \frac{1}{\sqrt{2}}\rangle$ cannot be similarly decomposed. We say that $\langle \frac{1}{\sqrt{2}}, 0 , 0, \frac{1}{\sqrt{2}}\rangle$ is an entangled state.

An important non-unitary quantum operator is \emph{projection} onto a subspace. For example, $\op{0}{0}$ (in matrix notation $\begin{psmallmatrix} 1 & 0 \\ 0 & 0 \end{psmallmatrix}$) projects a qubit onto the subspace where that qubit is in the $\ket{0}$ state. Projections are useful for describing quantum states after measurement has been performed. 
We sometimes use $\vert i \rangle_q \langle i \vert$ as shorthand for applying the projection $\ket{i}\bra{i}$ to qubit $q$ and an identity operation to every other qubit in the state.

\subsection{Quantum Circuits}
\label{sec:circuits}

\begin{figure}[t]
\centering
\begin{subfigure}[b]{.2\textwidth}
  \small
  \Qcircuit @C=0.5em @R=0.5em {
    \lstick{\ket{0}} & \gate{H} & \ctrl{1} & \qw & \qw \\
    \lstick{\ket{0}} & \qw & \targ & \ctrl{1} & \qw \\
    \lstick{\ket{0}} & \qw & \qw & \targ & \qw
    }
\par\vspace{0.7cm}
\caption{Quantum Circuit}
\end{subfigure} \qquad
\begin{subfigure}[b]{.15\textwidth}
\begin{lstlisting}
H 0
CNOT 0 1
CNOT 1 2
\end{lstlisting}
\par\vspace{0.6cm}
\caption{Quil}
\end{subfigure}
\begin{subfigure}[b]{.45\textwidth}
\begin{lstlisting}[language=python]
def ghz_state(qubits):
  program = Program()
  program += H(qubits[0])
  for q1,q2 in zip(qubits, qubits[1:]):
    program += CNOT(q1, q2)
  return program
\end{lstlisting}
\caption{PyQuil (arbitrary number of qubits)}
\end{subfigure}
\caption{Example quantum program: GHZ state preparation}
\label{fig:circuit-example}
\end{figure}
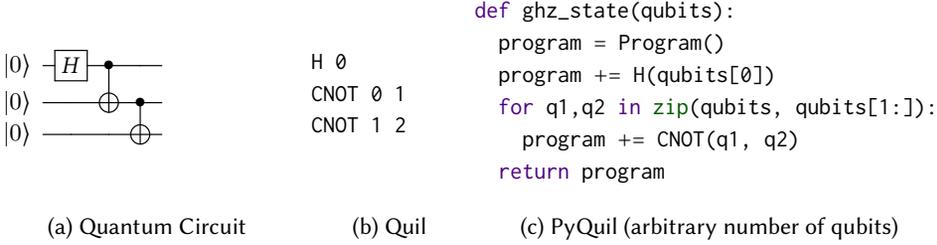

Quantum programs are typically expressed as circuits, as shown in \Cref{fig:circuit-example}(a). In these circuits, each horizontal wire represents a 
\emph{qubit} and boxes on these wires indicate quantum operators, or \emph{gates}. Gates can either be unitary operators (e.g., Hadamard, $\mathit{CNOT}$) or non-unitary ones (e.g., measurement). In software, quantum circuit programs are often represented using lists of instructions that describe the different gate applications. For example, \Cref{fig:circuit-example}(b) is the Quil~\cite{Smith2016} representation of the circuit in \Cref{fig:circuit-example}(a).

In the \emph{QRAM model}~\cite{Knill1996} quantum computers are used as co-processors to classical computers. The classical computer generates descriptions of circuits to send to the quantum computer and then processes the measurement results. High-level quantum programming languages are designed to follow this model. For example, \Cref{fig:circuit-example}(c) shows a program in PyQuil~\cite{Pyquil}, a quantum programming framework embedded in Python. The \lstinline{ghz_state} function takes an array \lstinline{qubits} and constructs a circuit that prepares the Greenberger-Horne-Zeilinger (GHZ) state \cite{Greenberger1989}, which is an $n$-qubit entangled quantum state of the form
\begin{align*}
    \ket{\text{GHZ}^n} = \frac{1}{\sqrt{2}}(\ket{0}^{\otimes n}+\ket{1}^{\otimes n}).
\end{align*}
Calling \lstinline{ghz_state([0,1,2])} returns the Quil program in \Cref{fig:circuit-example}(b), which produces the quantum state $\frac{1}{\sqrt{2}}(\ket{000}+\ket{111})$. The high-level language may provide facilities to optimize constructed circuits, e.g., to reduce gate count, circuit depth, and qubit usage. It may also perform transformations to account for hardware-specific details like the number of qubits, available set of gates, or connectivity between physical qubits.

\subsection{\sqir: A Small Quantum Intermediate Representation Supporting Verification}
\label{sec:sqir-overview}

What if we want to formally verify that \lstinline{ghz_state}, when passed an array of indices \lstinline{[0, ...,}$n-1$\lstinline{]}, returns a circuit that produces the quantum state $\ket{\text{GHZ}^n}$? What steps are necessary?

First, we need a way to formally define quantum states as matrices of complex numbers. Indeed, we need a way to define indexed \emph{families} of states---$\ket{\text{GHZ}^n}$ is a function from an index $n$ to a quantum state. Second, we need a formal language in which to express quantum programs; to this language we must ascribe a mathematical semantics in terms of quantum states. A program like \lstinline{ghz_state} is a function from an index (a list of length $n$) to a circuit (of size $n$), and this circuit's denotation is its equivalent (unitary) matrix (of size $2^n$). Finally, we need a way to mechanically reason that, for arbitrary $n$, the semantics of \lstinline{ghz_state([0,1,...,}$n-1$\lstinline{])} applied to the zero state ($\ket{0}^{\otimes n}$) is equal to the state $\ket{\text{GHZ}^n}$.

We designed \sqir, a \emph{small quantum intermediate representation}, to do all of these things. \sqir is a simple circuit-oriented language deeply embedded in the Coq proof assistant in a manner similar to how Quil is embedded in Python via PyQuil. We use \sqir's host language, Coq, to define the syntax and semantics of \sqir programs and to express properties about quantum states. We developed a library of lemmas and tactic-based automation to assist in writing proofs about quantum programs; such proofs make heavy use of complex numbers and linear algebra. These proofs are aided by isolating \sqir's \emph{unitary core} from primitives for measurement, which require consideration of probability distributions of outcomes (represented as \emph{density matrices}); this means that (sub-)programs that lack measurement can have simpler proofs. Either way, in \sqir we perform reasoning \emph{symbolically}. For example, we can prove that every circuit generated by the \sqir-equivalent of \lstinline{ghz_state} produces the expected state $\ket{\text{GHZ}^n}$ when applied to input lists of length $n$, \emph{for any} $n$. 

\sqir is implemented in just over 3500 lines of Coq. 
We started with Coq libraries for complex numbers and matrices developed for the \qwire language \cite{Paykin2017}; over the course of our work we have extended these libraries with around 3000 lines of code providing more automation for linear algebra and better support for complex phases.
We present \sqir's syntax and semantics along with an example program and verified property of correctness in \Cref{sec:sqire}.

\subsection{\voqc: A Verified Optimizer for Quantum Circuits}

\begin{figure}[t]
\centering
\includegraphics[width=0.5\textwidth]{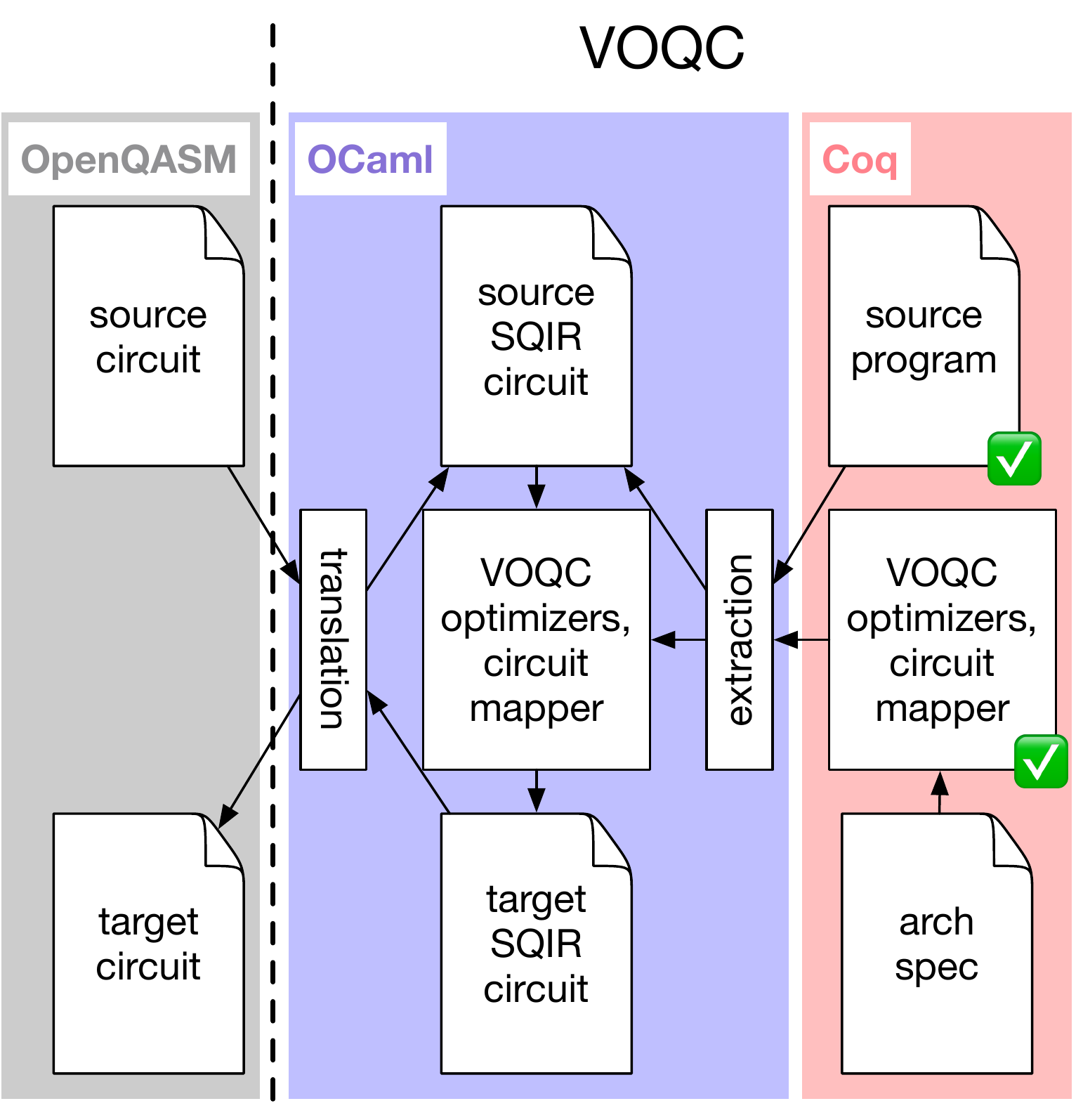}
\caption{The \voqc architecture. Input circuits can be described in OpenQASM, OCaml, or Coq (top). The verified \voqc optimizer is extracted to an executable OCaml optimizer (middle), which can produce an optimized \sqir or OpenQASM circuit (bottom). The dashed line indicates the separation between the standard quantum compiler stack (left) and our contribution (right).}
\label{fig:arch}
\end{figure}

While \sqir is suitable for proving correctness properties about source programs like \coqe{ghz_state}, its primary use has been as the intermediate representation of \voqc, our verified optimizer for quantum circuits, and the signature achievement of this paper. An optimizer is a function from programs to programs, with the intention that the output program has the same semantics as the input. In \voqc, we prove this is always the case: a \voqc optimization $f$ is a Coq function over \sqir circuit $C$, and we prove that the semantics of input circuit $C$ is always equivalent to the semantics of the output $f(C)$. 

The \voqc approach stands in contrast to prior work that relies on translation validation \cite{Smith2019, Kissinger2019, Amy2018}, which may fail to identify latent bugs in the optimizer, while adding compile-time overhead. 
By proving correctness with respect to an explicit semantics for input/output programs (i.e., that of \sqir), \voqc optimizations are flexible in their expression. Prior work has been limited to peephole optimizations \cite{Shi2019}, leaving highly effective, full-circuit optimizations we have proved correct in \voqc out of reach. Such global (non-peephole) proofs are aided by the design of \sqir (notably, the isolation of a unitary core) and accompanying proof automation.

The structure of \voqc is summarized in \Cref{fig:arch}. The \voqc transformations themselves are shown at the middle right, and are described in~\Cref{sec:voqc,sec:nonunitary-opt}. In addition to performing circuit optimizations, \voqc also performs \emph{circuit mapping}, transforming a \sqir program to an equivalent one that respects constraints imposed by the target architecture. Once again, we prove that it does so correctly (\Cref{sec:circuitmap}).

Using Coq's standard code extraction mechanism, we can extract \voqc into a standalone OCaml program. This program takes as input a \sqir program in an OCaml representation. This input can be extracted from a Coq-hosted (and proved correct) \sqir program (upper right), or from a program expressed in OpenQASM~\cite{Cross2017}, a standard representation for quantum circuits (upper left). Since a number of quantum programming frameworks, including Qiskit~\cite{Qiskit2019}, t$\vert$ket$\rangle$~\cite{tket}, Project Q~\cite{Steiger2018} and Cirq~\cite{Cirq}, can output OpenQASM, this allows us to run \voqc on a variety of generated circuits, without requiring the user to program in OCaml or Coq.

\voqc is implemented in about 7200 lines of Coq, with roughly 2100 lines for circuit mapping, 2100 lines for general-purpose \sqir program manipulation, 2200 lines for unitary program optimizations, and 800 lines for non-unitary program optimizations. We use about 400 lines of standalone OCaml code for running \voqc on our benchmarks in \Cref{sec:experiments}.
We started work on \sqir and \voqc in March 2019 and concluded work for this paper in May 2020. The majority of the optimizations were implemented and verified within a span of four months.


\section{\sqir: A Small Quantum Intermediate Representation}
\label{sec:sqire}
\label{sec:unitary}

Here we present the syntax and semantics of \sqir, a \emph{small quantum intermediate representation}. The \sqir language is composed of two parts: a core language of unitary operators and a full language that incorporates measurement. This section focuses on the former; \Cref{sec:general-sqire} presents the latter. 

As we will show, the semantics of a unitary \sqir program is expressed directly as a matrix, 
in contrast to the full \sqir, which treats programs as functions over density matrices.
This matrix semantics greatly simplifies proofs, both of the correctness of unitary optimizations (the bulk of \voqc) and of source programs, many of which are essentially unitary (measurement is the very last step). Other aspects of \sqir's design also make proofs easier, as we will discuss and demonstrate with an example at the end of the section.

\subsection{Unitary \sqir: Syntax}
\label{sec:sqir-syntax}

A unitary \sqir program $U$ is a sequence of applications of gates $G$ to qubits $q$.
\begin{align*}
    U~:=&~~U_1 ;~U_2~ \vert~G~ q~ \vert~G~ q_1~ q_2
\end{align*}
Qubits are referred to by natural numbers that index into a \emph{global register} of quantum bits.  
Each \sqir program is parameterized by a set of unitary one- and two-qubit gates (from which $G$ is drawn) and the dimension of the global register (i.e., the number of available qubits). In Coq, a unitary \sqir program 
\coqe{U} has type \coqe{ucom g n}, where \coqe{g} identifies the gate set and \coqe{n} is the size of the global register.

\begin{wrapfigure}{R}{.5\textwidth}
\begin{coq}
Fixpoint ghz (n : nat) : ucom base n :=
  match n with
  | 0 => I 0
  | 1 => H 0
  | S n' => ghz n'; CNOT (n'-1) n'
  end.
\end{coq}
\end{wrapfigure}
As an example, consider the program to the right, which is equivalent to PyQuil's \lstinline{ghz_state} from \Cref{fig:circuit-example}(c).
 The Coq function \coqe{ghz} recursively constructs a \sqir program, i.e., a Coq value of type \coqe{ucom base n}. This program, when run, prepares the GHZ state.
When $n$ is $0$, \coqe{ghz} produces a \sqir program that is just the identity gate $I$ applied to qubit $0$. When $n$ is $1$, the result is the Hadamard gate $H$ applied to qubit $0$. When $n$ is greater than $1$, \coqe{ghz} constructs the program $U_1; U_2$, where $U_1$ is the \coqe{ghz} circuit on \coqe{n'} (i.e., $n-1$) qubits, and $U_2$ is the appropriate $\mathit{CNOT}$ gate. The result of \coqe{ghz 3} is equivalent to the circuit shown in \Cref{fig:circuit-example}(a).

\subsection{Semantics}

\begin{figure}[t]
  \centering
  \begin{align*}
    \denote{U_1;~U_2}_d&=~\denote{U_2}_d \times \denote{U_1}_d \\
    \denote{G_1~ q}_d&=\begin{cases}
                          apply_1(G_1,~q,~d) &\text{well-typed} \\
                          0_{2^{d}} &\text{otherwise}
                          \end{cases} \\
    \denote{G_2~ q_1~ q_2}_d&=\begin{cases}
                          apply_2(G_2,~q_1,~q_2,~d) &\text{well-typed} \\
                          0_{2^{d}} &\text{otherwise}
                          \end{cases} 
  \end{align*}
  
\caption{Semantics of unitary \sqir programs, assuming a global register of dimension $d$. The $apply_k$ function maps a gate name to its corresponding unitary matrix and extends the intended operation to the given dimension by applying an identity operation on every other qubit in the system.}
  \label{fig:sqire-semantics}
\end{figure}

Suppose that $M_1$ and $M_2$ are the matrices corresponding to unitary gates $U_1$ and $U_2$, which we want to apply to a quantum state vector $\ket{\psi}$. Matrix multiplication is associative, so $M_2(M_1 \ket{\psi})$ is equivalent to $(M_2 M_1) \ket{\psi}$. Moreover, multiplying two unitary matrices yields a unitary matrix. As such, the semantics of \sqir program $U_1;~ U_2$ is naturally described by the unitary matrix $M_2 M_1$.

This semantics is shown in \Cref{fig:sqire-semantics}. There are two things to notice. First, if a program is not \emph{well-typed} its denotation is the zero matrix (of size $2^d \times 2^d$). A program $U$ is well-typed if every gate application is \emph{valid}, meaning that its index arguments are within the bounds of the global register, and no index is repeated.
The latter requirement enforces linearity and thereby quantum mechanics' \emph{no-cloning theorem}, which says that it is impossible to create a copy of an arbitrary quantum state.

Otherwise, the program's denotation follows from the composition of the matrices that correspond to each of the applications of its unitary gates, $G$. The only wrinkle is that a full program consists of many gates, each operating on 1 or 2 of the total qubits; thus, a gate application's matrix needs to apply the identity operation to the qubits not being operated on. This is what $apply_1$ and $apply_2$ do.  For example, $apply_1(G_u,~q,~d) = I_{2^q} \otimes u \otimes I_{2^{(d - q - 1)}}$ where $u$ is the matrix interpretation of the gate $G_u$ and $I_k$ is the $k \times k$ identity matrix. The $apply_2$ function requires us to decompose the two-qubit unitary into a sum of tensor products: for instance, $\mathit{CNOT}$ can be written as $\ket{0}\bra{0} \otimes I_2 + \ket{1}\bra{1} \otimes \sigma_x$ where $\sigma_x = \begin{psmallmatrix} 0 & 1 \\ 1 &0 \end{psmallmatrix}$. We then have \[apply_2(\mathit{CNOT},~q_1,~q_2,~d) = I_{2^{q_1}} \otimes \ket{0}\bra{0} \otimes I_{2^r} \otimes I_2 \otimes I_{2^s} + I_{2^{q_1}} \otimes \ket{1}\bra{1} \otimes I_{2^r} \otimes \sigma_x \otimes I_{2^s} \]
where $r = q_2 - q_1 - 1$ and $s = d - q_2 - 1$, assuming $q_1 < q_2$.

In our development we define the semantics of \sqir programs over the gate set $G \in \{R_{\theta, \phi, \lambda},~ \mathit{CNOT}\}$ where $R_{\theta, \phi, \lambda}$ is a general single-qubit rotation parameterized by three real-valued rotation angles and $\mathit{CNOT}$ is the standard two-qubit controlled-not gate. This is our \emph{base set} of gates. It is the same as the underlying set used by OpenQASM \cite{Cross2017} and is \emph{universal}, meaning that it can approximate any unitary operation to within arbitrary error.
The matrix interpretation of the single-qubit $R_{\theta, \phi, \lambda}$ gate is
\[\begin{pmatrix} 
    \cos(\theta/2) & -e^{i\lambda}\sin(\theta/2) \\
    e^{i\phi}\sin(\theta/2) & e^{i(\phi+\lambda)}\cos(\theta/2) 
  \end{pmatrix}\]
and the matrix interpretation of the $\mathit{CNOT}$ gate is given in \Cref{sec:basics}.

Common single-qubit gates can be defined in terms of $R_{\theta, \phi, \lambda}$. For example, the two single-qubit gates used in our GHZ example---identity $I$ and Hadamard $H$---are respectively defined as $R_{0, 0, 0}$ and $R_{\pi/2, 0, \pi}$. The Pauli $X$ ("NOT") gate is $R_{\pi, 0, \pi}$ and the Pauli $Z$ gate is $R_{0, 0, \pi}$.
We can also define more complex operations as \sqir programs.
For example, the $\mathit{SWAP}$ operation, which swaps two qubits, can be defined as a sequence of three $\mathit{CNOT}$ gates.

\subsection{Design for Proofs}
\label{sec:discussion}

We designed \sqir's unitary core to simplify formal proofs, in three ways. 

\paragraph{Zero Matrix for Ill-typed Programs} By giving a denotation of the zero matrix to ill-typed gate applications (and thereby ill-typed programs), we do not need to explicitly assume or prove that a program is well-typed in order to state a property about its semantics, thereby removing clutter from theorems and proofs.
For example, in our proof of the \coqe{ghz} program below we do not need to explicitly prove that \coqe{ghz n} is well-typed (although this is true). 

\paragraph{Phantom Types for Matrix Indices} We do not use dependent types to represent matrices in the semantics. Following \citet{Rand2017,Rand2018a}, we define matrices as functions from pairs of natural numbers to complex numbers. 
\begin{coq}
Definition Matrix (m n : nat) := nat -> nat -> \C.
\end{coq}
The arguments \coqe{m} and \coqe{n}, which are the dimensions of the matrix, are \emph{phantom types}---they do not appear in the definition. These phantom types are useful to define certain operations on matrices that depend on these dimensions, such as the tensor product and matrix multiplication. However, there is no proof burden internal to the matrices themselves. Instead, it is possible to show a matrix is well-formed within its specified bounds by means of an external predicate:
\begin{coq}
Definition WF_Matrix {m n} (M : Matrix m n) : P := $\forall$ i j, i $\geq$ m $\vee$ j $\geq$ n -> M i j = 0.
\end{coq}
Phantom types occupy a convenient middle ground in allowing information to be stored in the types, while pushing the majority of the work to external predicates.
For instance, we can define $\ket{i}^{\otimes n}$ (for $i \in \{0,1\}$) recursively as $\ket{i} \otimes \ket{i} \otimes \dots \otimes \ket{i}$, with $n$ repetitions. Coq has no way of inferring a type for this, so we declare that is has type \coqe{Vector 2^n}, and Coq will allow us to use it in any context where a vector is expected. However, before using it in rewrite rules like $I_{2^n} \times \ket{i}^{\otimes n} = \ket{i}^{\otimes n}$ (which says that multiplication by the identity matrix is an identity operation), we will need to show that $\ket{i}^{\otimes n}$ is a well-formed vector of length $2^n$, for which we provide convenient automation.

\paragraph{Qubits are Concrete, Not Abstract, Indices}

When we first set out to build \voqc, we thought to do it using \qwire~\cite{Paykin2017}, another formally verified quantum programming language embedded in Coq. However, we were surprised to find that we had tremendous difficulty proving that even simple transformations were correct. This experience led to the development of \sqir, and raised the question: Why does \sqir seem to make proofs easier, and what do we lose by using it rather than \qwire?

The fundamental difference between \sqir and \qwire is that \sqir programs use \emph{concrete (numeric) indices into a global register} to refer to qubits. As such, the semantics can naturally map qubits to rows and columns in the denoted matrix. In addition, qubit disjointness in a \sqir program is obvious---$G_1~m$ operates on a different qubit than $G_2~n$ when $m \not= n$. Both elements are important for easily proving equivalences, e.g., that gates acting on disjoint qubits commute (a property that allows us to reason about gates acting on different parts of the circuit in isolation). 

In \qwire, variables are implemented using higher-order abstract syntax~\cite{Pfenning1988} and refer to \emph{abstract} qubits. This approach eases programmability---larger circuits can be built by composing smaller ones, connecting inputs and outputs by normal variable binding, indifferent to the physical identity of a qubit. This approach is also used in the language Quipper~\cite{Green2013}.
However, we find that this approach complicates formal proof. To denote the semantics of a program that uses abstract qubits requires deciding how abstract qubits will be represented concretely, as rows and columns in the denotation matrix. Reasoning about this translation can be laborious, especially for recursive circuits and those that allocate and deallocate qubits (entailing de Bruijn-style index shifting~\cite{RandThesis}). Moreover, notions like disjointness are no longer obvious---$G_1~x$ and $G_2~y$ for variables $x \not= y$ may not be disjoint if $x$ and $y$ could be allocated to the same concrete qubit. 

From a proof-engineering standpoint all of the above benefits have been pivotal in allowing our proofs to scale up. 
\iftoggle{submission}{
  The full version of this paper \cite{VOQC}
}{
  \Cref{app:sqir-v-qwire}
}
presents a detailed comparison of \sqir and \qwire, exploring the tradeoffs of concrete versus abstract qubits at a lower level.

\begin{wrapfigure}{R}{.5\textwidth}
\begin{coq}
Definition GHZ (n : nat) : Vector (2 ^ n) :=
  match n with 
  | 0    => I 1 
  | S n' => $\frac{1}{\sqrt{2}}$ * $\ket{0}^{\otimes n}$ + $\frac{1}{\sqrt{2}}$ * $\ket{1}^{\otimes n}$
  end.
\end{coq}
\end{wrapfigure}

\subsection{Source-program Proofs}
\label{sec:source-proofs}

This paper focuses on \sqir's use in proving circuit optimizations correct, but \sqir was designed to support source-program proofs too. As an illustration, we present a \sqir proof of correctness for \emph{GHZ state preparation}. We close with some discussion of ongoing efforts to prove more sophisticated algorithms correct in \sqir.

\paragraph{GHZ Proof} 
As an example of a proof we can carry out using \sqir, we show that \coqe{ghz}, \sqir's Greenberger-Horne-Zeilinger (GHZ) state \cite{Greenberger1989} preparation circuit given in \Cref{sec:sqir-syntax}, correctly produces the GHZ state.
The GHZ state is an $n$-qubit entangled quantum state of the form $\frac{1}{\sqrt{2}}(\ket{0}^{\otimes n}+\ket{1}^{\otimes n})$. This vector can be defined in Coq as shown.
Like our definition of $\ket{i}^{\otimes n}$ discussed above, we declare that this expression has type \coqe{Vector 2^n}, which will allow us to use it in any context where Coq expects a vector, deferring the proof that it is well-formed.

Our goal is to show that for any \coqe{n} $ > 0$ the circuit generated by \coqe{ghz n} produces the corresponding \coqe{GHZ n} vector when applied to $\ket{0}^{\otimes n}$.
\begin{coq}
Lemma ghz_correct : forall n : nat, 
  n > 0 -> [[ghz n]]${}_n$ $\times$  $\ket{0}^{\otimes n}$ = GHZ n.
\end{coq}

The proof proceeds by induction on $n$. The $n=0$ case is trivial as it contradicts our hypothesis.
For $n = 1$ we show that $H$ applied to $\ket{0}$ produces the $\ket{+}$ state.
In the inductive step, the induction hypothesis says that the result of applying \texttt{ghz n'} to the input state \texttt{nket n' $\ket{0}$} is the state
\coqe{($\frac{1}{\sqrt{2}}$ * $\ket{0}^{\otimes n'}$ + $\frac{1}{\sqrt{2}}$ * $\ket{1}^{\otimes n'}$) $\otimes$ $\ket{0}$}.
By applying \coqe{CNOT (n' - 1) n'} to this state, we show that \coqe{ghz (n' + 1) = GHZ (n' + 1)}.
Our use of concrete indices allows us to easily describe the semantics of \coqe{CNOT (n'-1) n'}. If we had instead used abstract wires (e.g. variables \coqe{x} and \coqe{y}), then to reason about the semantics of \coqe{CNOT x y} we would also need to reason about the conversion of \coqe{x} and \coqe{y} to concrete indices, 
showing that in the inductive case \coqe{x} refers to a qubit in the GHZ state prepared by the recursive call and \coqe{y} references a fresh $\ket{0}$ qubit.

\paragraph{Further Proofs}
It turns out that with the right abstractions, \sqir is capable of verifying a range of quantum algorithms, from Simon's~\citeyearpar{Simon1994} and Grover's~\citeyearpar{Grover1996} algorithms to quantum phase estimation, a key component of Shor's factoring algorithm~\citeyearpar{Shor94}.
All in all, the \sqir development contains about 3500 lines of example proofs and programs including GHZ state preparation, superdense coding, quantum teleportation, the Deutsch-Jozsa algorithm, Simon's algorithm, Grover's algorithm, and quantum phase estimation.
As this paper's focus is \voqc, we refer the interested reader to a separate paper~\cite{CPPsub} for detailed discussion of these source-program proofs and proof techniques. We summarize some of the key takeaways here.

The textbook proofs of the algorithms listed above argue correctness by considering the behavior of the program on a basis vector of the form $\ket{i_1 i_2\dots i_n}$ for $i_k \in \{0,1\}$, or a weighted sum over such vectors. To match this style of reasoning, we developed a \sqir framework for describing quantum states as vectors. 
We also found that we needed more sophisticated math lemmas when reasoning about quantum source programs. 
For example, though the trigonometric lemmas in Coq's standard library are sufficient for verifying \voqc optimizations, our proof of quantum phase estimation relies on the Coq Interval package \cite{intervals} to prove bounds on trigonometric functions.\footnote{Laurent Th\'{e}ry helpfully pointed us to Interval and provided proofs of our \texttt{sin\_sublinear} and \texttt{sin\_PIx\_ge\_2x} lemmas.}

These extensions aside, we were able to reuse much of what we developed for \voqc. For example, we directly use the \sqir unitary semantics presented in this section, benefiting from its various language simplifications.
We also benefit from automation for matrices and complex numbers added to \qwire as part of our work on \voqc. 
In particular, we were able to re-purpose the \coqe{gridify} tactic we developed for proving low-level matrix equivalences (described in \Cref{sec:gridify}) to prove statements about the effects of different gates on vector states. 



\section{Optimizing Unitary \sqir Programs}
\label{sec:voqc}

This section and the next describe \voqc, our verified optimizer for quantum circuits. \voqc primarily implements optimizations inspired by the state-of-the-art circuit optimizer of \citet{Nam2018}. As such, we do not claim credit for the optimizations themselves. Rather, our contribution is a framework that is sufficiently flexible that it can be used to prove such state-of-the-art optimizations correct. This section focuses on \voqc's optimizations for unitary \sqir programs and mapping to connectivity-constrained architectures; the next section discusses how \voqc optimizes non-unitary \sqir programs.

\subsection{Program Equivalence}

The \voqc optimizer takes as input a \sqir program and attempts to reduce its total gate count by applying a series of
optimizations. 
For each optimization, we verify that it is \emph{semantics preserving} (or \emph{sound}), meaning that the output program is guaranteed to be equivalent to the input program.

We say that two unitary programs of dimension $d$ are equivalent, written $U_1 \equiv U_2$, if their denotation is the same, i.e., $\denote{U_1}_d = \denote{U_2}_d$. 
We also support a more general version of equivalence: We say that two circuits are \emph{equivalent up to a global phase}, written $U_1 \cong U_2$, when there exists a $\theta$ such that $\denote{U_1}_d = e^{i\theta}\denote{U_2}_d$.
This is useful in the quantum setting because $\ket{\psi}$ and $e^{i\theta}\ket{\psi}$ (for $\theta \in \mathbb{R}$) represent the same physical state. Note that the latter notion of equivalence matches the former when $\theta = 0$.

Given this definition of equivalence we can write our soundness condition for optimization function \coqe{optimize} as follows.
\begin{coq}
Definition sound {G} (optimize : forall {d : nat}, ucom G d -> ucom G d) :=
  forall (d : nat) (u : ucom G d), [[optimize u]]${}_d$ $\cong$ [[u]]${}_d$.
\end{coq}
This property is quantified over \coqe{G}, \coqe{d}, and \coqe{u}, meaning that the property holds for \emph{any program that uses any set of gates and any number of qubits}. The optimizations in our development are defined over a particular gate set, defined below, but still apply to programs that use any number of qubits. Our statements of soundness also occasionally have an additional precondition that requires program \coqe{u} to be well typed.

\subsection{\voqc Optimization Overview}

\voqc implements two basic kinds of optimizations: \emph{replacement} and \emph{propagate-cancel}. The former simply identifies a pattern of gates and replaces it with an equivalent pattern. The latter works by commuting sets of gates when doing so produces an equivalent quantum program---often with the effect of ``propagating'' a particular gate rightward in the program---until two adjacent gates can be removed because they cancel out. 

To ease the implementation of and proofs about these optimizations, we developed a framework of supporting library functions that operate on \sqir programs as lists of gate applications, rather than on the native \sqir representation. The conversion code takes a sequence of gate applications in the original \sqir program and \emph{flattens} it so that a program like $(G_1~p; G_2~q); G_3~r$ is represented as the Coq list $[G_1~p; G_2~q; G_3~r]$. The denotation of the list representation is the denotation of its corresponding \sqir program. 
Examples of the list operations our framework provides include:
\begin{itemize}
    \item Finding the next gate acting on a qubit that satisfies some predicate $f$.
    \item Propagating a gate using a set of cancellation and commutation rules (see \Cref{sec:prop-opt}).
    \item Replacing a sub-program with an equivalent program (see \Cref{sec:repl-opt}).
    \item Computing the maximal matching prefix of two programs.
\end{itemize}
We verify that these functions have the intended behavior (e.g., in the last example, that the returned sub-program is indeed a prefix of both input programs).

Our framework supports arbitrary gate sets (for example, the functions listed above are all parameterized by choice of gate set). However, in the optimizations described below we use a specific, universal gate set $\{H,~X,~Rz,~\mathit{CNOT}\}$ where $Rz(k)$ describes rotation about the $z$-axis by $k \cdot \pi$ for $k \in \mathbb{Q}$\@. 
This gate set is more convenient than \sqir's base gate set for two reasons. First, using a discrete gate set makes it possible to define optimizations using Coq's built-in pattern matching (with occasional equality checks between rational values). Second, using rational parameters instead of real parameters allows us to extract to OCaml rational numbers rather than floating point numbers, which would render verification unsound.
Most existing tools (e.g., Qiskit \cite{Qiskit2019} and \citet{Nam2018}) allow gates parameterized by floats, which invites rounding error and can lead to unsound optimization.

To compute a program's denotation, \voqc's gates $H$, $X$, and $Rz(k)$ are translated into $R_{\pi/2,0,\pi}$, $R_{\pi,0,\pi}$, and $R_{0,0,k\pi}$ in \sqir's base gate set. ($\mathit{CNOT}$ translates to itself.) 
\voqc's gate set is identical to \citeauthor{Nam2018}'s, with the exception of the $z$-axis rotation parameter type (rational, not float). 

\subsection{Optimization by Propagation and Cancellation}
\label{sec:prop-opt}

Our \emph{propagate-cancel} optimizations have two steps. First we localize a set of gates by repeatedly applying commutation rules. Then we apply a circuit equivalence to replace that set of gates. In \voqc, most optimizations of this form use a library of code patterns, but one---\emph{not propagation}---is slightly different, so we discuss it first.

\paragraph{Not Propagation}

The goal of not propagation is to remove cancelling $X$ (``not'') gates. 
Two $X$ gates cancel when they are adjacent or they are separated by a circuit that commutes with $X$. 
We find $X$ gates separated by commuting circuits by repeatedly applying the propagation rules in \Cref{fig:not-rules}.
An example application of the not propagation algorithm is shown in \Cref{fig:not-prop-ex}.

\begin{figure}[t]
\[
\begin{array}{r c l}
X~ q;~ H~ q & \equiv & H~ q;~ Z~ q \\
X~ q;~ Rz(k)~ q & \cong &  Rz(2 - k)~ q;~ X~ q \\
X~ q_1;~ \mathit{CNOT}~ q_1~ q_2 & \equiv &  \mathit{CNOT}~ q_1~ q_2;~ X~ q_1;~ X~ q_2 \\
X~ q_2;~ \mathit{CNOT}~ q_1~ q_2 &  \equiv & \mathit{CNOT}~ q_1~ q_2;~ X~ q_2
\end{array}
\]
\caption{Equivalences used in not propagation.}
\label{fig:not-rules}
\end{figure}

\begin{figure}[t]
\begin{center}
    \begin{tabular}{clclcl}
  &
  \begin{minipage}{0.3\linewidth}
  \small
  \Qcircuit @C=0.5em @R=0.5em {
     & \gate{X} & \ctrl{1} & \gate{H} & \qw \\
     & \qw & \targ & \gate{X} & \qw \\
    }
  \end{minipage}%
  &
  $\rightarrow$
  &
  \begin{minipage}{0.3\linewidth}
  \small
  \Qcircuit @C=0.5em @R=0.5em {
     & \ctrl{1} & \gate{X} &  \gate{H} & \qw \\
     & \targ & \gate{X} & \gate{X} & \qw \\
    }
  \end{minipage}%
  &
  $\rightarrow$
  &
  \begin{minipage}{0.3\linewidth}
  \small
    \Qcircuit @C=0.5em @R=0.5em {
     & \ctrl{1} & \gate{H} &  \gate{Z} & \qw \\
     & \targ & \qw & \qw & \qw \\
    }
  \end{minipage}%
  \end{tabular}
\end{center}
  \caption{An example of not propagation. In the first step the leftmost $X$ gate propagates through the $\mathit{CNOT}$ gate to become two $X$ gates. In the second step the upper $X$ gate propagates through the $H$ gate and the lower $X$ gates cancel.}
  \label{fig:not-prop-ex}
\end{figure}
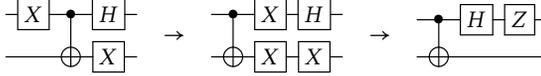

This implementation may introduce extra $X$ gates at the end of a circuit or extra $Z$ gates in the interior of the circuit.
Extra $Z$ gates are likely to be cancelled by the gate cancellation and rotation merging passes that follow, and moving $X$ gates to the end of a circuit makes the rotation merging optimization more likely to succeed.

We note that our version of this optimization is a simplification of \citeauthor{Nam2018}'s, which is specialized to a three-qubit $\mathit{TOFF}$ gate; this gate can be decomposed into a $\{H, Rz, CNOT\}$ program. 
In our experiments, we did not observe any difference in performance between \voqc and \citeauthor{Nam2018} due to this simplification.

\paragraph{Gate Cancellation}

The single- and two-qubit gate cancellation optimizations rely on the same propagate-cancel pattern used in not propagation, except that gates are returned to their original location if they fail to cancel. To support this pattern, we provide a general \coqe{propagate} function in \voqc. 
This function takes as inputs (i) an instruction list, (ii) a gate to propagate, and (iii) a set of rules for commuting and cancelling that gate. At each iteration, \coqe{propagate} performs the following actions:
\begin{enumerate}
    \item Check if a cancellation rule applies. If so, apply that rule and return the modified list.
    \item Check if a commutation rule applies. If so, commute the gate and recursively call \coqe{propagate} on the remainder of the list.
    \item Otherwise, return the gate to its original position.
\end{enumerate}
We have proved that our \coqe{propagate} function is sound when provided with valid commutation and cancellation rules.

\begin{figure}[t]
\begin{center}
\begin{tabular}{c@{$\quad\equiv\quad$}c@{$\qquad\qquad$}c@{$\quad\equiv\quad$}c}
  \begin{minipage}{0.3\linewidth}
  \Small
  \Qcircuit @C=0.5em @R=0.5em {
     & \qw & \qw & \ctrl{1} & \qw & \qw \\
     & \gate{Rz(k)} & \gate{H} & \targ & \gate{H} & \qw
    }
  \end{minipage}%
  &
  \begin{minipage}{0.3\linewidth}
  \Small
  \Qcircuit @C=0.5em @R=0.5em {
     & \qw & \ctrl{1} & \qw & \qw & \qw \\
     & \gate{H} & \targ & \gate{H} & \gate{Rz(k)} & \qw
    }
  \end{minipage}%
  &
  \begin{minipage}{0.3\linewidth}
  \Small
  \Qcircuit @C=0.7em @R=0.7em {
     & \ctrl{2} & \qw & \qw \\
     & \qw & \ctrl{1} & \qw \\
     & \targ & \targ & \qw
    }
  \end{minipage}%
  &
  \begin{minipage}{0.3\linewidth}
  \Small
  \Qcircuit @C=0.7em @R=0.7em {
     & \qw & \ctrl{2} & \qw \\
     & \ctrl{1} & \qw & \qw \\
     & \targ & \targ & \qw
    }
  \end{minipage}%
  \\[0.5cm]
  \begin{minipage}{0.3\linewidth}
  \Small
  \Qcircuit @C=0.5em @R=0.5em {
     & \qw & \ctrl{1} & \qw & \ctrl{1} & \qw \\
     & \gate{Rz(k)} & \targ & \gate{Rz(k')} & \targ & \qw
    }
  \end{minipage}%
  &
  \begin{minipage}{0.3\linewidth}
  \Small
  \Qcircuit @C=0.5em @R=0.5em {
     & \ctrl{1} & \qw & \ctrl{1} & \qw & \qw \\
     & \targ & \gate{Rz(k')} & \targ & \gate{Rz(k)} & \qw
    }
  \end{minipage}%
  &
  \begin{minipage}{0.3\linewidth}
  \Small
  \Qcircuit @C=0.5em @R=0.5em {
     & \ctrl{2} & \ctrl{1} & \qw \\
     & \qw & \targ & \qw \\
     & \targ & \qw & \qw
    }
  \end{minipage}%
  &
  \begin{minipage}{0.3\linewidth}
  \Small
  \Qcircuit @C=0.5em @R=0.5em {
     & \ctrl{1} & \ctrl{2} & \qw \\
     & \targ & \qw & \qw \\
     & \qw & \targ & \qw
    }
  \end{minipage}%
  \\[0.5cm]
  \begin{minipage}{0.3\linewidth}
  \Small
  \Qcircuit @C=0.5em @R=0.5em {
     & \gate{Rz(k)} & \ctrl{1} & \qw \\
     & \qw & \targ & \qw
    }
  \end{minipage}%
  &
  \begin{minipage}{0.3\linewidth}
  \Small
  \Qcircuit @C=0.5em @R=0.5em {
     & \ctrl{1} & \gate{Rz(k)} & \qw \\
     & \targ & \qw & \qw
    }
  \end{minipage}%
  &
  \begin{minipage}{0.3\linewidth}
  \Small
  \Qcircuit @C=0.5em @R=0.5em {
     & \ctrl{1} & \qw & \qw & \qw & \qw \\
     & \targ & \gate{H} & \ctrl{1} & \gate{H} & \qw \\
     & \qw & \qw & \targ & \qw & \qw
    }
  \end{minipage}%
  &
  \begin{minipage}{0.3\linewidth}
  \Small
  \Qcircuit @C=0.5em @R=0.5em {
     & \qw & \qw & \qw & \ctrl{1} & \qw \\
     & \gate{H} & \ctrl{1} & \gate{H} & \targ & \qw \\
     & \qw & \targ & \qw & \qw & \qw
    }
  \end{minipage}%
\end{tabular}
\end{center}
  \caption{Commutation equivalences for single- and two-qubit gates adapted from Nam et al. \cite[Figure 5]{Nam2018}. We use the second and third rules for propagating both single- and two-qubit gates.}
  \label{fig:comm-rules}
\end{figure}
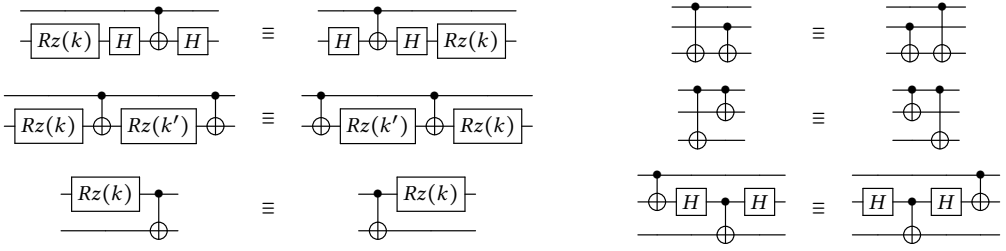

Each commutation or cancellation rule is implemented as a partial Coq function from an input circuit to an output circuit. A common pattern in these rules is to identify one gate (e.g., an $X$ gate), and then to look for an adjacent gate it might commute with (e.g., $\mathit{CNOT})$ or cancel with (e.g., $X$). 
For commutation rules, we use the rewrite rules shown \Cref{fig:comm-rules}.
For cancellation rules, we use the fact that $H$, $X$, and $\mathit{CNOT}$ are all self-cancelling and $Rz(k)$ and $Rz(k')$ combine to become $Rz(k + k')$. 

\subsection{Circuit Replacement}
\label{sec:repl-opt}

We have implemented two optimizations---Hadamard reduction and rotation merging---that work by replacing one pattern of gates with an equivalent one; no preliminary propagation is necessary. These aim either to reduce the gate count directly, or to set the stage for additional optimizations.

\paragraph{Hadamard Reduction}

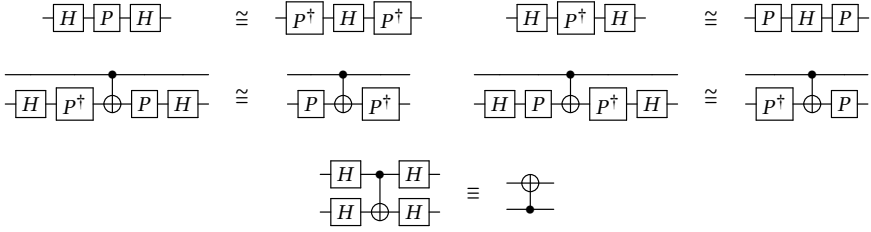
\begin{figure}[t]
\begin{center}
\begin{tabular}{c@{$\quad\cong\quad$}c@{$\qquad$}c@{$\quad\cong\quad$}c}
  \begin{minipage}{0.3\linewidth}
  \Small
  \Qcircuit @C=0.5em @R=0.5em {
     & \gate{H} & \gate{P} & \gate{H} & \qw 
    }
  \end{minipage}%
  & 
  \begin{minipage}{0.3\linewidth}
  \Small
  \Qcircuit @C=0.5em @R=0.5em {
     & \gate{P^\dagger} & \gate{H} & \gate{P^\dagger} & \qw
    }
  \end{minipage}%
  &
  \begin{minipage}{0.3\linewidth}
  \Small
  \Qcircuit @C=0.5em @R=0.5em {
     & \gate{H} & \gate{P^\dagger} & \gate{H} & \qw 
    }
  \end{minipage}%
  & 
  \begin{minipage}{0.3\linewidth}
  \Small
  \Qcircuit @C=0.5em @R=0.5em {
     & \gate{P} & \gate{H} & \gate{P} & \qw
    }
  \end{minipage}%
  \\[0.5cm]
  \begin{minipage}{0.3\linewidth}
  \Small
  \Qcircuit @C=0.5em @R=0.5em {
     & \qw & \qw & \ctrl{1} & \qw & \qw & \qw \\
     & \gate{H} & \gate{P^\dagger} & \targ & \gate{P} & \gate{H} & \qw
    }
  \end{minipage}%
  &
  \begin{minipage}{0.3\linewidth}
  \Small
  \Qcircuit @C=0.5em @R=0.5em {
     & \qw & \ctrl{1} & \qw & \qw \\
     & \gate{P} & \targ & \gate{P^\dagger} & \qw
    }
  \end{minipage}%
  &
  \begin{minipage}{0.3\linewidth}
  \Small
  \Qcircuit @C=0.5em @R=0.5em {
     & \qw & \qw & \ctrl{1} & \qw & \qw & \qw \\
     & \gate{H} & \gate{P} & \targ & \gate{P^\dagger} & \gate{H} & \qw
    }
  \end{minipage}%
  &
  \begin{minipage}{0.3\linewidth}
  \Small
  \Qcircuit @C=0.5em @R=0.5em {
     & \qw & \ctrl{1} & \qw & \qw \\
     & \gate{P^\dagger} & \targ & \gate{P} & \qw
    }
  \end{minipage}%
\end{tabular}\\[0.5cm]
\begin{tabular}{c@{$\quad\equiv\quad$}c}
  \begin{minipage}{0.3\linewidth}
  \Small
  \Qcircuit @C=0.5em @R=0.5em {
     & \gate{H} & \ctrl{1} & \gate{H} & \qw \\
     & \gate{H} & \targ & \gate{H} & \qw
    }
  \end{minipage}%
  &
  \begin{minipage}{0.3\linewidth}
  \Small
  \Qcircuit @C=0.7em @R=0.7em {
     & \targ & \qw \\
     & \ctrl{-1} & \qw
    }
  \end{minipage}%
\end{tabular}
\end{center}
  \caption{Equivalences for removing Hadamard gates adapted from \citet[Figure 4]{Nam2018}. $P$ is the phase gate $Rz(1/2)$ and $P^\dagger$ is its inverse $Rz(3/2)$.}
  \label{fig:had-red-rules}
\end{figure}

The Hadamard reduction routine employs the equivalences shown in \Cref{fig:had-red-rules} to reduce the number of $H$ gates in the program.
Removing $H$ gates is useful because $H$ gates limit the size of the $\{Rz, \mathit{CNOT}\}$ subcircuits used in the rotation merging optimization. 

\paragraph{Rotation Merging}

The rotation merging optimization allows for combining $Rz$ gates that are not physically adjacent in the circuit. 
This optimization is more sophisticated than the previous optimizations because it does not rely on small structural patterns (e.g., that adjacent $X$ gates cancel), but rather on more general (and non-local) circuit behavior.
The basic idea behind rotation merging is to (i) identify subcircuits consisting of only $\mathit{CNOT}$ and $Rz$ gates and (ii) merge $Rz$ gates within those subcircuits that are applied to qubits in the same logical state.

The argument for the correctness of this optimization relies on the \emph{phase polynomial} representation of a circuit.
Let $C$ be a circuit consisting of $\mathit{CNOT}$ gates and rotations about the $z$-axis.
Then on basis state $\ket{x_1, ..., x_n}$, $C$ will produce the state 
\[e^{ip(x_1, ..., x_n)}\ket{h(x_1, ..., x_n)}\]
where $h : \{0,1\}^n \rightarrow \{0,1\}^n$ is an affine reversible function and 
\[p(x_1, ..., x_n) = \sum_{i=1}^l (\theta_i \!\!\!\mod 2\pi) \dot\; f_i(x_1, ..., x_n)\]
is a linear combination of affine boolean functions. $p(x_1, ..., x_n)$ is called the phase polynomial of circuit $C$\@. 
Each rotation gate in the circuit is associated with one term of the sum and if two terms of the phase polynomial satisfy $f_i(x_1, ..., x_n) = f_j (x_1, ... , x_n )$ for some $i \neq j$, then the corresponding $i$ and $j$ rotations can be merged.

As an example, consider the two circuits shown below.
\begin{center}
\begin{tabular}{c@{$\quad\equiv\quad$}c}
  \begin{minipage}{0.3\linewidth}
  \Small
  \Qcircuit @C=0.5em @R=0.5em {
 & \qw & \ctrl{1} & \targ & \gate{Rz(k')} & \qw \\
 & \gate{Rz(k)} & \targ & \ctrl{-1} & \qw & \qw }
  \end{minipage}%
  &
  \begin{minipage}{0.3\linewidth}
  \Small
  \Qcircuit @C=0.5em @R=0.5em {
 & \ctrl{1} & \targ & \gate{Rz(k + k')} & \qw \\
 & \targ & \ctrl{-1} & \qw & \qw }
  \end{minipage}%
\end{tabular}
\end{center}
To prove that these circuits are equivalent, we can consider their behavior on basis state $\ket{x_1,x_2}$.
Recall that applying $Rz(k)$ to the basis state $\ket{x}$ produces the state $e^{ik\pi x}\ket{x}$ and $\mathit{CNOT} \ket{x,y}$ produces the state $\ket{x,x \oplus y}$ where $\oplus$ is the xor operation.
Evaluation of the left-hand circuit proceeds as follows:
\[ \ket{x_1, x_2} \rightarrow e^{ik\pi x_2}\ket{x_1, x_2} \rightarrow e^{ik\pi x_2}\ket{x_1, x_1 \oplus x_2} \rightarrow e^{ik\pi x_2}\ket{x_2, x_1 \oplus x_2} \rightarrow e^{ik\pi x_2}e^{ik'\pi x_2}\ket{x_2, x_1 \oplus x_2}. \]
Whereas evaluation of the right-hand circuit produces
\[ \ket{x_1, x_2} \rightarrow \ket{x_1, x_1 \oplus x_2} \rightarrow \ket{x_2, x_1 \oplus x_2} \rightarrow e^{i(k + k')\pi x_2}\ket{x_2, x_1 \oplus x_2}. \]
The two resulting states are equal because $e^{ik\pi x_2}e^{ik'\pi x_2} = e^{i(k + k')\pi x_2}$.
This implies that the unitary matrices corresponding to the two circuits are the same.
We can therefore replace the circuit on the left with the one on the right, removing one gate from the circuit.

Our rotation merging optimization follows the reasoning above for arbitrary $\{Rz, \mathit{CNOT}\}$ circuits.
For every gate in the program, it tracks the Boolean function associated with every qubit (the Boolean functions above are $x_1$, $x_2$, $x_1 \oplus x_2$), and merges $Rz$ rotations when they are applied to qubits associated with the same Boolean function.
To prove equivalence over $\{Rz, \mathit{CNOT}\}$ circuits, we show that the original and optimized circuits produce the same output on every basis state. We have found evaluating behavior on basis states to be useful for proving equivalences that are not as direct as those listed in \Cref{fig:comm-rules,fig:had-red-rules}.

Although our merge operation is identical to \citeauthor{Nam2018}'s, our approach to constructing $\{Rz, \mathit{CNOT}\}$ subcircuits differs. We construct a $\{Rz, \mathit{CNOT}\}$ subcircuit beginning from a $Rz$ gate whereas \citeauthor{Nam2018} begin from a $\mathit{CNOT}$ gate. The result of this simplification is that we may miss some opportunities for merging. However, in our experiments (\Cref{sec:experiments}) we found that this choice impacted only one benchmark.

\subsection{Proving Low-Level Circuit Equivalences}
\label{sec:gridify}

\voqc optimizations make heavy use of circuit equivalences such as those shown in \Cref{fig:not-rules,fig:comm-rules,fig:had-red-rules}. To prove that \voqc optimizations are sound, we must formally verify these equivalences are correct. Such proofs require showing equality between two matrix expressions, which can be tedious in the case where the matrix size is left symbolic.
For example, consider the following equivalence used in \emph{not propagation}:
\[ X~n;~\mathit{CNOT}~m~n \equiv \mathit{CNOT}~m~n;~X~n \]
for arbitrary $n, m$ and dimension $d$. Applying our definition of equivalence, this amounts to proving
\begin{equation} \label{eqn:x-cnot-comm}
\begin{split}
apply_1(X, n, d) ~\times~ apply_2(\mathit{CNOT}, m, n, d) = apply_2(\mathit{CNOT}, m, n, d) ~\times~ apply_1(X, n, d),
\end{split}
\end{equation}
per the semantics in \Cref{fig:sqire-semantics}.
Suppose both sides of the equation are well typed ($m < d$ and $n < d$ and $m \not= n$), and consider the case where $m < n$ (the $n < m$ case is similar). We expand $apply_1$ and $apply_2$ as follows with $p = n - m - 1$ and $q = d - n - 1$:
\begin{align*} 
apply_1(X, n, d) &= I_{2^n} \otimes \sigma_x \otimes I_{2 ^ {q}}\\
apply_2(\mathit{CNOT}, m, n, d) &= I_{2^m} \otimes \op{1}{1} \otimes I_{2^{p}} \otimes \sigma_x \otimes I_{2^{q}} + I_{2^m} \otimes \op{0}{0} \otimes I_{2^{p}} \otimes I_{2} \otimes I_{2^{q}}
\end{align*}
Here, $\sigma_x$ is the matrix interpretation of the $X$ gate and $\op{1}{1} \otimes \sigma_x + \op{0}{0} \otimes I_2$ is the matrix interpretation of the $\mathit{CNOT}$ gate (in Dirac notation). 
We can complete the proof of equivalence by normalizing and simplifying each side of \Cref{eqn:x-cnot-comm}, showing both sides to be the same. 

\paragraph{Automation}
We address the tedium of such proofs in \voqc by almost entirely automating the matrix normalization and simplification steps. We provide a Coq tactic called \coqe{gridify} for proving general equivalences correct.  
Rather than assuming $m < n < d$ as above, the \coqe{gridify} tactic does case analysis, immediately solving all cases where the circuit is ill-typed (e.g., $m = n$ or $d \leq m$) and thus has the zero matrix as its denotation. 
In the remaining cases ($m<n$ and $n<m$ above), it  puts the expressions into a form we call \emph{grid normal} and applies a set of matrix identities. 

In grid normal form, each arithmetic expression has addition on the outside, followed by tensor product, with multiplication on the inside, i.e., $((..\times..)\otimes(..\times..)) + ((..\times..)\otimes(..\times..))$. The \coqe{gridify} tactic rewrites an expression into this form by using the following rules of matrix arithmetic: 
\begin{itemize}
    \item $I_{mn} = I_m \otimes I_n$
    \item $A \times (B + C) = A \times B + A \times C$
    \item $(A + B) \times C = A \times C + B \times C$
    \item $A \otimes (B + C) = A \otimes B + A \otimes C$
    \item $(A + B) \otimes C = A \otimes C + B \otimes C$
    \item $(A \otimes B) \times (C \otimes D) = (A \times C) \otimes (B \times D)$
\end{itemize}
The first rule is applied to facilitate application of the other rules. (For instance, in the example above, $I_{2^n}$ would be replaced by $I_{2^m} \otimes I_2 \otimes I_{2^{p}}$ to match the structure of the $apply_2$ term.) After expressions are in grid normal form, \coqe{gridify} simplifies them by removing multiplication by the identity matrix and rewriting simple matrix products (e.g. $\sigma_x\sigma_x = I_2$).

In our example, normalization and simplification by \coqe{gridify} rewrites each side of the equality in \Cref{eqn:x-cnot-comm} to be the following
\begin{gather*}
 I_{2^m} \otimes \op{1}{1} \otimes I_{2^p} \otimes I_2 \otimes I_{2^q} + I_{2^m} \otimes \op{0}{0} \otimes I_{2^p} \otimes \sigma_x \otimes I_{2^q},
\end{gather*}
thus proving that the two expressions are equal.

We use \coqe{gridify} to verify most of the equivalences used in the optimizations given in \Cref{sec:prop-opt,sec:repl-opt}. 
The tactic is most effective when equivalences are small: The equivalences used in \emph{gate cancellation} and \emph{Hadamard reduction} apply to patterns of at most five gates applied to up to three qubits within an arbitrary circuit.
For equivalences over large sets of qubits, like the one used in \emph{rotation merging}, we do not use \coqe{gridify} directly, but still rely on our automation for matrix simplification.

\subsection{Scheduling}

The \voqc \coqe{optimize} function applies each of the optimizations we have discussed one after the other, in the following order (due to \citeauthor{Nam2018}):
\[0, 1, 3, 2, 3, 1, 2, 4, 3, 2\]
where 0 is not propagation, 1 is Hadamard reduction, 2 is single-qubit gate cancellation, 3 is two-qubit gate cancellation, and 4 is rotation merging. 
\citeauthor{Nam2018} justify this ordering at length, though they do not prove that it is optimal. 
In brief, removing $X$ and $H$ gates (0,1) allows for more effective application of the gate cancellation (2,3) and rotation merging (4) optimizations.
In our experiments (\Cref{sec:experiments}), we observed that single-qubit gate cancellation and rotation merging were the most effective at reducing gate count.

\subsection{Circuit Mapping}
\label{sec:circuitmap}

We have also implemented  and verified a transformation that maps a circuit to a connectivity-constrained architecture. Similar to how optimization aims to reduce qubit and gate usage to make programs more feasible to run on near-term machines, \emph{circuit mapping} aims to address the connectivity constraints of near-term machines \cite{Saeedi2011, Zulehner2017}.
Circuit mapping algorithms take as input an arbitrary circuit and output a circuit that respects the connectivity constraints of some underlying architecture.
 
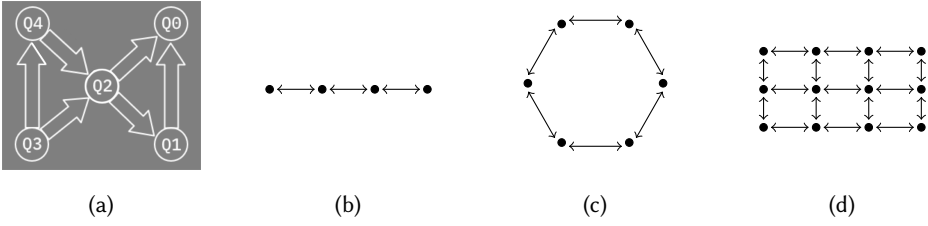
\begin{figure}
\centering
\begin{subfigure}[b]{0.23\linewidth}
\begin{center}
\includegraphics[scale=0.28]{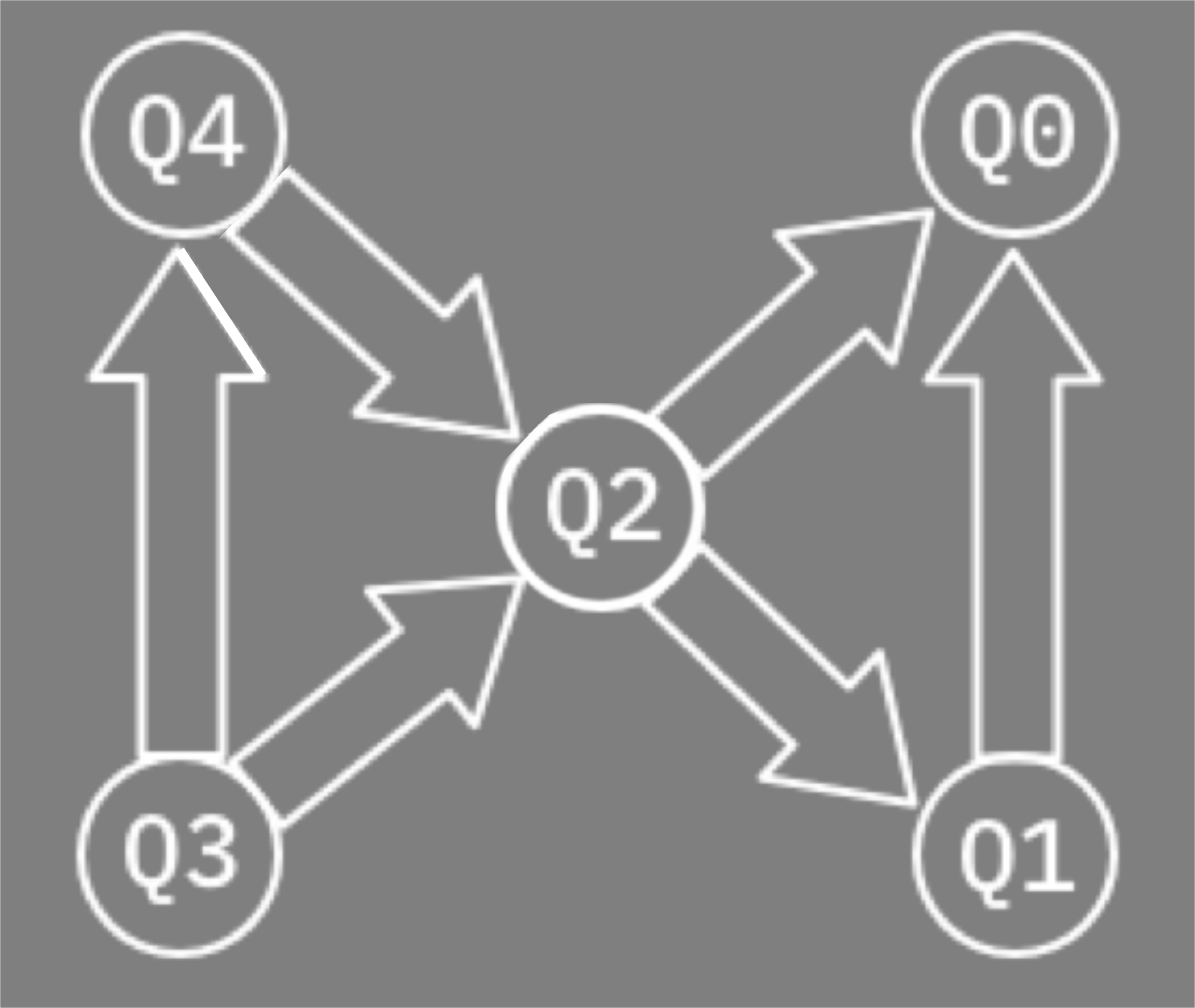}
\end{center}
\caption{}
\end{subfigure}
\centering
\begin{subfigure}[b]{0.23\linewidth}
\begin{center}
\begin{tikzpicture}
  \draw [fill] (0,0) circle [radius=0.05];
  \draw [<->] (.1,0) -- (.6,0);
  \draw [fill] (0.7,0) circle [radius=0.05];
  \draw [<->] (.8,0) -- (1.3,0);
  \draw [fill] (1.4,0) circle [radius=0.05];
  \draw [<->] (1.5,0) -- (2,0);
  \draw [fill] (2.1,0) circle [radius=0.05];
\end{tikzpicture}
\par\vspace{1cm}
\end{center}
\caption{}
\end{subfigure}
\begin{subfigure}[b]{0.23\linewidth}
\begin{center}
\begin{tikzpicture}
  \newdimen\R
  \R=0.9cm
  \draw [<->] (7:\R) -- (53:\R);
  \draw [<->] (67:\R) -- (113:\R);
  \draw [<->] (127:\R) -- (173:\R);
  \draw [<->] (187:\R) -- (233:\R);
  \draw [<->] (247:\R) -- (293:\R);
  \draw [<->] (307:\R) -- (353:\R);
  \foreach \x in {60,120,...,360} 
  \draw [fill] (\x:\R) circle [radius=0.05];
\end{tikzpicture}
\par\vspace{0.3cm}
\end{center}
\caption{}
\end{subfigure}
\begin{subfigure}[b]{0.23\linewidth}
\begin{center}
\begin{tikzpicture}
  \draw [fill] (0,0) circle [radius=0.05];
  \draw [<->] (.1,0) -- (.6,0);
  \draw [fill] (0.7,0) circle [radius=0.05];
  \draw [<->] (.8,0) -- (1.3,0);
  \draw [fill] (1.4,0) circle [radius=0.05];
  \draw [<->] (1.5,0) -- (2,0);
  \draw [fill] (2.1,0) circle [radius=0.05];
  \draw [<->] (0,0.1) -- (0,0.4);
  \draw [<->] (0.7,0.1) -- (0.7,0.4);
  \draw [<->] (1.4,0.1) -- (1.4,0.4);
  \draw [<->] (2.1,0.1) -- (2.1,0.4);
  \draw [fill] (0,0.5) circle [radius=0.05];
  \draw [<->] (.1,0.5) -- (.6,0.5);
  \draw [fill] (0.7,0.5) circle [radius=0.05];
  \draw [<->] (.8,0.5) -- (1.3,0.5);
  \draw [fill] (1.4,0.5) circle [radius=0.05];
  \draw [<->] (1.5,0.5) -- (2,0.5);
  \draw [fill] (2.1,0.5) circle [radius=0.05];
  \draw [<->] (0,0.6) -- (0,0.9);
  \draw [<->] (0.7,0.6) -- (0.7,0.9);
  \draw [<->] (1.4,0.6) -- (1.4,0.9);
  \draw [<->] (2.1,0.6) -- (2.1,0.9);
  \draw [fill] (0,1) circle [radius=0.05];
  \draw [<->] (.1,1) -- (.6,1);
  \draw [fill] (0.7,1) circle [radius=0.05];
  \draw [<->] (.8,1) -- (1.3,1);
  \draw [fill] (1.4,1) circle [radius=0.05];
  \draw [<->] (1.5,1) -- (2,1);
  \draw [fill] (2.1,1) circle [radius=0.05];
\end{tikzpicture}
\par\vspace{0.5cm}
\end{center}
\caption{}
\end{subfigure}
  \caption{Examples of two-qubit gate connections on near-term quantum machines. From left to right: IBM’s Tenerife machine~\cite{tenerife}, LNN, LNN ring, and 2D grid. The last three architectures are shown with a fixed number vertices, but in our implementation the number of vertices is a parameter. Double-ended arrows indicate that two-qubit gates are possible in both directions. }
  \label{fig:connectivity}
\end{figure}

For example, consider the connectivity of IBMs's five-qubit Tenerife machine~\cite{tenerife} shown in \Cref{fig:connectivity}(a). This is a representative example of a superconducting qubit system, where qubits are laid out in a 2-dimensional grid and possible interactions are described by directed edges between them. The direction of the edge indicates which qubit can be the control of a two-qubit gate and which can be the target. For instance, a $\mathit{CNOT}$ gate may be applied with Q4 as the control and Q2 as the target, but not the reverse. No two-qubit gate is possible between physical qubits Q4 and Q1. 

We have implemented a simple circuit mapper for unitary SQIR programs and verified that it is sound and produces programs that satisfy the relevant hardware constraints. Our circuit mapper is parameterized by two functions that describe the connectivity of an architecture: one function determines whether an edge is in the connectivity graph and another function finds an undirected path between any two nodes.
Our mapping algorithm takes as input (i) these functions, (ii) a program referencing \emph{logical} qubits, and (iii) a map expressing the initial correspondence between the program's logical qubits and the \emph{physical} qubits available on the machine. 
The algorithm produces a program referencing physical qubits as well as an updated correspondence.
Every time a $\mathit{CNOT}$ occurs between two logical qubits whose corresponding physical qubits are not adjacent in the underlying architecture, we insert $\mathit{SWAP}$ operations to move the target and control into adjacent positions and update the physical-logical qubit correspondence accordingly.
To apply a $\mathit{CNOT}$ when an edge points in the wrong direction, we make use of the equivalence \coqe{CNOT b a $\equiv$ H a; H b; CNOT a b; H a; H b}.
For soundness, we prove that the mapped circuit is equivalent to the original up to a permutation of the qubits.

Although our mapping algorithm is simple, it allows for some flexibility in design because we do not specify the method for choosing the initial physical-logical qubit correspondence (called ``placement'') or the implementation of the function that finds paths in the connectivity graph (``routing''). This allows, for example, placement and routing strategies that take into account error characteristics of the machine \cite{Tannu2019}.
We expect that our verification framework can be applied to more sophisticated mapping algorithms such a those that partition the circuit into layers and insert $\mathit{SWAP}$s between layers rather than na\"ively inserting $\mathit{SWAP}$s before $\mathit{CNOT}$ gates \cite{Zulehner2017}.
We have used our framework to implement and verify mapping functions for the Tenerife architecture pictured in \Cref{fig:connectivity}(a) as well as the linear nearest neighbor (LNN), LNN ring, and 2D nearest neighbor architectures pictured in \Cref{fig:connectivity}(b-d).


\section{Full \sqir: Adding Measurement}
\label{sec:general-sqire}

While the bulk of \voqc proofs only use the unitary core of \sqir, we also support programs with measurement. Measurement plays a key role in protocols from quantum teleportation and quantum key distribution~\cite{bennett2020quantum} to repeat-until-success loops~\cite{paetznick2014repeat} and error-correcting codes~\cite{gottesman2010introduction}. It also enables a number of interesting optimizations, which we discuss in \Cref{sec:nonunitary-opt}. We begin with the syntax and semantics of full \sqir, and 
a proof that makes use of the non-unitary semantics.


\subsection{Syntax and Semantics}

To describe general quantum programs $P$, we extend unitary \sqir with a \emph{branching measurement} operation.
\begin{align*}
    P~:=~~\code{skip} \mid P_1 ;~P_2 \mid U \mid \code{meas}~q~P_1~P_2 
\end{align*}
The command \coqe{meas}$~q~P_1~P_2$ (inspired by a similar construct in QPL~\cite{Selinger2004}) measures the qubit $q$ and either performs program $P_1$ or $P_2$ depending on the result. 
We define non-branching measurement and resetting a qubit to $\ket{0}$ in terms of branching measurement:
\begin{align*}
    \code{measure}~q &= \code{meas}~q~\code{skip}~\code{skip} \\
    \code{reset}~q &= \code{meas}~q~(X~q)~\code{skip}
\end{align*}

\Cref{fig:sqire-syntax-semantics-general} defines the semantics of a non-unitary program as a function from \emph{density matrices} $\rho$ to density matrices, following the approach of several previous efforts \cite{Paykin2017, Ying2011}. 
Density matrices provide a way to describe arbitrary quantum states, including \emph{mixed states} which are probability distributions over \emph{quantum pure states} and arise in the analysis of general quantum programs.
For example, $\frac{1}{2}\begin{psmallmatrix}1 & 0 \\ 0 & 1\end{psmallmatrix}$ represents a $50\%$ chance of $\ket{0}$ and a $50\%$ chance of $\ket{1}$.

\begin{figure}[t]
\centering
  \begin{align*}
    \pdenote{\code{skip}}_d(\rho)&=\rho \\
    \pdenote{P_1;~P_2}_d(\rho)&=~(\pdenote{P_2}_d \circ \pdenote{P_1}_d) (\rho) \\
    \pdenote{U}_d(\rho)&= \denote{U}_d \times \rho \times \denote{U}_d^\dagger \\
   \pdenote{\code{meas}~q~P_1~P_2}_d(\rho)&= \pdenote{P_2}_d(\vert 0 \rangle_q \langle 0 \vert \times \rho \times \vert 0 \rangle_q \langle 0 \vert) \\ 
   &+ ~ \pdenote{P_1}_d(\vert 1 \rangle_q \langle 1 \vert \times \rho \times \vert 1 \rangle_q \langle 1 \vert)
  \end{align*}

  \caption{\sqir density matrix semantics, assuming a global register of size $d$.}
  \label{fig:sqire-syntax-semantics-general}

\end{figure}

\subsection{Example: Quantum Teleportation}
\label{sec:teleport}

The goal of quantum teleportation is to transmit a state $\ket{\psi}$ from one party (Alice) to another (Bob) using a shared entangled state. The circuit for quantum teleportation is shown in \Cref{fig:teleport-circ} and the corresponding \sqir program is given below.
\begin{coq}
Definition bell : ucom base 3 := H 1; CNOT 1 2.
Definition alice : com base 3 := CNOT 0 1 ; H 0; measure 0; measure 1.
Definition bob : com base 3 := CNOT 1 2; CZ 0 2; reset 0; reset 1. 
Definition teleport : com base 3 := bell; alice; bob.
\end{coq}
The \coqe{bell} circuit prepares a Bell pair on qubits 1 and 2, which are respectively sent to Alice and Bob. Alice applies \coqe{CNOT} from qubit 0 to qubit 1 and then measures both qubits and (implicitly) sends them to Bob. Finally, Bob performs operations controlled by the (now classical) values on qubits 0 and 1 and then resets them to the zero state.

\begin{figure}[t]
\centering
\includegraphics[trim=0 9cm 0 8cm,clip,width=8cm]{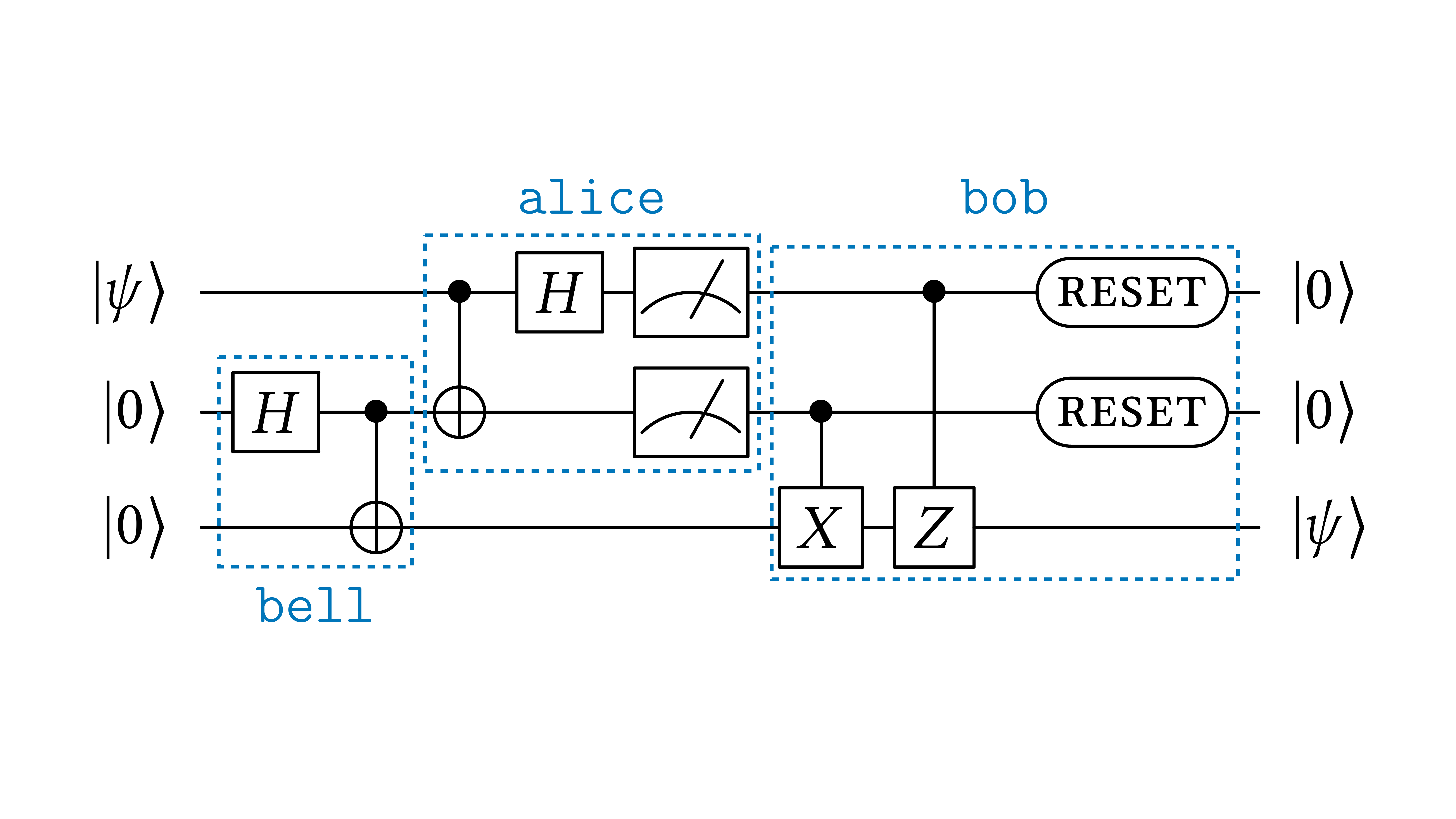}
  \caption{Circuit for quantum teleportation. In the standard presentation Bob only acts on the last qubit, given two classical bits as input. In our presentation, Bob (equivalently) performs operations controlled by the first two qubits, which are in a post-measurement classical state. We include the reset operations to simplify our statement of correctness.}
  \label{fig:teleport-circ}
\end{figure}

The correctness property for this program says that for any (well-formed)
density matrix $\rho$, \coqe{teleport} takes the state $\rho \otimes \op{0}{0} \otimes \op{0}{0}$ to the state $\op{0}{0} \otimes \op{0}{0} \otimes \rho$.
\begin{coq}
Lemma teleport_correct : forall (ρ : Density 2),
 WF_Matrix ρ -> $\pdenote{\text{teleport}}_3$ (ρ ⊗ ∣0⟩⟨0∣ ⊗ ∣0⟩⟨0∣) = ∣0⟩⟨0∣ ⊗ ∣0⟩⟨0∣ ⊗ ρ.
\end{coq}
The proof is simple: We perform (automated) arithmetic to show that the output matrix has the desired form.

Quantum teleportation is a rare case in which we prove something about a fixed-size program (i.e., $n$ is fixed to 3 qubits) and not one of arbitrary dimension. Using the density matrix semantics to prove properties about programs of arbitrary dimension $n$ is more involved. Stating the property requires introducing a symbolic density matrix $\rho$, which will be multiplied on the left and right by $2^n \times 2^n$ matrices in the denotation. In our experience this results in complicated proof terms that are tedious to manipulate, even with automation. By contrast, when reasoning about the equivalence of two unitary programs we can simply compare their unitary matrices, without carrying around a symbolic $\rho$ (or input vector).

\subsection{Non-unitary Optimizations} 
\label{sec:nonunitary-opt}

We have implemented two verified optimizations of non-unitary programs in \voqc, inspired by optimizations in IBM's Qiskit compiler \cite{Qiskit2019}: removing pre-measurement $z$-rotations, and classical state propagation. For these optimizations, we represent a non-unitary program $P$ as a list of \emph{blocks}. A block is a binary tree whose leaves are unitary programs (in list form) and nodes are measurements $\code{meas}~q~P_1~P_2$ whose children $P_1$ and $P_2$ are lists of blocks. Since the density matrix semantics denotes programs as functions over matrices, we say that programs $P_1$ and $P_2$ of dimension $d$ are equivalent if for every input $\rho$, $\pdenote{P_1}_d(\rho) = \pdenote{P_2}_d(\rho)$.

\paragraph{$z$-rotations Before Measurement}

$z$-axis rotations (or, more generally, diagonal unitary operations) before a measurement will have no effect on the measurement outcome, so they can safely be removed from the program.
This optimization locates $Rz$ gates before measurement operations and removes them.
It was inspired by Qiskit's RemoveDiagonalGatesBeforeMeasure pass.

\paragraph{Classical State Propagation}

Once a qubit has been measured, the subsequent branch taken provides information about the qubit's (now classical) state, which may allow pre-computation of some values. For example, in the branch where qubit $q$ has been measured to be in the $\ket{0}$ state, any $\mathit{CNOT}$ with $q$ as the control will be a no-op and any subsequent measurements of $q$ will still produce zero.

In detail, given a qubit $q$ in classical state $\ket{i}$, our optimization applies these propagation rules:
\begin{itemize}
    \item $Rz(k)~ q$ preserves the classical state of $q$. 
    \item $X~ q$ flips the classical state of $q$.
    \item If $i = 0$ then $\mathit{CNOT}~ q~ q'$ is removed, and if $i = 1$ then $\mathit{CNOT}~ q~ q'$ becomes $X~ q'$.
    \item $\code{meas}~q~P_1~P_0$ becomes $P_i$.
    \item $H~ q$ and $\mathit{CNOT}~ q'~ q$ make $q$ non-classical and terminate analysis.
\end{itemize}

Our statement of correctness for one round of propagation says that if qubit $q$ is in a classical state in the input, then the optimized program will have the same denotation as the unoptimized original. We express the requirement that qubit $q$ be in classical state $i \in \{0, 1\}$ with the condition
$\ket{i}_q\bra{i} \times \rho \times \ket{i}_q\bra{i} = \rho,$
which says that projecting state $\rho$ onto the subspace where $q$ is in state $\ket{i}$ results in no loss of information.

This optimization is not implemented directly in Qiskit, but Qiskit contains passes that have a similar effect.
For example, the RemoveResetInZeroState pass removes adjacent reset gates, as the second has no effect.


\section{Experimental Evaluation}
\label{sec:experiments}

The value of \voqc (and \sqir) is determined by the quality of the verified optimizations we can write with it. We can judge optimization quality empirically. In particular, we can run \voqc on a benchmark of circuit programs and see how well it optimizes those programs, compared to (non-verified) state-of-the-art compilers.

To this end, we compared the performance of \voqc's verified (unitary) optimizations against IBM's Qiskit compiler \cite{Qiskit2019}, CQC's \tket \cite{tket, Sivarajah2020}, PyZX \cite{Kissinger2019}, and the optimizers presented by \citet{Nam2018} and \citet{Amy2013} on a set of benchmarks developed by \citeauthor{Amy2013} We find that \voqc has comparable performance to all of these: it generally beats all but \citeauthor{Nam2018} in terms of both total gate count reduction and $T$-gate reduction, and often matches \citeauthor{Nam2018} However, our aim is not to claim superiority over these tools (after all, we have implemented a subset of the unitary optimizations available in \citeauthor{Nam2018}, and Qiskit and  \tket contain many features that \citeauthor{Nam2018} does not), but to demonstrate that the optimizations we have implemented are on par with existing \emph{unverified} tools.

\paragraph{Benchmarks}

We evaluated performance by applying \voqc and the other compilers to the benchmark of \citet{Amy2013}, which consists of programs written in the ``Clifford+T'' gate set ($CNOT$, $H$, $S$ and $T$, where $S$ and $T$ are $z$-axis rotations by $\pi/2$ and $\pi/4$, respectively). The benchmark programs contain arithmetic circuits, implementations of multiple-control $X$ gates, and Galois field multiplier circuits, ranging from 45 to 13,593 gates and 5 to 96 qubits. All of these circuits are unitary, so they only serve to evaluate our unitary circuit optimizations. We do not evaluate circuit mapping transformations.

We measure the reduction in total gate count and $T$-gate count. 
Total gate count is a useful metric for near-term quantum computing, where the length of the computation must be minimized to reduce error.
$T$-gate count is relevant in the \emph{fault-tolerant regime} where qubits are encoded using quantum error correcting codes and operations are performed fault-tolerantly.
In this regime the standard method for making Clifford+T circuits fault tolerant produces particularly expensive translations for $T$ gates, so reducing $T$-count is a common optimization goal. The Clifford+T set is a subset of \voqc's gate set where each $z$-axis rotation is restricted to be a multiple of $\pi/4$ (an odd multiple of $\pi/4$ corresponds to one $T$ gate).

Some of the benchmarks contain doubly-controlled $Z$, or $CCZ$, gates. 
Before applying optimizations we convert $CCZ$ gates to \voqc's gate set using the following standard decomposition, where \coqe{T} is the \coqe{Rz(1/4)} gate and \coqe{T$^\dagger$} is its inverse \coqe{Rz(7/4)}.
\begin{coq}
CCZ a b c := [ CNOT b c ; T$^\dagger$ c ; CNOT a c ; T c ; CNOT b c ; T$^\dagger$ c ; CNOT a c ; 
               CNOT a b ; T$^\dagger$ b ; CNOT a b ; T a ; T b ; T c ].
\end{coq}

We also evaluated the performance of \voqc on the benchmark used by \citeauthor{Nam2018}, of which \citeauthor{Amy2013} is a subset. \citeauthor{Nam2018}'s benchmark includes additional unitary circuits for simulating Hamiltonian dynamics as well as quantum Fourier transform and adder circuits, which are subroutines in Shor's integer factoring algorithm \cite{Shor94}. \voqc's results on \citeauthor{Amy2013}'s benchmark are representative of \voqc's behavior on the full set; details are given in \aref{app:extended-eval}.

\paragraph*{Baseline: Total Gate Count}

To evaluate reduction in total gate count, we compare \voqc's performance with that of \citeauthor{Nam2018}, Qiskit Terra version 0.15.2 (release date September 8, 2020), and \tket version 0.6.0 (September 18, 2020). We do not include the results from \citeauthor{Amy2013} or PyZX because their optimizations are aimed at reducing $T$-count, and often result in a higher total gate count.
\Cref{tab:list-of-opts} performs a direct comparison of functionality provided by \voqc versus \citeauthor{Nam2018}, Qiskit, and \tket.
For the Qiskit optimizations, L${}_i$ indicates that a routine is used by optimization level $i$. 
For \citeauthor{Nam2018}, P stands for ``preprocessing'' and L and H indicate whether the routine is in the ``light'' or ``heavy'' versions of the optimizer. \voqc provides the complete and verified functionality of the routines marked with $\checkmark$; we write $\checkmark^*$ to indicate that \voqc contains a verified optimization with similar, although not identical, behavior.

Compared to \citeauthor{Nam2018}'s rotation merging, \voqc performs a slightly less powerful optimization (as discussed in \Cref{sec:repl-opt}). 
Conversely, \voqc's one- and two-qubit gate cancellation routines generalize Qiskit's Optimize1qGates and CXCancellation when restricted to \voqc's gate set. 
For CommutativeCancellation, Qiskit's routine follows the same pattern as our gate cancellation routines, but uses matrix multiplication to determine whether gates commute while we use a rule-based approach; neither is strictly more effective than the other.
\tket's FullPeepholeOptimise performs local rewrites similar to those applied by Qiskit.

\begin{table}[t]
 \caption{Summary of unitary optimizations for reducing total gate count.}
\centering
\begin{tabular}{l l}
 \underline{Nam et al.} & \\
 Not propagation (P) & $\checkmark^*$ \\
 Hadamard gate reduction (L, H) & $\checkmark$ \\
 Single-qubit gate cancellation (L, H) & $\checkmark$ \\
 Two-qubit gate cancellation (L, H) & $\checkmark$ \\
 Rotation merging using phase polynomials (L) & $\checkmark^*$ \\
 Floating $R_z$ gates (H) & \\
 Special-purpose optimizations (L, H) & \\
 \quad \textbullet ~ LCR optimizer & $\checkmark$ \\
 \quad \textbullet ~ Toffoli decomposition & \\
 \hline
 \underline{Qiskit Terra 0.15.2} & \\
 CXCancellation (L${}_1$) & $\checkmark^*$ \\
 Optimize1qGates (L${}_1$, L${}_2$, L${}_3$) & $\checkmark^*$ \\
 CommutativeCancellation (L${}_2$, L${}_3$) & $\checkmark^*$ \\
 ConsolidateBlocks (L${}_3$) &  \\
 \hline
 \underline{\tket 0.6.0} & \\
 FullPeepholeOptimise & $\checkmark^*$ 
 \end{tabular}
 \label{tab:list-of-opts}
\end{table}

When evaluating Qiskit, we include all unitary optimizations up to level 3.
In our evaluation, both Qiskit and \tket use the gate set $\{u_1, u_2, u_3, \mathit{CNOT}\}$ where $u_3$ is $R_{\theta, \phi, \lambda}$ from \voqc's base set and $u_1$ and $u_2$ are $u_3$ with certain arguments fixed. This gate set gives these industrial compilers an advantage over \voqc because they can, for example, represent $X$ followed by $H$ with a single gate.

\paragraph*{Baseline: $T$-Gate Count}

To evaluate reduction in $T$-gate count, we compare \voqc against \citeauthor{Nam2018}, \citeauthor{Amy2013}, and PyZX version 0.6.0 (release date June 16, 2020). We do not include results from Qiskit or \tket because these compilers produce circuits that do not use the Clifford+T set.
When evaluating PyZX, we use the full\_reduce method, which applies an optimization similar in intent to rotation merging, but implemented in terms of the ZX-calculus.

\begin{table*}[t]
\caption{Reduced total gate counts on \citet{Amy2013} benchmarks. Red cells indicate programs optimized incorrectly. Bold results mark the best performing optimizer.}
\centering
\begin{tabular}{c|cccccc}
& \multicolumn{5}{c}{\textbf{Total Gate Count}} \\
\textbf{Name} & \textbf{Original} & \textbf{Qiskit} & \textbf{t$\vert$ket$\rangle$} & \textbf{Nam (L)} & \textbf{Nam (H)}  & \textbf{VOQC}  \\ \hline
adder\_8                      & 900               & 805             & 775         & 646  & \textbf{606}  & 682 \\
barenco\_tof\_3               & 58                & 51              & 51          & 42  & \textbf{40}   & 50 \\
barenco\_tof\_4               & 114               & 100             & 100         & 78  & \textbf{72}   & 95 \\
barenco\_tof\_5               & 170               & 149             & 149         & 114 & \textbf{104}  & 140 \\
barenco\_tof\_10              & 450               & 394             & 394         & 294 & \textbf{264}  & 365 \\
csla\_mux\_3                  & 170               & 156             & \textbf{155} & 161 & \textbf{155}  & 158 \\
csum\_mux\_9                  & 420               & 382             & 361          & 294 & \textbf{266}  & 308 \\
gf2\textasciicircum{}4\_mult  & 225               & 206             & 206          & \textbf{187} & \textbf{187}  & 192 \\
gf2\textasciicircum{}5\_mult  & 347               & 318             & 319          & 296 & 296           & \textbf{291} \\
gf2\textasciicircum{}6\_mult  & 495               & 454             & 454          & \textbf{403} & \textbf{403}  & 410 \\
gf2\textasciicircum{}7\_mult  & 669               & 614             & 614          & 555 & 555           & \textbf{549} \\
gf2\textasciicircum{}8\_mult  & 883               & 804             & 806          & 712 & 712           & \textbf{705} \\
gf2\textasciicircum{}9\_mult  & 1095              & 1006            & 1009         & 891 & 891           & \textbf{885} \\
gf2\textasciicircum{}10\_mult & 1347              & 1238            & 1240         & \textbf{1070} & \textbf{1070} & 1084 \\
gf2\textasciicircum{}16\_mult & 3435              & 3148            & 3150         & 2707 & 2707          & \textbf{2695} \\
gf2\textasciicircum{}32\_mult & 13593             & 12506           & 12507        & 10601 & 10601         & \textbf{10577} \\
mod5\_4                       & 63                & 58              & 58           & \textbf{51} & \textbf{51}   & 56 \\
mod\_mult\_55                 & 119               & 106             & 102          & 91 & 91            & \textbf{90} \\
mod\_red\_21                  & 278               & 227             & 224          & 184 & \textbf{180}  & 214 \\
qcla\_adder\_10               & 521               & 469             & 460          & 411 & \textbf{399}  & 438 \\
qcla\_com\_7                  & 443               & 398             & 392          & \textbf{284} & \textbf{284}  & 314 \\
qcla\_mod\_7                  & 884               & 793             & 780          & \cellcolor{red!25}636 & \cellcolor{red!25}624  & \textbf{723} \\
rc\_adder\_6                  & 200               & 170             & 172          & 142 & \textbf{140}  & 157 \\
tof\_3                        & 45                & 40              & 40           & \textbf{35} & \textbf{35}   & 40 \\
tof\_4                        & 75                & 66              & 66           & \textbf{55} & \textbf{55}   & 65 \\
tof\_5                        & 105               & 92              & 92           & \textbf{75} & \textbf{75}   & 90 \\
tof\_10                       & 255               & 222             & 222          & \textbf{175} & \textbf{175}  & 215 \\
vbe\_adder\_3                 & 150               & 138             & 139          & \textbf{89} & \textbf{89}   & 101 \\ \hline
\textbf{Geo. Mean Reduction}    & --                & 10.1\%          & 10.6\%       & 23.3\% & 24.8\%        & 17.8\%     
\end{tabular}
\label{tab:total-counts}
\end{table*}
\begin{table*}[t]
\caption{Reduced $T$-gate counts on the \citet{Amy2013} benchmarks. Red cells indicate programs optimized incorrectly. Bold results mark the best performing optimizer.}
\centering
\begin{tabular}{c|cccccc} & \multicolumn{5}{c}{\textbf{$T$-Gate Count}} \\
\textbf{Name} & \textbf{Original} & \textbf{Amy} & \textbf{PyZX} & \textbf{Nam (L)} & \textbf{Nam (H)} & \textbf{VOQC}  \\ \hline
adder\_8                      & 399               & 215            & \textbf{173}   & 215 & 215            & 215            \\
barenco\_tof\_3               & 28                & \textbf{16}    & \textbf{16}    & \textbf{16} & \textbf{16}    & \textbf{16}             \\
barenco\_tof\_4               & 56                & \textbf{28}    & \textbf{28}    & \textbf{28} & \textbf{28}    & \textbf{28}             \\
barenco\_tof\_5               & 84                & \textbf{40}    & \textbf{40}    & \textbf{40} & \textbf{40}    & \textbf{40}            \\
barenco\_tof\_10              & 224               & \textbf{100}   & \textbf{100}   & \textbf{100} & \textbf{100}   & \textbf{100}            \\
csla\_mux\_3                  & 70                & \cellcolor{red!25}62    & \textbf{62}    & 64 & 64             & 64             \\
csum\_mux\_9                  & 196               & 112            & \textbf{84}    & \textbf{84} & \textbf{84}    & \textbf{84}    \\
gf2\textasciicircum{}4\_mult  & 112               & \textbf{68}    & \textbf{68}    & \textbf{68} & \textbf{68}    & \textbf{68}    \\
gf2\textasciicircum{}5\_mult  & 175               & \textbf{111}   & 115            & 115 & 115            & 115            \\
gf2\textasciicircum{}6\_mult  & 252               & \textbf{150}   & \textbf{150}   &  \textbf{150} & \textbf{150}   & \textbf{150}   \\
gf2\textasciicircum{}7\_mult  & 343               & \textbf{217}   & \textbf{217}   & \textbf{217} & \textbf{217}   & \textbf{217}   \\
gf2\textasciicircum{}8\_mult  & 448               & \textbf{264}   & \textbf{264}   & \textbf{264} & \textbf{264}   & \textbf{264}   \\
gf2\textasciicircum{}9\_mult  & 567               & \textbf{351}   & \textbf{351}   & \textbf{351} & \textbf{351}   & \textbf{351}   \\
gf2\textasciicircum{}10\_mult & 700               & \textbf{410}   & \textbf{410}   & \textbf{410} & \textbf{410}   & \textbf{410}   \\
gf2\textasciicircum{}16\_mult & 1792              & \textbf{1040}  & \textbf{1040}  & \textbf{1040} & \textbf{1040}  & \textbf{1040}  \\
gf2\textasciicircum{}32\_mult & 7168              & \textbf{4128}  & \textbf{4128}  & \textbf{4128} & \textbf{4128}  & \textbf{4128}  \\
mod5\_4                       & 28                & 16             & \textbf{8}     & 16 & 16             & 16             \\
mod\_mult\_55                 & 49                & 37             & \textbf{35}    & \textbf{35} & \textbf{35}    & \textbf{35}    \\
mod\_red\_21                  & 119               & \textbf{73}    & \textbf{73}    & \textbf{73} & \textbf{73}    & \textbf{73}    \\
qcla\_adder\_10               & 238               & \textbf{162}   & \textbf{162}   & \textbf{162} & \textbf{162}   & 164            \\
qcla\_com\_7                  & 203               & \textbf{95}    & \textbf{95}    & \textbf{95} & \textbf{95}    & \textbf{95}    \\
qcla\_mod\_7                  & 413               & 249            & \textbf{237}            & \cellcolor{red!25}237 & \cellcolor{red!25}235   & 249            \\
rc\_adder\_6                  & 77                & 63             & \textbf{47}    & \textbf{47} & \textbf{47}    & \textbf{47}    \\
tof\_3                        & 21                & \textbf{15}    & \textbf{15}    & \textbf{15} & \textbf{15}    & \textbf{15}    \\
tof\_4                        & 35                & \textbf{23}    & \textbf{23}    & \textbf{23} & \textbf{23}    & \textbf{23}    \\
tof\_5                        & 49                & \textbf{31}    & \textbf{31}    & \textbf{31} & \textbf{31}    & \textbf{31}    \\
tof\_10                       & 119               & \textbf{71}    & \textbf{71}    & \textbf{71} & \textbf{71}    & \textbf{71}    \\
vbe\_adder\_3                 & 70                & \textbf{24}    & \textbf{24}    & \textbf{24} & \textbf{24}    & \textbf{24}    \\ \hline
\textbf{Geo. Mean Reduction}    & --             & 39.7\%         & 42.6\%         & 41.4\% & 41.4\%         & 41.4\%      
\end{tabular}
\label{tab:t-counts}
\end{table*}

\paragraph{Results}

The results are shown in \Cref{tab:total-counts} and \Cref{tab:t-counts}. 
In each row, we have marked in bold the gate count of the best-performing optimizer. The geometric mean of the reduction in each benchmarks is given in the last row.
Shaded cells mark that the resulting optimized circuit has been found to be inequivalent to the original circuit, indicating a bug in the relevant optimizer.\footnote{The bug in csla\_mux\_3 was found by \citet{Nam2018} and the bug in qcla\_com\_7 was found by \citet{Kissinger2019b}; both were discovered using translation validation.} Results for incorrectly-optimized circuits are n
ot included in the averages on the last line.
We do not re-run \citeauthor{Nam2018} (which is proprietary software) or \citeauthor{Amy2013}. We report results from \citet{Nam2018}.

On average, Qiskit reduces the total gate count by 10.1\%, \tket by 10.6\%, \citeauthor{Nam2018} by 23.3\% (light) and 24.8\% (heavy), and \voqc by 17.8\%.
\voqc outperforms or matches the performance of Qiskit and \tket on all benchmarks but one.
In 8 out of 28 cases \voqc outperforms \citeauthor{Nam2018} 
The gap in performance between \voqc and the industrial compilers is due to \voqc's rotation merging optimization, which has no analogue in Qiskit or \tket.
The gap in performance between \citeauthor{Nam2018} and \voqc is due to the fact that we have not yet implemented all their optimization passes (per \Cref{tab:list-of-opts}). 
In particular, \citeauthor{Nam2018}'s ``special-purpose Toffoli decomposition'' (which affects how $CCZ$ gates are decomposed) enables rotation merging and single-qubit gate cancellation to cancel two gates (e.g. cancel $T$ and $T^\dagger$) where we instead combine two gates into one (e.g. $T$ and $T$ becomes $P$).
Interestingly, the cases where \voqc outperforms \citeauthor{Nam2018} can also be attributed to their Toffoli decomposition heuristics, which sometimes result in fewer cancellations than the na\"ive decomposition that we use.
We do not expect adding and verifying this form of Toffoli decomposition to pose a challenge in \voqc.

\voqc's performance is closer to \citeauthor{Nam2018}'s when considering $T$-count. On average, \citeauthor{Amy2013} reduce the $T$-gate count by 39.7\%, PyZX by 42.6\%, and \citeauthor{Nam2018} and \voqc by 41.4\%. \voqc matches \citeauthor{Nam2018} on all benchmarks but two. The first case (qcla\_adder\_10) is due to our simplification in rotation merging. In the second case (qcla\_mod\_7), \citeauthor{Nam2018}'s optimized circuit was later found to be inequivalent to the original circuit \cite{Kissinger2019b}, so the lower $T$-count is spurious.
On 16 benchmarks, all optimizers produce the same $T$-count. This is somewhat surprising since, although all these optimizers rely on some form of rotation merging, their implementations differ substantially. \citet{Kissinger2019b} posit that these results indicate a local optimum in the ancilla-free case for some of the benchmarks (in particular the tof benchmarks, whose $T$-count is not reduced by applying additional techniques \cite{Heyfron2018}).

To compare the running times of the different tools, we ran 11 trials of \voqc, Qiskit, \tket, and PyZX (taking the median time for each benchmark) on a standard laptop with a 2.9 GHz Intel Core i5 processor and 16 GB of 1867 MHz DDR3 memory, running macOS Catalina. We consider the timings for \citeauthor{Amy2013} and \citeauthor{Nam2018} given in \citet[Table 4]{Nam2018}, which were measured on a similar machine with 8 GB RAM running OS X El Capitan. We show the geometric mean running times over all 28 benchmarks below. 
\begin{center}
    \begin{tabular}{c|c|c|c|c|c|c}
        \voqc & Nam (L) & Nam (H) & Qiskit & \tket & Amy & PyZX \\
        \hline
        0.013s & 0.002s & 0.018s & 0.812s & 0.129s & 0.007s & 0.384s
    \end{tabular}
\end{center}

All the tools are fast; \citeauthor{Nam2018} light optimization tends to be the fastest and Qiskit tends to be the slowest. However, these means are not the entire story: the tools' performances scale differently with increasing qubit and gate count. For example, on gf2\textasciicircum{}32\_mult (the largest benchmark) Qiskit and \voqc are comparable with running times of 31.6s and 27.4s respectively; \citeauthor{Nam2018} light optimization and \tket are very fast with running times of 1.8s and 7.0s; and \citeauthor{Nam2018} heavy optimization, \citeauthor{Amy2013}, and PyZX are fairly slow with running times of 275.7s, 602.6s, and 577.1s.
\iftoggle{submission}{ }
{ For more details on \voqc's performance, see \aref{app:extended-eval}. }

These results are encouraging evidence that \voqc supports useful and interesting verified optimizations, and that we have faithfully implemented \citeauthor{Nam2018}'s optimizations.
Furthermore, despite having been written with verification in mind, \voqc's running times are not significantly worse than (and sometimes better than) that of current tools.

\paragraph{Trusted Code}
For performance, \voqc uses OCaml primitives for describing rational numbers, maps and sets, rather than the code extracted from Coq. 
Thus we implicitly trust that the OCaml implementation of these data types is consistent with Coq's; we believe that this is a reasonable assumption.
Furthermore, our translation from OpenQASM to \sqir and extraction from Coq to OCaml are not formally verified.


\section{Related Work}
\label{sec:related}

Our work on \voqc and \sqir is primarily related to work on quantum program compilation, especially work aiming to add assurance to the compilation process. It is also related to work on quantum source-program verification.

\paragraph*{Verified Quantum Compilation}

Quantum compilation is an active area. In addition to Qiskit, \tket, and \citet{Nam2018} (discussed in \Cref{sec:experiments}), other recent compiler efforts include quilc~\cite{quilc}, ScaffCC \cite{JavadiAbhari2014}, and Project Q~\cite{Steiger2018}. 
Due to resource limits on near-term quantum computers, most compilers for quantum programs contain some degree of optimization, and nearly all place an emphasis on satisfying architectural requirements, like mapping to a particular gate set or qubit topology.
None of the optimization or mapping code in these compilers is formally verified.

However, \voqc is not the only quantum compiler to which automated reasoning or formal verification has been applied.
\citet{amy18reversible} developed a certified optimizing compiler from source Boolean expressions to reversible circuits, but did not handle general quantum programs. \citet{Rand2018} developed a similar compiler for quantum circuits but without optimizations (using the \qwire language). 

The problem of optimization verification has also been considered in the context of the ZX-calculus~\cite{Coecke2011}, which is a formalism for describing quantum tensor networks (which generalize quantum circuits) based on categorical quantum mechanics~\cite{Abramsky2009}. The ZX-calculus is characterized by a small set of rewrite rules that allow translation of a diagram to any other diagram representing the same computation~\cite{Jeandel2018}. \citet{Fagan2018} verified an optimizer for ZX diagrams representing Clifford circuits 
(which use the non-universal gate set $\{\mathit{CNOT}, H, S\}$) 
in the Quantomatic graphical proof assistant~\cite{Kissinger2015}. 
PyZX~\cite{Kissinger2019} uses ZX diagrams as an intermediate representation for compiling quantum circuits, and generally achieves performance comparable to leading compilers~\cite{Kissinger2019b}. While PyZX is not verified in a proof assistant like Coq (the ``Py'' stands for Python), it does rely on a small, well-studied equational theory. 
Additionally, PyZX can perform translation validation to check if a compiled circuit is equivalent to the original. However, PyZX's translation validator is not guaranteed to succeed for any two equivalent circuits. 

A recent paper from \citet{Smith2019} presents a compiler with built-in translation validation via QMDD equivalence checking \cite{Miller2006}. However the optimizations they consider are much simpler than \voqc's and the QMDD approach scales poorly with increasing number of qubits. Our optimizations are all verified for arbitrary dimension.

Concurrently with our work, \citet{Shi2019} developed CertiQ, an approach to verifying properties of circuit transformations in the Qiskit compiler, which is implemented in Python. Their approach has two steps. First, it uses matrix multiplication to check that the unitary semantics of two concrete gate patterns are equivalent. Second, it uses symbolic execution to generate verification conditions for parts of Qiskit that manipulate circuits. These are given to an SMT solver to verify that pattern equivalences are applied correctly according to programmer-provided function specifications and invariants. That CertiQ can analyze Python code directly in a mostly automated fashion is appealing.
However, it is limited in the optimizations it can verify. 
For example, equivalences that range over arbitrary indices, like $\mathit{CNOT}~m~x;~\mathit{CNOT}~n~x \equiv \mathit{CNOT}~n~x;~\mathit{CNOT}~m~x$ cannot be verified by matrix multiplication; CertiQ checks a concrete instance of this pattern and then applies it to more general circuits. 
More complex optimizations like rotation merging (the most powerful optimization in our experiments) cannot be generalized from simple, concrete circuits. CertiQ may also fail to prove an optimization correct, e.g., because of complicated control code; in this case it falls back to translation validation, which adds extra cost and the possibility of failure at run-time. By contrast, every optimization in \voqc has been proved correct.
Finally, CertiQ does not directly represent the semantics of quantum programs, so it cannot be used as a tool for verifying general properties of a program's semantics (as we do in \Cref{sec:sqire}).

\paragraph*{Verified Quantum Programming}

We designed \sqir primarily as the intermediate language for \voqc's verified optimizations, but it can be used for verified source programming as well, per \Cref{sec:sqire} and ongoing work \cite{CPPsub}. Early attempts to formally verify aspects of a quantum computation in a proof assistant were an Agda implementation of the Quantum IO Monad \cite{Green2010} and a small Coq quantum library by \citet{Boender2015}. Later, \citet{Rand2017} embedded the higher-level \qwire programming language in the Coq proof assistant, and used it to verify a variety of simple programs~\citep{Rand2017}, assertions regarding ancilla qubits~\citep{Rand2018}, and its own metatheory~\citep{RandThesis}. 
\voqc and \sqir reuse parts of \qwire's Coq development, and take inspiration and lessons from its design. However, as discussed in \Cref{sec:discussion} and \aref{app:sqir-v-qwire}, \qwire's higher-level abstractions (notably, its representation of variables using higher-order abstract syntax) complicate verification.

Concurrently with this work, \citet{qbricks} introduced \qbricks, a tool implemented in Why3 \cite{filliatre13esop} whose aim is to support mostly-automated verification of complex quantum algorithms. Their design in many ways mirrors \sqir's: both tools provide special support for reasoning about quantum programs and the languages are simplified so that programs have a straightforward translation to their semantics. A \qbricks program specifies a circuit without variables, instead using operators for parallel and sequential composition. \qbricks defines the meaning of its programs using a ``higher order'' path-sum semantics \cite{Amy2018}, which limits it to unitary programs, but substantially enhances automation. \sqir and \qbricks have been used to verify similar algorithms (e.g., Grover's algorithm and quantum phase estimation~\cite{CPPsub}), but \qbricks' approach reduces manual effort.




\section{Conclusions and Future Work}

This paper has presented \voqc, the first fully verified optimizer for quantum circuits. A key component of \voqc is \sqir, a simple, low-level quantum language deeply embedded in the the Coq proof assistant, which gives a semantics to quantum programs that is amenable to proof. Optimization passes are expressed as Coq functions which are proved to preserve the semantics of their input \sqir programs. \voqc's optimizations are mostly based on local circuit equivalences, implemented by replacing one pattern of gates with another, or commuting a gate rightward until it can be cancelled. Others, like rotation merging, are more complex. These were inspired by, and in some cases generalize, optimizations in industrial compilers, but in \voqc are proved correct. When applied to a benchmark suite of 28 circuit programs, we found \voqc performed comparably to state-of-the-art compilers, reducing gate count on average by 17.8\% compared to 10.1\% for IBM's Qiskit compiler, 10.6\% for CQC's \tket, and 24.8\% for the cutting-edge research optimizer by \citet{Nam2018}.
Furthermore, \voqc reduced $T$-gate count on average by 41.4\% compared to 39.7\% by \citet{Amy2013}, 41.4\% by \citeauthor{Nam2018}, and 42.6\% by the PyZX optimizer.

Moving forward, we plan to incorporate \voqc into a full-featured verified compilation stack for quantum programs, following the vision of a recent Computing Community Consortium report \cite{Martonosi2019}.
We can verify compilation from high-level languages with formal semantics like Silq \cite{silq} to \sqir circuits.
We can implement validated parsers~\cite{Jourdan:2012:VLP:2259248.2259268} for languages like OpenQASM and verify their translation to \sqir (e.g., using metaQASM's semantics~\cite{Amy2019}); this work is already in progress~\cite{Singhal2020}.
We can also add support for hardware-specific transformations that compile to a particular gate set. Indeed, most of the sophisticated code in Qiskit is devoted to efficiently mapping programs to IBM's architecture, and IBM's 2018 Developer Challenge centered around designing new circuit mapping algorithms \cite{ibm-dev-challenge-2018}.
We leave it as future work to incorporate optimizations and mapping algorithms from additional compilers into \voqc. Our experience so far makes us optimistic about the prospects for doing so successfully.


\begin{acks}
We thank Leonidas Lampropoulos, Kartik Singhal, and anonymous reviewers for their helpful comments on drafts of this paper.
This material is based upon work supported by the \grantsponsor{ascr}{U.S. Department of Energy, Office of Science, Office of Advanced Scientific Computing Research}{http://dx.doi.org/10.13039/100006192}, Quantum Testbed Pathfinder Program under Award Number \grantnum{ascr}{DE-SC0019040}.
  
\end{acks}

\balance
\bibliography{references}


\begin{thebibliography}{66}


\ifx \showCODEN    \undefined \def \showCODEN     #1{\unskip}     \fi
\ifx \showDOI      \undefined \def \showDOI       #1{#1}\fi
\ifx \showISBNx    \undefined \def \showISBNx     #1{\unskip}     \fi
\ifx \showISBNxiii \undefined \def \showISBNxiii  #1{\unskip}     \fi
\ifx \showISSN     \undefined \def \showISSN      #1{\unskip}     \fi
\ifx \showLCCN     \undefined \def \showLCCN      #1{\unskip}     \fi
\ifx \shownote     \undefined \def \shownote      #1{#1}          \fi
\ifx \showarticletitle \undefined \def \showarticletitle #1{#1}   \fi
\ifx \showURL      \undefined \def \showURL       {\relax}        \fi
\providecommand\bibfield[2]{#2}
\providecommand\bibinfo[2]{#2}
\providecommand\natexlab[1]{#1}
\providecommand\showeprint[2][]{arXiv:#2}

\bibitem[\protect\citeauthoryear{Abramsky and Coecke}{Abramsky and
  Coecke}{2009}]%
        {Abramsky2009}
\bibfield{author}{\bibinfo{person}{Samson Abramsky} {and} \bibinfo{person}{Bob
  Coecke}.} \bibinfo{year}{2009}\natexlab{}.
\newblock \showarticletitle{Categorical quantum mechanics}.
\newblock \bibinfo{journal}{\emph{Handbook of quantum logic and quantum
  structures}}  \bibinfo{volume}{2} (\bibinfo{year}{2009}),
  \bibinfo{pages}{261--325}.
\newblock


\bibitem[\protect\citeauthoryear{Aleksandrowicz, Alexander, Barkoutsos, Bello,
  Ben-Haim, Bucher, Cabrera-Hern{\'a}ndez, Carballo-Franquis, Chen, Chen, Chow,
  C{\'o}rcoles-Gonzales, Cross, Cross, Cruz-Benito, Culver, Gonz{\'a}lez,
  Torre, Ding, Dumitrescu, Duran, Eendebak, Everitt, Sertage, Frisch, Fuhrer,
  Gambetta, Gago, Gomez-Mosquera, Greenberg, Hamamura, Havlicek, Hellmers,
  Herok, Horii, Hu, Imamichi, Itoko, Javadi-Abhari, Kanazawa, Karazeev,
  Krsulich, Liu, Luh, Maeng, Marques, Mart{\'\i}n-Fern{\'a}ndez, McClure,
  McKay, Meesala, Mezzacapo, Moll, Rodr{\'\i}guez, Nannicini, Nation,
  Ollitrault, O'Riordan, Paik, P{\'e}rez, Phan, Pistoia, Prutyanov, Reuter,
  Rice, Davila, Rudy, Ryu, Sathaye, Schnabel, Schoute, Setia, Shi, Silva,
  Siraichi, Sivarajah, Smolin, Soeken, Takahashi, Tavernelli, Taylor, Taylour,
  Trabing, Treinish, Turner, Vogt-Lee, Vuillot, Wildstrom, Wilson, Winston,
  Wood, Wood, W{\"o}rner, Akhalwaya, and Zoufal}{Aleksandrowicz
  et~al\mbox{.}}{2019}]%
        {Qiskit2019}
\bibfield{author}{\bibinfo{person}{Gadi Aleksandrowicz},
  \bibinfo{person}{Thomas Alexander}, \bibinfo{person}{Panagiotis Barkoutsos},
  \bibinfo{person}{Luciano Bello}, \bibinfo{person}{Yael Ben-Haim},
  \bibinfo{person}{David Bucher}, \bibinfo{person}{Francisco~Jose
  Cabrera-Hern{\'a}ndez}, \bibinfo{person}{Jorge Carballo-Franquis},
  \bibinfo{person}{Adrian Chen}, \bibinfo{person}{Chun-Fu Chen},
  \bibinfo{person}{Jerry~M. Chow}, \bibinfo{person}{Antonio~D.
  C{\'o}rcoles-Gonzales}, \bibinfo{person}{Abigail~J. Cross},
  \bibinfo{person}{Andrew Cross}, \bibinfo{person}{Juan Cruz-Benito},
  \bibinfo{person}{Chris Culver}, \bibinfo{person}{Salvador De La~Puente
  Gonz{\'a}lez}, \bibinfo{person}{Enrique De~La Torre}, \bibinfo{person}{Delton
  Ding}, \bibinfo{person}{Eugene Dumitrescu}, \bibinfo{person}{Ivan Duran},
  \bibinfo{person}{Pieter Eendebak}, \bibinfo{person}{Mark Everitt},
  \bibinfo{person}{Ismael~Faro Sertage}, \bibinfo{person}{Albert Frisch},
  \bibinfo{person}{Andreas Fuhrer}, \bibinfo{person}{Jay Gambetta},
  \bibinfo{person}{Borja~Godoy Gago}, \bibinfo{person}{Juan Gomez-Mosquera},
  \bibinfo{person}{Donny Greenberg}, \bibinfo{person}{Ikko Hamamura},
  \bibinfo{person}{Vojtech Havlicek}, \bibinfo{person}{Joe Hellmers},
  \bibinfo{person}{{\L}ukasz Herok}, \bibinfo{person}{Hiroshi Horii},
  \bibinfo{person}{Shaohan Hu}, \bibinfo{person}{Takashi Imamichi},
  \bibinfo{person}{Toshinari Itoko}, \bibinfo{person}{Ali Javadi-Abhari},
  \bibinfo{person}{Naoki Kanazawa}, \bibinfo{person}{Anton Karazeev},
  \bibinfo{person}{Kevin Krsulich}, \bibinfo{person}{Peng Liu},
  \bibinfo{person}{Yang Luh}, \bibinfo{person}{Yunho Maeng},
  \bibinfo{person}{Manoel Marques}, \bibinfo{person}{Francisco~Jose
  Mart{\'\i}n-Fern{\'a}ndez}, \bibinfo{person}{Douglas~T. McClure},
  \bibinfo{person}{David McKay}, \bibinfo{person}{Srujan Meesala},
  \bibinfo{person}{Antonio Mezzacapo}, \bibinfo{person}{Nikolaj Moll},
  \bibinfo{person}{Diego~Moreda Rodr{\'\i}guez}, \bibinfo{person}{Giacomo
  Nannicini}, \bibinfo{person}{Paul Nation}, \bibinfo{person}{Pauline
  Ollitrault}, \bibinfo{person}{Lee~James O'Riordan}, \bibinfo{person}{Hanhee
  Paik}, \bibinfo{person}{Jes{\'u}s P{\'e}rez}, \bibinfo{person}{Anna Phan},
  \bibinfo{person}{Marco Pistoia}, \bibinfo{person}{Viktor Prutyanov},
  \bibinfo{person}{Max Reuter}, \bibinfo{person}{Julia Rice},
  \bibinfo{person}{Abd{\'o}n~Rodr{\'\i}guez Davila}, \bibinfo{person}{Raymond
  Harry~Putra Rudy}, \bibinfo{person}{Mingi Ryu}, \bibinfo{person}{Ninad
  Sathaye}, \bibinfo{person}{Chris Schnabel}, \bibinfo{person}{Eddie Schoute},
  \bibinfo{person}{Kanav Setia}, \bibinfo{person}{Yunong Shi},
  \bibinfo{person}{Adenilton Silva}, \bibinfo{person}{Yukio Siraichi},
  \bibinfo{person}{Seyon Sivarajah}, \bibinfo{person}{John~A. Smolin},
  \bibinfo{person}{Mathias Soeken}, \bibinfo{person}{Hitomi Takahashi},
  \bibinfo{person}{Ivano Tavernelli}, \bibinfo{person}{Charles Taylor},
  \bibinfo{person}{Pete Taylour}, \bibinfo{person}{Kenso Trabing},
  \bibinfo{person}{Matthew Treinish}, \bibinfo{person}{Wes Turner},
  \bibinfo{person}{Desiree Vogt-Lee}, \bibinfo{person}{Christophe Vuillot},
  \bibinfo{person}{Jonathan~A. Wildstrom}, \bibinfo{person}{Jessica Wilson},
  \bibinfo{person}{Erick Winston}, \bibinfo{person}{Christopher Wood},
  \bibinfo{person}{Stephen Wood}, \bibinfo{person}{Stefan W{\"o}rner},
  \bibinfo{person}{Ismail~Yunus Akhalwaya}, {and} \bibinfo{person}{Christa
  Zoufal}.} \bibinfo{year}{2019}\natexlab{}.
\newblock \bibinfo{title}{Qiskit: An open-source framework for quantum
  computing}.
\newblock
\newblock
\urldef\tempurl%
\url{https://doi.org/10.5281/zenodo.2562110}
\showDOI{\tempurl}


\bibitem[\protect\citeauthoryear{Altenkirch and Green}{Altenkirch and
  Green}{2010}]%
        {Altenkirch2010}
\bibfield{author}{\bibinfo{person}{Thorsten Altenkirch} {and}
  \bibinfo{person}{Alexander~S Green}.} \bibinfo{year}{2010}\natexlab{}.
\newblock \showarticletitle{The quantum {IO} monad}.
\newblock \bibinfo{journal}{\emph{Semantic Techniques in Quantum Computation}}
  (\bibinfo{year}{2010}), \bibinfo{pages}{173--205}.
\newblock


\bibitem[\protect\citeauthoryear{Amy}{Amy}{2018}]%
        {Amy2018}
\bibfield{author}{\bibinfo{person}{Matthew Amy}.}
  \bibinfo{year}{2018}\natexlab{}.
\newblock \bibinfo{title}{Towards large-scale functional verification of
  universal quantum circuits}.
\newblock
  \bibinfo{howpublished}{\url{https://www.mathstat.dal.ca/qpl2018/papers/QPL_2018_paper_30.pdf}}.
\newblock
\newblock
\shownote{Presented at QPL 2018.}


\bibitem[\protect\citeauthoryear{Amy}{Amy}{2019}]%
        {Amy2019}
\bibfield{author}{\bibinfo{person}{Matthew Amy}.}
  \bibinfo{year}{2019}\natexlab{}.
\newblock \showarticletitle{Sized types for low-level quantum metaprogramming}.
  In \bibinfo{booktitle}{\emph{Reversible Computation}},
  \bibfield{editor}{\bibinfo{person}{Michael~Kirkedal Thomsen} {and}
  \bibinfo{person}{Mathias Soeken}} (Eds.). \bibinfo{publisher}{Springer
  International Publishing}, \bibinfo{address}{Cham}, \bibinfo{pages}{87--107}.
\newblock
\urldef\tempurl%
\url{https://doi.org/10.1007/978-3-030-21500-2_6}
\showDOI{\tempurl}


\bibitem[\protect\citeauthoryear{Amy, Azimzadeh, and Mosca}{Amy
  et~al\mbox{.}}{2018}]%
        {Amy2018b}
\bibfield{author}{\bibinfo{person}{Matthew Amy}, \bibinfo{person}{Parsiad
  Azimzadeh}, {and} \bibinfo{person}{Michele Mosca}.}
  \bibinfo{year}{2018}\natexlab{}.
\newblock \showarticletitle{On the controlled-{NOT} complexity of
  controlled-{NOT}{\textendash}phase circuits}.
\newblock \bibinfo{journal}{\emph{Quantum Science and Technology}}
  \bibinfo{volume}{4}, \bibinfo{number}{1} (\bibinfo{year}{2018}).
\newblock


\bibitem[\protect\citeauthoryear{Amy, Maslov, and Mosca}{Amy
  et~al\mbox{.}}{2013}]%
        {Amy2013}
\bibfield{author}{\bibinfo{person}{Matthew Amy}, \bibinfo{person}{Dmitri
  Maslov}, {and} \bibinfo{person}{Michele Mosca}.}
  \bibinfo{year}{2013}\natexlab{}.
\newblock \showarticletitle{Polynomial-time {T}-depth optimization of
  {C}lifford+{T} circuits via matroid partitioning}.
\newblock \bibinfo{journal}{\emph{IEEE Transactions on Computer-Aided Design of
  Integrated Circuits and Systems}}  \bibinfo{volume}{33} (\bibinfo{date}{03}
  \bibinfo{year}{2013}).
\newblock
\urldef\tempurl%
\url{https://doi.org/10.1109/TCAD.2014.2341953}
\showDOI{\tempurl}


\bibitem[\protect\citeauthoryear{Amy, Roetteler, and Svore}{Amy
  et~al\mbox{.}}{2017}]%
        {amy18reversible}
\bibfield{author}{\bibinfo{person}{Matthew Amy}, \bibinfo{person}{Martin
  Roetteler}, {and} \bibinfo{person}{Krysta~M. Svore}.}
  \bibinfo{year}{2017}\natexlab{}.
\newblock \showarticletitle{Verified compilation of space-efficient reversible
  circuits}. In \bibinfo{booktitle}{\emph{Proceedings of the 28th International
  Conference on Computer Aided Verification (CAV 2017)}}.
  \bibinfo{publisher}{Springer}.
\newblock
\urldef\tempurl%
\url{https://www.microsoft.com/en-us/research/publication/verified-compilation-of-space-efficient-reversible-circuits/}
\showURL{%
\tempurl}


\bibitem[\protect\citeauthoryear{Bennett and Brassard}{Bennett and
  Brassard}{2020}]%
        {bennett2020quantum}
\bibfield{author}{\bibinfo{person}{Charles~H Bennett} {and}
  \bibinfo{person}{Gilles Brassard}.} \bibinfo{year}{2020}\natexlab{}.
\newblock \showarticletitle{Quantum cryptography: Public key distribution and
  coin tossing}.
\newblock \bibinfo{journal}{\emph{arXiv preprint arXiv:2003.06557}}
  (\bibinfo{year}{2020}).
\newblock


\bibitem[\protect\citeauthoryear{Bichsel, Baader, Gehr, and Vechev}{Bichsel
  et~al\mbox{.}}{2020}]%
        {silq}
\bibfield{author}{\bibinfo{person}{Benjamin Bichsel},
  \bibinfo{person}{Maximilian Baader}, \bibinfo{person}{Timon Gehr}, {and}
  \bibinfo{person}{Martin Vechev}.} \bibinfo{year}{2020}\natexlab{}.
\newblock \showarticletitle{Silq: A high-level quantum language with safe
  uncomputation and intuitive semantics}. In
  \bibinfo{booktitle}{\emph{Proceedings of the 41st ACM SIGPLAN Conference on
  Programming Language Design and Implementation}} \emph{(\bibinfo{series}{PLDI
  2020})}. \bibinfo{publisher}{Association for Computing Machinery},
  \bibinfo{address}{New York, NY, USA}.
\newblock
\urldef\tempurl%
\url{https://doi.org/10.1145/3385412.3386007}
\showDOI{\tempurl}


\bibitem[\protect\citeauthoryear{Boender, Kamm{\"u}ller, and Nagarajan}{Boender
  et~al\mbox{.}}{2015}]%
        {Boender2015}
\bibfield{author}{\bibinfo{person}{Jaap Boender}, \bibinfo{person}{Florian
  Kamm{\"u}ller}, {and} \bibinfo{person}{Rajagopal Nagarajan}.}
  \bibinfo{year}{2015}\natexlab{}.
\newblock \showarticletitle{Formalization of quantum protocols using {Coq}}. In
  \bibinfo{booktitle}{\emph{Proceedings of the 12th International Workshop on
  Quantum Physics and Logic, Oxford, U.K., July 15-17, 2015}}
  \emph{(\bibinfo{series}{Electronic Proceedings in Theoretical Computer
  Science}, Vol.~\bibinfo{volume}{195})},
  \bibfield{editor}{\bibinfo{person}{Chris Heunen}, \bibinfo{person}{Peter
  Selinger}, {and} \bibinfo{person}{Jamie Vicary}} (Eds.).
  \bibinfo{publisher}{Open Publishing Association}, \bibinfo{pages}{71--83}.
\newblock
\urldef\tempurl%
\url{https://doi.org/10.4204/EPTCS.195.6}
\showDOI{\tempurl}


\bibitem[\protect\citeauthoryear{{Cambridge Quantum Computing Ltd}}{{Cambridge
  Quantum Computing Ltd}}{2019}]%
        {tket}
\bibfield{author}{\bibinfo{person}{{Cambridge Quantum Computing Ltd}}.}
  \bibinfo{year}{2019}\natexlab{}.
\newblock \bibinfo{title}{pytket}.
\newblock
\newblock
\urldef\tempurl%
\url{https://cqcl.github.io/pytket/build/html/index.html}
\showURL{%
\tempurl}


\bibitem[\protect\citeauthoryear{{Chareton}, {Bardin}, {Bobot}, {Perrelle}, and
  {Valiron}}{{Chareton} et~al\mbox{.}}{2020}]%
        {qbricks}
\bibfield{author}{\bibinfo{person}{Christophe {Chareton}},
  \bibinfo{person}{S{\'e}bastien {Bardin}}, \bibinfo{person}{Fran{\c{c}}ois
  {Bobot}}, \bibinfo{person}{Valentin {Perrelle}}, {and}
  \bibinfo{person}{Benoit {Valiron}}.} \bibinfo{year}{2020}\natexlab{}.
\newblock \showarticletitle{Toward certified quantum programming}.
\newblock \bibinfo{journal}{\emph{arXiv e-prints}} (\bibinfo{year}{2020}).
\newblock
\showeprint[arxiv]{2003.05841}~[cs.PL]


\bibitem[\protect\citeauthoryear{Coecke and Duncan}{Coecke and Duncan}{2009}]%
        {Coecke2011}
\bibfield{author}{\bibinfo{person}{Bob Coecke} {and} \bibinfo{person}{Ross
  Duncan}.} \bibinfo{year}{2009}\natexlab{}.
\newblock \showarticletitle{Interacting quantum observables: Categorical
  algebra and diagrammatics}.
\newblock \bibinfo{journal}{\emph{New Journal of Physics}}
  \bibinfo{volume}{13} (\bibinfo{date}{06} \bibinfo{year}{2009}).
\newblock
\urldef\tempurl%
\url{https://doi.org/10.1088/1367-2630/13/4/043016}
\showDOI{\tempurl}


\bibitem[\protect\citeauthoryear{{Coq Development Team}}{{Coq Development
  Team}}{2019}]%
        {coq}
\bibfield{author}{\bibinfo{person}{The {Coq Development Team}}.}
  \bibinfo{year}{2019}\natexlab{}.
\newblock \bibinfo{title}{The {C}oq proof assistant, version 8.10.0}.
\newblock
\newblock
\urldef\tempurl%
\url{https://doi.org/10.5281/zenodo.3476303}
\showDOI{\tempurl}


\bibitem[\protect\citeauthoryear{{Cross}, {Bishop}, {Smolin}, and
  {Gambetta}}{{Cross} et~al\mbox{.}}{2017}]%
        {Cross2017}
\bibfield{author}{\bibinfo{person}{Andrew~W. {Cross}}, \bibinfo{person}{Lev~S.
  {Bishop}}, \bibinfo{person}{John~A. {Smolin}}, {and} \bibinfo{person}{Jay~M.
  {Gambetta}}.} \bibinfo{year}{2017}\natexlab{}.
\newblock \showarticletitle{{Open quantum assembly language}}.
\newblock \bibinfo{journal}{\emph{arXiv e-prints}} (\bibinfo{date}{Jul}
  \bibinfo{year}{2017}).
\newblock
\showeprint[arxiv]{1707.03429}~[quant-ph]


\bibitem[\protect\citeauthoryear{de~Bruijn}{de~Bruijn}{1972}]%
        {deBruijn1972}
\bibfield{author}{\bibinfo{person}{Nicolaas~Govert de Bruijn}.}
  \bibinfo{year}{1972}\natexlab{}.
\newblock \showarticletitle{Lambda calculus notation with nameless dummies, a
  tool for automatic formula manipulation, with application to the
  Church-Rosser theorem}. In \bibinfo{booktitle}{\emph{Indagationes
  Mathematicae (Proceedings)}}, Vol.~\bibinfo{volume}{75}. Elsevier,
  \bibinfo{pages}{381--392}.
\newblock
\urldef\tempurl%
\url{https://doi.org/10.1016/1385-7258(72)90034-0}
\showDOI{\tempurl}


\bibitem[\protect\citeauthoryear{{Fagan} and {Duncan}}{{Fagan} and
  {Duncan}}{2018}]%
        {Fagan2018}
\bibfield{author}{\bibinfo{person}{Andrew {Fagan}} {and} \bibinfo{person}{Ross
  {Duncan}}.} \bibinfo{year}{2018}\natexlab{}.
\newblock \showarticletitle{Optimising {C}lifford circuits with {Q}uantomatic}.
  In \bibinfo{booktitle}{\emph{Proceedings of the 15th International Conference
  on Quantum Physics and Logic, {QPL} 2018, Halifax, Nova Scotia, 3-7 June
  2018}}.
\newblock
\urldef\tempurl%
\url{https://doi.org/10.4204/EPTCS.287.5}
\showDOI{\tempurl}


\bibitem[\protect\citeauthoryear{Filli\^atre and Paskevich}{Filli\^atre and
  Paskevich}{2013}]%
        {filliatre13esop}
\bibfield{author}{\bibinfo{person}{Jean-Christophe Filli\^atre} {and}
  \bibinfo{person}{Andrei Paskevich}.} \bibinfo{year}{2013}\natexlab{}.
\newblock \showarticletitle{Why3 --- Where programs meet provers}. In
  \bibinfo{booktitle}{\emph{Proceedings of the 22nd European Symposium on
  Programming}} \emph{(\bibinfo{series}{Lecture Notes in Computer Science})}.
\newblock


\bibitem[\protect\citeauthoryear{Gottesman}{Gottesman}{2010}]%
        {gottesman2010introduction}
\bibfield{author}{\bibinfo{person}{Daniel Gottesman}.}
  \bibinfo{year}{2010}\natexlab{}.
\newblock \showarticletitle{An introduction to quantum error correction and
  fault-tolerant quantum computation}. In \bibinfo{booktitle}{\emph{Quantum
  information science and its contributions to mathematics, Proceedings of
  Symposia in Applied Mathematics}}, Vol.~\bibinfo{volume}{68}.
  \bibinfo{pages}{13--58}.
\newblock


\bibitem[\protect\citeauthoryear{Green, Lumsdaine, Ross, Selinger, and
  Valiron}{Green et~al\mbox{.}}{2013}]%
        {Green2013}
\bibfield{author}{\bibinfo{person}{Alexander Green},
  \bibinfo{person}{Peter~LeFanu Lumsdaine}, \bibinfo{person}{Neil~J. Ross},
  \bibinfo{person}{Peter Selinger}, {and} \bibinfo{person}{Beno{\^i}t
  Valiron}.} \bibinfo{year}{2013}\natexlab{}.
\newblock \showarticletitle{Quipper: A scalable quantum programming language}.
  In \bibinfo{booktitle}{\emph{Proceedings of the 34th ACM SIGPLAN Conference
  on Programming Language Design and Implementation}}
  \emph{(\bibinfo{series}{PLDI 2013})}. \bibinfo{pages}{333--342}.
\newblock
\urldef\tempurl%
\url{https://doi.org/10.1145/2491956.2462177}
\showDOI{\tempurl}


\bibitem[\protect\citeauthoryear{Green}{Green}{2010}]%
        {Green2010}
\bibfield{author}{\bibinfo{person}{Alexander~S Green}.}
  \bibinfo{year}{2010}\natexlab{}.
\newblock \emph{\bibinfo{title}{Towards a formally verified functional quantum
  programming language}}.
\newblock \bibinfo{thesistype}{Ph.D. Dissertation}. \bibinfo{school}{University
  of Nottingham}.
\newblock


\bibitem[\protect\citeauthoryear{Greenberger, Horne, and Zeilinger}{Greenberger
  et~al\mbox{.}}{1989}]%
        {Greenberger1989}
\bibfield{author}{\bibinfo{person}{Daniel~M. Greenberger},
  \bibinfo{person}{Michael~A. Horne}, {and} \bibinfo{person}{Anton Zeilinger}.}
  \bibinfo{year}{1989}\natexlab{}.
\newblock \bibinfo{booktitle}{\emph{Going beyond {B}ell's {T}heorem}}.
\newblock \bibinfo{publisher}{Springer Netherlands},
  \bibinfo{address}{Dordrecht}, \bibinfo{pages}{69--72}.
\newblock
\showISBNx{978-94-017-0849-4}
\urldef\tempurl%
\url{https://doi.org/10.1007/978-94-017-0849-4_10}
\showDOI{\tempurl}


\bibitem[\protect\citeauthoryear{Grover}{Grover}{1996}]%
        {Grover1996}
\bibfield{author}{\bibinfo{person}{Lov~K Grover}.}
  \bibinfo{year}{1996}\natexlab{}.
\newblock \showarticletitle{A fast quantum mechanical algorithm for database
  search}. In \bibinfo{booktitle}{\emph{Proceedings of the twenty-eighth annual
  ACM Symposium on Theory of Computing}}. \bibinfo{pages}{212--219}.
\newblock


\bibitem[\protect\citeauthoryear{Heyfron and T.~Campbell}{Heyfron and
  T.~Campbell}{2018}]%
        {Heyfron2018}
\bibfield{author}{\bibinfo{person}{Luke Heyfron} {and} \bibinfo{person}{Earl
  T.~Campbell}.} \bibinfo{year}{2018}\natexlab{}.
\newblock \showarticletitle{An efficient quantum compiler that reduces $T$
  count}.
\newblock \bibinfo{journal}{\emph{Quantum Science and Technology}}
  \bibinfo{volume}{4} (\bibinfo{year}{2018}).
\newblock
\urldef\tempurl%
\url{https://doi.org/10.1088/2058-9565/aad604}
\showDOI{\tempurl}


\bibitem[\protect\citeauthoryear{Hietala, Rand, Hung, Li, and Hicks}{Hietala
  et~al\mbox{.}}{2020}]%
        {CPPsub}
\bibfield{author}{\bibinfo{person}{Kesha Hietala}, \bibinfo{person}{Robert
  Rand}, \bibinfo{person}{Shih-Han Hung}, \bibinfo{person}{Liyi Li}, {and}
  \bibinfo{person}{Michael Hicks}.} \bibinfo{year}{2020}\natexlab{}.
\newblock \bibinfo{title}{Proving Quantum Programs Correct}.
\newblock
\newblock
\showeprint[arxiv]{2010.01240}~[cs.PL]


\bibitem[\protect\citeauthoryear{IBM}{IBM}{[n.d.]}]%
        {tenerife}
\bibfield{author}{\bibinfo{person}{IBM}.} \bibinfo{year}{[n.d.]}\natexlab{}.
\newblock \bibinfo{title}{{IBM Q5 Tenerife} V1.x.x version log}.
\newblock
\newblock
\urldef\tempurl%
\url{https://github.com/Qiskit/ibmq-device-information/blob/master/backends/tenerife/V1/version_log.md}
\showURL{%
\tempurl}


\bibitem[\protect\citeauthoryear{{IBM Research Editorial Staff}}{{IBM Research
  Editorial Staff}}{2018}]%
        {ibm-dev-challenge-2018}
\bibfield{author}{\bibinfo{person}{{IBM Research Editorial Staff}}.}
  \bibinfo{year}{2018}\natexlab{}.
\newblock \bibinfo{title}{We have winners! ... of the {IBM Qiskit} developer
  challenge}.
\newblock
\newblock
\urldef\tempurl%
\url{https://www.ibm.com/blogs/research/2018/08/winners-qiskit-developer-challenge/}
\showURL{%
\tempurl}


\bibitem[\protect\citeauthoryear{Javadi-Abhari, Patil, Kudrow, Heckey, Lvov,
  Chong, and Martonosi}{Javadi-Abhari et~al\mbox{.}}{2014}]%
        {JavadiAbhari2014}
\bibfield{author}{\bibinfo{person}{Ali Javadi-Abhari}, \bibinfo{person}{Shruti
  Patil}, \bibinfo{person}{Daniel Kudrow}, \bibinfo{person}{Jeff Heckey},
  \bibinfo{person}{Alexey Lvov}, \bibinfo{person}{Frederic~T. Chong}, {and}
  \bibinfo{person}{Margaret Martonosi}.} \bibinfo{year}{2014}\natexlab{}.
\newblock \showarticletitle{ScaffCC: A framework for compilation and analysis
  of quantum computing programs}. In \bibinfo{booktitle}{\emph{Proceedings of
  the 11th ACM Conference on Computing Frontiers}} (Cagliari, Italy)
  \emph{(\bibinfo{series}{CF '14})}. \bibinfo{publisher}{ACM},
  \bibinfo{address}{New York, NY, USA}, Article \bibinfo{articleno}{1},
  \bibinfo{numpages}{10}~pages.
\newblock
\showISBNx{978-1-4503-2870-8}
\urldef\tempurl%
\url{https://doi.org/10.1145/2597917.2597939}
\showDOI{\tempurl}


\bibitem[\protect\citeauthoryear{Jeandel, Perdrix, and Vilmart}{Jeandel
  et~al\mbox{.}}{2018}]%
        {Jeandel2018}
\bibfield{author}{\bibinfo{person}{Emmanuel Jeandel}, \bibinfo{person}{Simon
  Perdrix}, {and} \bibinfo{person}{Renaud Vilmart}.}
  \bibinfo{year}{2018}\natexlab{}.
\newblock \showarticletitle{A complete axiomatisation of the {ZX}-calculus for
  {Clifford+ T} quantum mechanics}. In \bibinfo{booktitle}{\emph{Proceedings of
  the 33rd Annual ACM/IEEE Symposium on Logic in Computer Science}}. ACM,
  \bibinfo{pages}{559--568}.
\newblock
\urldef\tempurl%
\url{https://doi.org/10.1145/3209108.3209131}
\showDOI{\tempurl}


\bibitem[\protect\citeauthoryear{Jourdan, Pottier, and Leroy}{Jourdan
  et~al\mbox{.}}{2012}]%
        {Jourdan:2012:VLP:2259248.2259268}
\bibfield{author}{\bibinfo{person}{Jacques-Henri Jourdan},
  \bibinfo{person}{Fran{\c{c}}ois Pottier}, {and} \bibinfo{person}{Xavier
  Leroy}.} \bibinfo{year}{2012}\natexlab{}.
\newblock \showarticletitle{Validating LR(1) parsers}. In
  \bibinfo{booktitle}{\emph{Programming Languages and Systems}},
  \bibfield{editor}{\bibinfo{person}{Helmut Seidl}} (Ed.).
  \bibinfo{publisher}{Springer Berlin Heidelberg}, \bibinfo{address}{Berlin,
  Heidelberg}, \bibinfo{pages}{397--416}.
\newblock
\showISBNx{978-3-642-28869-2}


\bibitem[\protect\citeauthoryear{{Kissinger} and {van de Wetering}}{{Kissinger}
  and {van de Wetering}}{2019}]%
        {Kissinger2019b}
\bibfield{author}{\bibinfo{person}{Aleks {Kissinger}} {and}
  \bibinfo{person}{John {van de Wetering}}.} \bibinfo{year}{2019}\natexlab{}.
\newblock \showarticletitle{Reducing {T}-count with the {ZX}-calculus}.
\newblock \bibinfo{journal}{\emph{arXiv e-prints}} (\bibinfo{year}{2019}).
\newblock
\showeprint[arxiv]{1903.10477}~[quant-ph]


\bibitem[\protect\citeauthoryear{Kissinger and van~de Wetering}{Kissinger and
  van~de Wetering}{2020}]%
        {Kissinger2019}
\bibfield{author}{\bibinfo{person}{Aleks Kissinger} {and} \bibinfo{person}{John
  van~de Wetering}.} \bibinfo{year}{2020}\natexlab{}.
\newblock \showarticletitle{{PyZX}: Large scale automated diagrammatic
  reasoning}.
\newblock \bibinfo{journal}{\emph{Electronic Proceedings in Theoretical
  Computer Science}}  \bibinfo{volume}{318} (\bibinfo{date}{04}
  \bibinfo{year}{2020}), \bibinfo{pages}{230--242}.
\newblock
\urldef\tempurl%
\url{https://doi.org/10.4204/EPTCS.318.14}
\showDOI{\tempurl}


\bibitem[\protect\citeauthoryear{Kissinger and Zamdzhiev}{Kissinger and
  Zamdzhiev}{2015}]%
        {Kissinger2015}
\bibfield{author}{\bibinfo{person}{Aleks Kissinger} {and}
  \bibinfo{person}{Vladimir Zamdzhiev}.} \bibinfo{year}{2015}\natexlab{}.
\newblock \showarticletitle{Quantomatic: A proof assistant for diagrammatic
  reasoning}. In \bibinfo{booktitle}{\emph{Automated Deduction - CADE-25}},
  \bibfield{editor}{\bibinfo{person}{Amy~P. Felty} {and} \bibinfo{person}{Aart
  Middeldorp}} (Eds.). \bibinfo{publisher}{Springer International Publishing},
  \bibinfo{address}{Cham}, \bibinfo{pages}{326--336}.
\newblock


\bibitem[\protect\citeauthoryear{Knill}{Knill}{1996}]%
        {Knill1996}
\bibfield{author}{\bibinfo{person}{Emmanuel Knill}.}
  \bibinfo{year}{1996}\natexlab{}.
\newblock \bibinfo{booktitle}{\emph{Conventions for quantum pseudocode}}.
\newblock \bibinfo{type}{{T}echnical {R}eport}. \bibinfo{institution}{Los
  Alamos National Lab., NM (United States)}.
\newblock


\bibitem[\protect\citeauthoryear{Leroy}{Leroy}{2009}]%
        {compcert}
\bibfield{author}{\bibinfo{person}{Xavier Leroy}.}
  \bibinfo{year}{2009}\natexlab{}.
\newblock \showarticletitle{Formal verification of a realistic compiler}.
\newblock \bibinfo{journal}{\emph{Commun. ACM}} \bibinfo{volume}{52},
  \bibinfo{number}{7} (\bibinfo{date}{July} \bibinfo{year}{2009}),
  \bibinfo{pages}{107--115}.
\newblock
\showISSN{0001-0782}
\urldef\tempurl%
\url{https://doi.org/10/c9sb7q}
\showDOI{\tempurl}


\bibitem[\protect\citeauthoryear{Martonosi and Roetteler}{Martonosi and
  Roetteler}{2019}]%
        {Martonosi2019}
\bibfield{author}{\bibinfo{person}{Margaret Martonosi} {and}
  \bibinfo{person}{Martin Roetteler}.} \bibinfo{year}{2019}\natexlab{}.
\newblock \bibinfo{title}{Next steps in quantum computing: computer science's
  role}.
\newblock
\newblock
\showeprint[arxiv]{1903.10541}~[cs.ET]


\bibitem[\protect\citeauthoryear{Melquiond}{Melquiond}{2020}]%
        {intervals}
\bibfield{author}{\bibinfo{person}{Guillaume Melquiond}.}
  \bibinfo{year}{2020}\natexlab{}.
\newblock \bibinfo{title}{Interval package for {C}oq}.
\newblock
\newblock
\urldef\tempurl%
\url{https://gitlab.inria.fr/coqinterval/interval}
\showURL{%
\tempurl}


\bibitem[\protect\citeauthoryear{{Miller} and {Thornton}}{{Miller} and
  {Thornton}}{2006}]%
        {Miller2006}
\bibfield{author}{\bibinfo{person}{D.~M. {Miller}} {and} \bibinfo{person}{M.~A.
  {Thornton}}.} \bibinfo{year}{2006}\natexlab{}.
\newblock \showarticletitle{{QMDD}: A decision diagram structure for reversible
  and quantum circuits}. In \bibinfo{booktitle}{\emph{36th International
  Symposium on Multiple-Valued Logic (ISMVL'06)}}. \bibinfo{pages}{30--30}.
\newblock
\urldef\tempurl%
\url{https://doi.org/10.1109/ISMVL.2006.35}
\showDOI{\tempurl}


\bibitem[\protect\citeauthoryear{Nam, Ross, Su, Childs, and Maslov}{Nam
  et~al\mbox{.}}{2018}]%
        {Nam2018}
\bibfield{author}{\bibinfo{person}{Yunseong Nam}, \bibinfo{person}{Neil~J.
  Ross}, \bibinfo{person}{Yuan Su}, \bibinfo{person}{Andrew~M. Childs}, {and}
  \bibinfo{person}{Dmitri Maslov}.} \bibinfo{year}{2018}\natexlab{}.
\newblock \showarticletitle{Automated optimization of large quantum circuits
  with continuous parameters}.
\newblock \bibinfo{journal}{\emph{npj Quantum Information}}
  \bibinfo{volume}{4}, \bibinfo{number}{1} (\bibinfo{year}{2018}),
  \bibinfo{pages}{23}.
\newblock
\showISBNx{2056-6387}
\urldef\tempurl%
\url{https://doi.org/10.1038/s41534-018-0072-4}
\showDOI{\tempurl}


\bibitem[\protect\citeauthoryear{Paetznick and Svore}{Paetznick and
  Svore}{2014}]%
        {paetznick2014repeat}
\bibfield{author}{\bibinfo{person}{Adam Paetznick} {and}
  \bibinfo{person}{Krysta~M Svore}.} \bibinfo{year}{2014}\natexlab{}.
\newblock \showarticletitle{Repeat-until-success: non-deterministic
  decomposition of single-qubit unitaries}.
\newblock \bibinfo{journal}{\emph{Quantum Information \& Computation}}
  \bibinfo{volume}{14}, \bibinfo{number}{15-16} (\bibinfo{year}{2014}),
  \bibinfo{pages}{1277--1301}.
\newblock


\bibitem[\protect\citeauthoryear{Paykin, Rand, and Zdancewic}{Paykin
  et~al\mbox{.}}{2017}]%
        {Paykin2017}
\bibfield{author}{\bibinfo{person}{Jennifer Paykin}, \bibinfo{person}{Robert
  Rand}, {and} \bibinfo{person}{Steve Zdancewic}.}
  \bibinfo{year}{2017}\natexlab{}.
\newblock \showarticletitle{{QWIRE}: A core language for quantum circuits}. In
  \bibinfo{booktitle}{\emph{Proceedings of the 44th ACM SIGPLAN Symposium on
  Principles of Programming Languages}} (Paris, France)
  \emph{(\bibinfo{series}{POPL 2017})}. \bibinfo{publisher}{ACM},
  \bibinfo{address}{New York, NY, USA}, \bibinfo{pages}{846--858}.
\newblock
\urldef\tempurl%
\url{https://doi.org/10.1145/3009837.3009894}
\showDOI{\tempurl}


\bibitem[\protect\citeauthoryear{Pfenning and Elliott}{Pfenning and
  Elliott}{1988}]%
        {Pfenning1988}
\bibfield{author}{\bibinfo{person}{Frank Pfenning} {and} \bibinfo{person}{Conal
  Elliott}.} \bibinfo{year}{1988}\natexlab{}.
\newblock \showarticletitle{Higher-order abstract syntax}. In
  \bibinfo{booktitle}{\emph{Proceedings of the ACM SIGPLAN 1988 Conference on
  Programming Language Design and Implementation}} (Atlanta, Georgia, USA)
  \emph{(\bibinfo{series}{PLDI '88})}. \bibinfo{publisher}{ACM},
  \bibinfo{address}{New York, NY, USA}, \bibinfo{pages}{199--208}.
\newblock
\urldef\tempurl%
\url{https://doi.org/10.1145/53990.54010}
\showDOI{\tempurl}


\bibitem[\protect\citeauthoryear{Preskill}{Preskill}{2018}]%
        {Preskill2018}
\bibfield{author}{\bibinfo{person}{John Preskill}.}
  \bibinfo{year}{2018}\natexlab{}.
\newblock \showarticletitle{Quantum computing in the {NISQ} era and beyond}.
\newblock \bibinfo{journal}{\emph{{Quantum}}}  \bibinfo{volume}{2}
  (\bibinfo{date}{Aug.} \bibinfo{year}{2018}), \bibinfo{pages}{79}.
\newblock
\showISSN{2521-327X}
\urldef\tempurl%
\url{https://doi.org/10.22331/q-2018-08-06-79}
\showDOI{\tempurl}


\bibitem[\protect\citeauthoryear{Rand}{Rand}{2018}]%
        {RandThesis}
\bibfield{author}{\bibinfo{person}{Robert Rand}.}
  \bibinfo{year}{2018}\natexlab{}.
\newblock \emph{\bibinfo{title}{Formally verified quantum programming}}.
\newblock \bibinfo{thesistype}{Ph.D. Dissertation}. \bibinfo{school}{University
  of Pennsylvania}.
\newblock


\bibitem[\protect\citeauthoryear{Rand, Paykin, Lee, and Zdancewic}{Rand
  et~al\mbox{.}}{2018b}]%
        {Rand2018}
\bibfield{author}{\bibinfo{person}{Robert Rand}, \bibinfo{person}{Jennifer
  Paykin}, \bibinfo{person}{Dong-Ho Lee}, {and} \bibinfo{person}{Steve
  Zdancewic}.} \bibinfo{year}{2018}\natexlab{b}.
\newblock \showarticletitle{Re{QWIRE}: Reasoning about reversible quantum
  circuits}. In \bibinfo{booktitle}{\emph{Proceedings of the 15th International
  Conference on Quantum Physics and Logic, {QPL} 2018, Halifax, Nova Scotia,
  3-7 June 2018}}.
\newblock
\urldef\tempurl%
\url{https://doi.org/10.4204/EPTCS.287.17}
\showDOI{\tempurl}


\bibitem[\protect\citeauthoryear{Rand, Paykin, and Zdancewic}{Rand
  et~al\mbox{.}}{2017}]%
        {Rand2017}
\bibfield{author}{\bibinfo{person}{Robert Rand}, \bibinfo{person}{Jennifer
  Paykin}, {and} \bibinfo{person}{Steve Zdancewic}.}
  \bibinfo{year}{2017}\natexlab{}.
\newblock \showarticletitle{{QWIRE} practice: Formal verification of quantum
  circuits in {C}oq}. In \bibinfo{booktitle}{\emph{Proceedings 14th
  International Conference on Quantum Physics and Logic, {QPL} 2017, Nijmegen,
  The Netherlands, 3-7 July 2017.}} \bibinfo{pages}{119--132}.
\newblock
\urldef\tempurl%
\url{https://doi.org/10.4204/EPTCS.266.8}
\showDOI{\tempurl}


\bibitem[\protect\citeauthoryear{Rand, Paykin, and Zdancewic}{Rand
  et~al\mbox{.}}{2018a}]%
        {Rand2018a}
\bibfield{author}{\bibinfo{person}{Robert Rand}, \bibinfo{person}{Jennifer
  Paykin}, {and} \bibinfo{person}{Steve Zdancewic}.}
  \bibinfo{year}{2018}\natexlab{a}.
\newblock \bibinfo{title}{Phantom types for quantum programs}.
\newblock \bibinfo{howpublished}{The Fourth International Workshop on Coq for
  Programming Languages}.
\newblock


\bibitem[\protect\citeauthoryear{{Rigetti Computing}}{{Rigetti
  Computing}}{2019a}]%
        {Pyquil}
\bibfield{author}{\bibinfo{person}{{Rigetti Computing}}.}
  \bibinfo{year}{2019}\natexlab{a}.
\newblock \bibinfo{title}{Pyquil documentation}.
\newblock
\newblock
\urldef\tempurl%
\url{http://pyquil.readthedocs.io/en/latest/}
\showURL{%
\tempurl}


\bibitem[\protect\citeauthoryear{{Rigetti Computing}}{{Rigetti
  Computing}}{2019b}]%
        {quilc}
\bibfield{author}{\bibinfo{person}{{Rigetti Computing}}.}
  \bibinfo{year}{2019}\natexlab{b}.
\newblock \bibinfo{title}{The @rigetti optimizing {Quil} compiler}.
\newblock
\newblock
\urldef\tempurl%
\url{https://github.com/rigetti/quilc}
\showURL{%
\tempurl}


\bibitem[\protect\citeauthoryear{Saeedi, Wille, and Drechsler}{Saeedi
  et~al\mbox{.}}{2011}]%
        {Saeedi2011}
\bibfield{author}{\bibinfo{person}{Mehdi Saeedi}, \bibinfo{person}{Robert
  Wille}, {and} \bibinfo{person}{Rolf Drechsler}.}
  \bibinfo{year}{2011}\natexlab{}.
\newblock \showarticletitle{Synthesis of quantum circuits for linear nearest
  neighbor architectures}.
\newblock \bibinfo{journal}{\emph{Quantum Information Processing}}
  \bibinfo{volume}{10}, \bibinfo{number}{3} (\bibinfo{date}{01 Jun}
  \bibinfo{year}{2011}), \bibinfo{pages}{355--377}.
\newblock
\showISSN{1573-1332}
\urldef\tempurl%
\url{https://doi.org/10.1007/s11128-010-0201-2}
\showDOI{\tempurl}


\bibitem[\protect\citeauthoryear{Selinger}{Selinger}{2004}]%
        {Selinger2004}
\bibfield{author}{\bibinfo{person}{Peter Selinger}.}
  \bibinfo{year}{2004}\natexlab{}.
\newblock \showarticletitle{Towards a quantum programming language}.
\newblock \bibinfo{journal}{\emph{Mathematical Structures in Computer Science}}
  \bibinfo{volume}{14}, \bibinfo{number}{4} (\bibinfo{date}{Aug.}
  \bibinfo{year}{2004}), \bibinfo{pages}{527--586}.
\newblock
\urldef\tempurl%
\url{https://doi.org/10.1017/S0960129504004256}
\showDOI{\tempurl}


\bibitem[\protect\citeauthoryear{{Shi}, {Li}, {Tao}, {Javadi-Abhari}, {Cross},
  {Chong}, and {Gu}}{{Shi} et~al\mbox{.}}{2019}]%
        {Shi2019}
\bibfield{author}{\bibinfo{person}{Yunong {Shi}}, \bibinfo{person}{Xupeng
  {Li}}, \bibinfo{person}{Runzhou {Tao}}, \bibinfo{person}{Ali
  {Javadi-Abhari}}, \bibinfo{person}{Andrew~W. {Cross}},
  \bibinfo{person}{Frederic~T. {Chong}}, {and} \bibinfo{person}{Ronghui {Gu}}.}
  \bibinfo{year}{2019}\natexlab{}.
\newblock \showarticletitle{{Contract-based verification of a realistic quantum
  compiler}}.
\newblock \bibinfo{journal}{\emph{arXiv e-prints}} (\bibinfo{date}{Aug}
  \bibinfo{year}{2019}).
\newblock
\showeprint[arxiv]{1908.08963}~[quant-ph]


\bibitem[\protect\citeauthoryear{{Shor}}{{Shor}}{1994}]%
        {Shor94}
\bibfield{author}{\bibinfo{person}{P.~W. {Shor}}.}
  \bibinfo{year}{1994}\natexlab{}.
\newblock \showarticletitle{Algorithms for quantum computation: Discrete
  logarithms and factoring}. In \bibinfo{booktitle}{\emph{Proceedings 35th
  Annual Symposium on Foundations of Computer Science}}
  \emph{(\bibinfo{series}{FOCS '94})}.
\newblock


\bibitem[\protect\citeauthoryear{Simon}{Simon}{1994}]%
        {Simon1994}
\bibfield{author}{\bibinfo{person}{DR Simon}.} \bibinfo{year}{1994}\natexlab{}.
\newblock \showarticletitle{On the power of quantum computation}. In
  \bibinfo{booktitle}{\emph{Proceedings of the 35th Annual Symposium on
  Foundations of Computer Science}}. \bibinfo{pages}{116--123}.
\newblock


\bibitem[\protect\citeauthoryear{Singhal, Rand, and Hicks}{Singhal
  et~al\mbox{.}}{2020}]%
        {Singhal2020}
\bibfield{author}{\bibinfo{person}{Kartik Singhal}, \bibinfo{person}{Robert
  Rand}, {and} \bibinfo{person}{Michael Hicks}.}
  \bibinfo{year}{2020}\natexlab{}.
\newblock \bibinfo{title}{Verified translation between low-level quantum
  languages}.
\newblock \bibinfo{howpublished}{The First International Workshop on
  Programming Languages for Quantum Computing}.
\newblock


\bibitem[\protect\citeauthoryear{{Sivarajah}, {Dilkes}, {Cowtan}, {Simmons},
  {Edgington}, and {Duncan}}{{Sivarajah} et~al\mbox{.}}{2020}]%
        {Sivarajah2020}
\bibfield{author}{\bibinfo{person}{Seyon {Sivarajah}}, \bibinfo{person}{Silas
  {Dilkes}}, \bibinfo{person}{Alexander {Cowtan}}, \bibinfo{person}{Will
  {Simmons}}, \bibinfo{person}{Alec {Edgington}}, {and} \bibinfo{person}{Ross
  {Duncan}}.} \bibinfo{year}{2020}\natexlab{}.
\newblock \showarticletitle{{t$|$ket$\rangle$ : A retargetable compiler for
  {NISQ} Devices}}.
\newblock \bibinfo{journal}{\emph{arXiv e-prints}} (\bibinfo{year}{2020}).
\newblock
\showeprint[arxiv]{2003.10611}~[quant-ph]


\bibitem[\protect\citeauthoryear{Smith and Thornton}{Smith and
  Thornton}{2019}]%
        {Smith2019}
\bibfield{author}{\bibinfo{person}{Kaitlin~N. Smith} {and}
  \bibinfo{person}{Mitchell~A. Thornton}.} \bibinfo{year}{2019}\natexlab{}.
\newblock \showarticletitle{A quantum computational compiler and design tool
  for technology-specific targets}. In \bibinfo{booktitle}{\emph{Proceedings of
  the 46th International Symposium on Computer Architecture}}
  \emph{(\bibinfo{series}{ISCA '19})}.
\newblock
\urldef\tempurl%
\url{https://doi.org/10.1145/3307650.3322262}
\showDOI{\tempurl}


\bibitem[\protect\citeauthoryear{{Smith}, {Curtis}, and {Zeng}}{{Smith}
  et~al\mbox{.}}{2016}]%
        {Smith2016}
\bibfield{author}{\bibinfo{person}{Robert~S. {Smith}},
  \bibinfo{person}{Michael~J. {Curtis}}, {and} \bibinfo{person}{William~J.
  {Zeng}}.} \bibinfo{year}{2016}\natexlab{}.
\newblock \showarticletitle{{A practical quantum instruction set
  architecture}}.
\newblock \bibinfo{journal}{\emph{arXiv e-prints}} (\bibinfo{date}{Aug}
  \bibinfo{year}{2016}).
\newblock
\showeprint[arxiv]{1608.03355}~[quant-ph]


\bibitem[\protect\citeauthoryear{Steiger, H{\"a}ner, and Troyer}{Steiger
  et~al\mbox{.}}{2018}]%
        {Steiger2018}
\bibfield{author}{\bibinfo{person}{Damian~S. Steiger}, \bibinfo{person}{Thomas
  H{\"a}ner}, {and} \bibinfo{person}{Matthias Troyer}.}
  \bibinfo{year}{2018}\natexlab{}.
\newblock \showarticletitle{ProjectQ: An open source software framework for
  quantum computing}.
\newblock \bibinfo{journal}{\emph{Quantum}}  \bibinfo{volume}{2}
  (\bibinfo{year}{2018}), \bibinfo{pages}{49}.
\newblock
\urldef\tempurl%
\url{https://doi.org/10.22331/q-2018-01-31-49}
\showDOI{\tempurl}


\bibitem[\protect\citeauthoryear{Svore, Geller, Troyer, Azariah, Granade, Heim,
  Kliuchnikov, Mykhailova, Paz, and Roetteler}{Svore et~al\mbox{.}}{2018}]%
        {Svore2018}
\bibfield{author}{\bibinfo{person}{Krysta Svore}, \bibinfo{person}{Alan
  Geller}, \bibinfo{person}{Matthias Troyer}, \bibinfo{person}{John Azariah},
  \bibinfo{person}{Christopher Granade}, \bibinfo{person}{Bettina Heim},
  \bibinfo{person}{Vadym Kliuchnikov}, \bibinfo{person}{Mariia Mykhailova},
  \bibinfo{person}{Andres Paz}, {and} \bibinfo{person}{Martin Roetteler}.}
  \bibinfo{year}{2018}\natexlab{}.
\newblock \showarticletitle{Q\#: Enabling scalable quantum computing and
  development with a high-level {DSL}}. In
  \bibinfo{booktitle}{\emph{Proceedings of the Real World Domain Specific
  Languages Workshop 2018}}. ACM, \bibinfo{pages}{7}.
\newblock
\urldef\tempurl%
\url{https://doi.org/10.1145/3183895.3183901}
\showDOI{\tempurl}


\bibitem[\protect\citeauthoryear{Tannu and Qureshi}{Tannu and Qureshi}{2019}]%
        {Tannu2019}
\bibfield{author}{\bibinfo{person}{Swamit~S. Tannu} {and}
  \bibinfo{person}{Moinuddin~K. Qureshi}.} \bibinfo{year}{2019}\natexlab{}.
\newblock \showarticletitle{Not all qubits are created equal: A case for
  variability-aware policies for {NISQ}-era quantum computers}. In
  \bibinfo{booktitle}{\emph{Proceedings of the Twenty-Fourth International
  Conference on Architectural Support for Programming Languages and Operating
  Systems}} \emph{(\bibinfo{series}{ASPLOS ’19})}.
\newblock
\urldef\tempurl%
\url{https://doi.org/10.1145/3297858.3304007}
\showDOI{\tempurl}


\bibitem[\protect\citeauthoryear{{The Cirq Developers}}{{The Cirq
  Developers}}{2019}]%
        {Cirq}
\bibfield{author}{\bibinfo{person}{{The Cirq Developers}}.}
  \bibinfo{year}{2019}\natexlab{}.
\newblock \bibinfo{title}{Cirq: A python library for {NISQ} circuits}.
\newblock
\newblock
\urldef\tempurl%
\url{https://cirq.readthedocs.io/en/stable/}
\showURL{%
\tempurl}


\bibitem[\protect\citeauthoryear{Ying}{Ying}{2011}]%
        {Ying2011}
\bibfield{author}{\bibinfo{person}{Mingsheng Ying}.}
  \bibinfo{year}{2011}\natexlab{}.
\newblock \showarticletitle{Floyd--hoare logic for quantum programs}.
\newblock \bibinfo{journal}{\emph{ACM Transactions on Programming Languages and
  Systems (TOPLAS)}} \bibinfo{volume}{33}, \bibinfo{number}{6}
  (\bibinfo{year}{2011}), \bibinfo{pages}{19}.
\newblock
\urldef\tempurl%
\url{https://doi.org/10.1145/2049706.2049708}
\showDOI{\tempurl}


\bibitem[\protect\citeauthoryear{Zamdzhiev}{Zamdzhiev}{2016}]%
        {Zamdzhiev16talk}
\bibfield{author}{\bibinfo{person}{Vladimir Zamdzhiev}.}
  \bibinfo{year}{2016}\natexlab{}.
\newblock \bibinfo{title}{Quantum computing: The good, the bad, and the (not
  so) ugly!}
\newblock
\newblock
\newblock
\shownote{Invited talk, Tulane University.}


\bibitem[\protect\citeauthoryear{{Zulehner}, {Paler}, and {Wille}}{{Zulehner}
  et~al\mbox{.}}{2017}]%
        {Zulehner2017}
\bibfield{author}{\bibinfo{person}{Alwin {Zulehner}},
  \bibinfo{person}{Alexandru {Paler}}, {and} \bibinfo{person}{Robert {Wille}}.}
  \bibinfo{year}{2017}\natexlab{}.
\newblock \showarticletitle{{An efficient methodology for mapping quantum
  circuits to the {IBM} {QX} architectures}}.
\newblock \bibinfo{journal}{\emph{arXiv e-prints}} (\bibinfo{date}{Dec}
  \bibinfo{year}{2017}).
\newblock
\showeprint[arxiv]{1712.04722}~[quant-ph]


\end{thebibliography}

\iftoggle{submission}{ }{
  \clearpage
  \appendix

\pagenumbering{arabic}
\renewcommand*{\thepage}{A\arabic{page}}


\section{Additional Benchmark Results}\label{app:extended-eval}

In this section, we evaluate \voqc's performance on all 99 benchmark programs considered by \citet{Nam2018}, confirming our conclusion from \Cref{sec:experiments} that \voqc is a faithful implementation of a subset of the optimizations present in \citeauthor{Nam2018} (along with being proved correct!).
The benchmarks are divided into three categories, as described below.
Our versions of the benchmarks are available online.\footnote{\url{https://github.com/inQWIRE/VOQC-benchmarks}}
All results were obtained using a laptop with a 2.9 GHz Intel Core i5 processor and 16 GB of 1867 MHz DDR3 memory, running macOS Catalina. For timings, we take the median of three trials. \citeauthor{Nam2018}'s results are from a similar machine with 8 GB RAM running OS X El Capitan. Their implementation is written in Fortran.

Overall, the results are consistent with those presented in \Cref{sec:experiments}. In cases where Toffoli decomposition and heavy optimization are not used (the QFT, QFT-based adder, and product formula circuits), \voqc's results are identical to \citeauthor{Nam2018}'s. In the other cases, \voqc is slightly less effective than \citeauthor{Nam2018} for the reasons discussed in \Cref{sec:experiments}. 
In the worst case, \voqc's run time is four orders of magnitude worse than \citeauthor{Nam2018}'s. However, \voqc's run time is often less than a second. We view this performance as acceptable, given that benchmarks with more than 1000 two-qubit gates (the only programs for which \voqc optimization takes longer than one second) are well out of reach of current quantum hardware \cite{Preskill2018}. We are optimistic that \voqc's performance can be improved through more careful engineering.

\paragraph{Arithmetic and Toffoli}
These benchmarks are a superset of the arithmetic and Toffoli circuits discussed in \Cref{sec:experiments}. They range from 45 to 346,533 gates and 5 to 489 qubits. Results on all 32 benchmarks are given in \Cref{tab:aat_table}.
As discussed in \Cref{sec:experiments}, \voqc's performance does not match \citeauthor{Nam2018}'s because we have not yet implemented all of their transformations (in particular, we are missing ``Toffoli decomposition'' and ``Floating $R_z$ gates'').

\paragraph{QFT and Adders}
These benchmarks consist of components of Shor’s integer factoring algorithm, in particular the quantum Fourier transform (QFT) and integer adders. Two types of adders are considered: an in-place modulo 2$q$ adder implemented in the Quipper library and an in-place adder based on the QFT.
These benchmarks range from 148 to 381,806 gates and 8 to 4096 qubits.
Results on all 27 benchmarks are given in \Cref{tab:adder-table}, \Cref{tab:qft-table}, and \Cref{tab:qfa-table}.
The Quipper adder programs use similar gates to the arithmetic and Toffoli circuits, so the results are similar---\voqc is close to \citeauthor{Nam2018}, but under-performs due to our simplified Toffoli decomposition. 
The QFT circuits use rotations parameterized by $\pi/2^n$ for varying $n \in \mathbb{N}$ (and no Toffoli gates) so \voqc's results are identical to \citeauthor{Nam2018}'s. For consistency with \citeauthor{Nam2018}, on the QFT and QFT-based adder circuits we run a simplified version of our optimizer that does not include rotation merging.

\paragraph{Product Formula} 
These benchmarks implement product formula algorithms for simulating Hamiltonian dynamics. The benchmarks range from 260 to 127,500 gates and 10 to 100 qubits; they use rotations parameterized by floating point numbers, which we convert to OCaml rationals at parse time. The product formula circuits are intended to be repeated for a fixed number of iterations, and our resource estimates account for this. \voqc applies \citeauthor{Nam2018}'s ``LCR'' optimization routine to optimize programs across loop iterations.
On all 40 product formula benchmarks, our results are the same as those reported by \citet[Table 3]{Nam2018}. $H$ gate reductions range from 62.5\% to 75\%. Reductions in Clifford $z$-axis rotations (i.e. rotations by multiples of $\pi/2$) range from 75\% to 87.5\% while reductions in non-Clifford $z$-axis rotations range from 0\% to 28.6\%. $\mathit{CNOT}$ gate reductions range from 0\% to 33\%. Runtimes range from 0.01s for parsing and optimizing to 610.46s for parsing and 406.93s for optimizing. By comparison, \citeauthor{Nam2018}'s runtimes range from 0.004s to 0.137s.

\begin{table}[t]
\caption{Total gate count reduction on the ``Arithmetic and Toffoli'' circuits. Includes all programs listed in \Cref{tab:total-counts} and \Cref{tab:t-counts} as well as four gf programs omitted from that table. The reported \voqc time only includes optimization time. \voqc's parse time was less that 0.001s for all benchmarks except the larger gf programs; the largest, gf2\textasciicircum{}163\_mult, required 427s (~7.1min) to parse. Nam (H) results were not available for the large benchmarks.}
\begin{tabular}{c|c|cc|cc|cc}
& \multicolumn{1}{c|}{\textbf{Orig.}} & \multicolumn{2}{c|}{\textbf{Nam (L)}} & \multicolumn{2}{c|}{\textbf{Nam (H)}} & \multicolumn{2}{c}{\textbf{\voqc}} \\
\textbf{Name} & \textbf{Total} & \textbf{Total} & \textbf{t(s)} & \textbf{Total} & \textbf{t(s)} & \textbf{Total} & \textbf{t(s)} \\
\hline
adder\_8                       & 900     & 646     & 0.004            & \textbf{606}     & 0.101            & 682 & 0.048 \\
barenco\_tof\_3                & 58      & 42      & \textless{}0.001 & \textbf{40}      & 0.001            & 50 & 0.001 \\
barenco\_tof\_4                & 114     & 78      & \textless{}0.001 & \textbf{72}      & 0.001            & 95 & 0.002 \\
barenco\_tof\_5                & 170     & 114     & \textless{}0.001 & \textbf{104}     & 0.003            & 140 & 0.003 \\
barenco\_tof\_10               & 450     & 294     & 0.001            & \textbf{264}     & 0.012            & 365 & 0.019 \\
csla\_mux\_3                   & 170     & 161     & \textless{}0.001 & \textbf{155}     & 0.009            & 158 & 0.003 \\
csum\_mux\_9                   & 420     & 294     & \textless{}0.001 & \textbf{266}     & 0.009            & 308 & 0.006 \\
gf2\textasciicircum{}4\_mult   & 225     & \textbf{187}     & 0.001            & \textbf{187}     & 0.009            & 192 & 0.006 \\
gf2\textasciicircum{}5\_mult   & 347     & 296     & 0.001            & 296     & 0.020            & \textbf{291} & 0.012 \\
gf2\textasciicircum{}6\_mult   & 495     & \textbf{403}     & 0.003            & \textbf{403}     & 0.047            & 410 & 0.025  \\
gf2\textasciicircum{}7\_mult   & 669     & 555     & 0.004            & 555     & 0.105            & \textbf{549} & 0.045 \\
gf2\textasciicircum{}8\_mult   & 883     & 712     & 0.006            & 712     & 0.192            & \textbf{705} & 0.070 \\
gf2\textasciicircum{}9\_mult   & 1095    & 891     & 0.010            & 891     & 0.347            & \textbf{885} & 0.119 \\
gf2\textasciicircum{}10\_mult  & 1347    & \textbf{1070}    & 0.009            & \textbf{1070}    & 0.429            & 1084 & 0.183 \\
gf2\textasciicircum{}16\_mult  & 3435    & 2707    & 0.065            & 2707    & 5.566            & \textbf{2695} & 1.347 \\
gf2\textasciicircum{}32\_mult  & 13593   & 10601   & 1.834            & 10601   & 275.698          & \textbf{10577} & 26.808 \\
gf2\textasciicircum{}64\_mult  & 53691   & 41563   & 58.341           & --       & --                & \textbf{41515} & 546.887 \\
gf2\textasciicircum{}128\_mult & 213883 & 165051  & 1744.746         & --       & --                & \textbf{164955} & 9841.797 \\
gf2\textasciicircum{}131\_mult & 224265 & 173370  & 1953.353         & --       & --                & \textbf{173273} & 10877.112 \\
gf2\textasciicircum{}163\_mult & 346533 & 267558  & 4955.927         & --       & --                & \textbf{267437} & 27612.565 \\
mod5\_4                        & 63      & \textbf{51}      & \textless{}0.001 & \textbf{51}      & 0.001            & 56 & <0.001 \\
mod\_mult\_55                  & 119     & 91      & \textless{}0.001 & 91      & 0.002            & \textbf{90} & 0.002 \\
mod\_red\_21                   & 278     & 184     & \textless{}0.001 & \textbf{180}     & 0.008            & 214 & 0.005 \\
qcla\_adder\_10                & 521     & 411     & 0.002            & \textbf{399}     & 0.044            & 438 & 0.018 \\
qcla\_com\_7                   & 443     & \textbf{284}     & 0.001            & \textbf{284}     & 0.016            & 314 & 0.013 \\
qcla\_mod\_7                   & 884     & \cellcolor{red!25}636     & 0.004            & \cellcolor{red!25}624     & 0.077            & \textbf{723} & 0.058 \\
rc\_adder\_6                   & 200     & 142     & \textless{}0.001 & \textbf{140}     & 0.004            & 157 & 0.003 \\
tof\_3                         & 45      & \textbf{35}      & \textless{}0.001 & \textbf{35}      & \textless{}0.001 & 40 & <0.001 \\
tof\_4                         & 75      & \textbf{55}      & \textless{}0.001 & \textbf{55}      & \textless{}0.001 & 65 & 0.001 \\
tof\_5                         & 105     & \textbf{75}     & \textless{}0.001 & \textbf{75}      & 0.001            & 90 & 0.002 \\
tof\_10                        & 255     & \textbf{175}     & \textless{}0.001 & \textbf{175}     & 0.004            & 215 & 0.006 \\
vbe\_adder\_3                  & 150     & \textbf{89}      & \textless{}0.001 & \textbf{89}      & 0.001            & 101 & 0.002 \\
\hline
\textbf{Avg. Red.} & & 24.6\% & & 26.4\% & & 19.2\% &
\end{tabular}
\label{tab:aat_table}
\end{table}

\begin{table}[t]
\caption{Total gate count reduction on Quipper adder circuits. \voqc's $H$ and $T$ counts are identical to Nam (L) and (H), but the total $Rz$ and $CNOT$ counts are higher due to Nam et al.'s specialized Toffoli decomposition. The difference between Nam (L) and Nam (H) is entirely due to $CNOT$ count. Our initial gate counts are higher than those reported by Nam et al. because we do not have special handling for +/- control Toffoli gates; we simply consider the standard Toffoli gate conjugated by additional $X$ gates.}
\vspace{-5pt}
\begin{tabular}{c|c|cc|cc|ccc}
& \multicolumn{1}{c|}{\textbf{Original}} & \multicolumn{2}{c|}{\textbf{Nam (L)}} & \multicolumn{2}{c|}{\textbf{Nam (H)}} & \multicolumn{3}{c}{\textbf{\voqc}} \\
\textbf{n} & \textbf{Total} & \textbf{Total} & \textbf{t(s)} & \textbf{Total} & \textbf{t(s)} & \textbf{Total} & \textbf{Parse t(s)} & \textbf{Opt. t(s)} \\
\hline
8    & 585    & 239   & 0.001   & 190   & 0.006     & 352 & <0.01 & 0.02 \\
16   & 1321   & 527   & 0.003   & 414   & 0.018     & 784 & <0.01 & 0.12 \\
32   & 2793   & 1103  & 0.014   & 862   & 0.066     & 1648 & 0.01 & 0.63 \\
64   & 5737   & 2255  & 0.057   & 1758  & 0.598     & 3376 & 0.03 & 3.30 \\
128  & 11625  & 4559  & 0.244   & 3550  & 4.697     & 6832 & 0.16 & 16.37 \\
256  & 23401  & 9167  & 1.099   & 7134  & 34.431    & 13744 & 1.06 & 79.74 \\
512  & 46953  & 18383 & 5.292   & 14302 & 307.141   & 27568 & 7.50 & 394.74 \\
1024 & 94057  & 36815 & 25.987  & 28638 & 2446.336  & 55216 & 45.76 & 1894.41 \\
2048 & 188265 & 73679 & 145.972 & 57310 & 23886.841 & 110512 & 252.48 & 9307.36 \\
\hline
\textbf{Avg. Red.} & & 63.7\% & & 71.6\% & & 45.7\% & & 
\end{tabular}
\label{tab:adder-table}
\vspace{1em}
\end{table}

\begin{table}[t]
\caption{Results on QFT circuits. Exact timings and gate counts are not available for Nam (L) or Nam (H), but our results are consistent with those reported in \citet[Figure 1]{Nam2018}.}
\vspace{-5pt}
\begin{tabular}{c|ccc|ccccc}
& \multicolumn{3}{c|}{\textbf{Original}} & \multicolumn{5}{c}{\textbf{\voqc}}\\
\textbf{n} & $CNOT$ & $R_z$ &  $H$ & $CNOT$ & $R_z$ &  $H$ & \textbf{Parse t(s)} & \textbf{Opt. t(s)} \\
\hline
8 & 56 & 84 & 8 & 56 & 42 & 8 & <0.01 & <0.01 \\
16 & 228 & 342 & 16 & 228 & 144 & 16 & <0.01 & <0.01 \\
32 & 612 & 918 & 32 & 612 & 368 & 32 & 0.01 & 0.01 \\
64 & 1380 & 2070 & 64 & 1380 & 816 & 64 & 0.05 & 0.07 \\
128 & 2916 & 4374 & 128 & 2916 & 1712 & 128 & 0.29 & 0.39 \\
256 & 5988 & 8982 & 256 & 5988 & 3504 & 256 & 1.99 & 2.34 \\
512 & 12132 & 18198 & 512 & 12132 & 7088 & 512 & 13.57 & 15.69 \\
1024 & 24420 & 36630 & 1024 & 24420 & 14256 & 1024 & 93.45 & 106.71 \\
2048 & 48996 & 73494 & 2048 & 48996 & 28592 & 2048 & 562.18 & 674.11 \\
\hline
\textbf{Avg. Red.} & & & & 0\% & 59.3\% & 0\% & &
\end{tabular}
\label{tab:qft-table}
\vspace{1em}
\end{table}

\begin{table}[t]
\caption{Results on QFT-based adder circuits. Final gate counts are identical for \voqc and Nam (L).}
\vspace{-5pt}
\begin{tabular}{c|ccc|ccccc|c}
& \multicolumn{3}{c|}{\textbf{Original}} & \multicolumn{5}{c|}{\textbf{\voqc}} & \multicolumn{1}{c}{\textbf{Nam (L)}} \\
\textbf{n} & $CNOT$ & $R_z$ &  $H$ & $CNOT$ & $R_z$ &  $H$ & \textbf{Parse t(s)} & \textbf{Opt. t(s)} & \textbf{t(s)} \\
\hline
8    & 184 & 276 & 16 & 184 & 122 & 16 & <0.01 & <0.01 & \textless{}0.001 \\
16   & 716 & 1074 & 32 & 716 & 420 & 32 & 0.01 & 0.02 & 0.001 \\
32   & 1900 & 2850 & 64 & 1900 & 1076  & 64 & 0.10 & 0.13 & 0.002 \\
64   & 4268 & 6402 & 128 & 4268 & 2388 & 128 & 0.76 & 0.90 & 0.004 \\
128  & 9004 & 13506 & 256 & 9004 & 5012 & 256 & 5.80 & 5.52 & 0.08 \\
256  & 18476 & 27714 & 512 & 18476 & 10260 & 512 & 43.73 & 36.80 & 0.018 \\
512  & 37420 & 56130 & 1024 & 37420 & 20756 & 1024 & 293.19 & 255.20 & 0.045 \\
1024 & 75308 & 112962 & 2048 & 75308 & 41748 & 2048 & 1516.76 & 1695.65 & 0.115 \\
2048 & 151084 & 226626 & 4096 & 151084 & 83732 & 4096 & 7488.03 & 8481.66 & 0.215 \\
\hline
\textbf{Avg. Red.} & & & & 0\% & 61.8\% & 0\% & & &
\end{tabular}
\label{tab:qfa-table}
\end{table}


\section{\sqir vs. \qwire}
\label{app:sqir-v-qwire}

As discussed in \Cref{sec:discussion}, a key difference between \sqir and \qwire is how they refer to qubits: \sqir uses concrete indices into a global register, while \qwire uses abstract variables, implemented using higher-order abstract syntax~\cite{Pfenning1988}. The tradeoff between the two approaches is most evident in how they support composition.

\paragraph*{Composition in \qwire}
\qwire circuits have the following form:
\begin{coq}
Inductive Circuit (w : WType) : Set :=
| output : Pat w -> Circuit w
| gate   : forall {w1 w2}, Gate w1 w2 ->  Pat w1 -> (Pat w2 -> Circuit w) -> Circuit w
| lift   : Pat Bit -> (bool -> Circuit w) -> Circuit w.
\end{coq}
Patterns \coqe{Pat} type the variables in \qwire circuits; their type index \coqe{w} corresponds to some collection of bits and qubits. 
The circuit \coqe{output p} is simply a wire with the wire type associated with \coqe{p}. Note that this, like all \coqe{Circuit}s, is an open term. The definition of \coqe{gate} takes in a \coqe{Gate} parameterized by an input type \coqe{w1} and an output \coqe{w2}, an appropriately typed input pattern, and a \emph{continuation} of the form \coqe{Pat w2 -> Circuit w}, which describes how the gate's output is used in the rest of the circuit. 
Finally, \coqe{lift} takes a single \coqe{Bit} (or classical wire) and a continuation that constructs a circuit based on that bit's interpretation as a Boolean value. 

This use of continuations makes composition easy to define:
\begin{coq}
Fixpoint compose {w1 w2} (c : Circuit w1) (f : Pat w1 -> Circuit w2) : Circuit w2 :=
  match c with 
  | output p    => f p
  | gate g p c' => gate g p (fun p' => compose (c' p') f)
  | lift p c'   => lift p (fun bs => compose (c' bs) f)
  end.
\end{coq}
In each case, the continuation is applied directly to the output of the first circuit.

While circuits correspond to open terms, closed terms are represented by \emph{boxed} circuits:
\begin{coq}
Inductive Box w1 w2 : Set := box : (Pat w1 -> Circuit w2) -> Box w1 w2.
\end{coq}
For convenience, \coqe{box (fun w => c)} is written as simply \coqe{box w => c} and \coqe{let p ← c1 ; c2} is similarly defined as notation for \coqe{compose c1 (fun p => c2)}. One can \coqe{unbox} a boxed circuit \coqe{Box w1 w2} simply by providing a valid \coqe{Pat w1}, obtaining a \coqe{Circuit w2}.

This representation allows for easy sequential and parallel  composition of closed circuits. 
Running two circuits in sequence involves connecting the output of the first circuit to the input of the second circuit (where the types will guarantee compatibility) and running them in parallel gives the circuits disjoint inputs and outputs their results (leading to a tensor type).
\begin{coq}
Definition inSeq {w1 w2 w3} (c1 : Box w1 w2) (c2 : Box w2 w3): Box w1 w3 :=
  box p1 ⇒ 
    let p2 ← unbox c1 p1;
    unbox c2 p2.

Definition inPar {w1 w2 w1' w2'} (c1 : Box w1 w2) (c2 : Box w1' w2') 
    : Box (w1 ⊗ w1') (w2 ⊗ w2'):=
  box (p1,p2) ⇒ 
    let p1'     ← unbox c1 p1;
    let p2'     ← unbox c2 p2; 
    (p1',p2').
\end{coq}
Unfortunately, proving useful specifications for these functions is quite difficult. The denotation of a circuit is (in the unitary case) a square matrix of size $2^n$ for some $n$. To construct such a matrix we need to map all of a circuit's (abstract) variables to $0$ through $n - 1$, ensuring that the mapping function has no gaps even when we initialize or discard qubits. \qwire maintains this invariant through compiling to a de Bruijn-style variable representation~\cite{deBruijn1972}. Reasoning about the denotation of circuits, then, involves reasoning about this compilation procedure. In the case of open circuits (the most basic circuit type), we must also reason about the contexts that type the available variables, which change upon every gate application. 

\paragraph*{Composition in \sqir}

Composing two \sqir programs requires manually defining a mapping from the global registers of both programs to a new, combined global register. To do this, we provide two helper functions, which respectively renumber a unitary program's concrete indices according to a mapping \coqe{f}, and change the program's global register size.
\begin{coq}
Fixpoint map_qubits {U dim} (f : nat -> nat) (c : ucom U dim) : ucom U dim :=
  match c with
  | c1; c2 => map_qubits f c1; map_qubits f c2
  | uapp1 u n => uapp1 u (f n)
  | uapp2 u m n => uapp2 u (f m) (f n)
  end.
  
Fixpoint cast {U dim} (c : ucom U dim) dim' : ucom U dim' := 
  match c with 
  | c1; c2 => cast c1 dim' ; cast c2 dim'
  | uapp1 u n => uapp1 u n
  | uapp2 u m n => uapp2 u m n
  end.      
\end{coq}

With these, we can define parallel composition in \sqir:
\begin{coq}
Definition inPar {U dim1 dim2} (c1 : ucom U dim1) (c2 : ucom U dim2) :=
  cast c1 (dim1 + dim2); 
  cast (map_qubits (fun q => dim1 + q) c2) (dim1 + dim2).
\end{coq}

The correctness property for \coqe{inPar} says that the denotation of \coqe{inPar c1 c2} can be constructed from the denotations of \coqe{c1} and \coqe{c2}.
\begin{coq}
Lemma inPar_correct : forall c1 c2 d1 d2,
  uc_well_typed d1 c1 -> [[inPar c1 c2 d1]]${}_{d1 + d2}$ = [[c1]]${}_{d1}$ ⊗ [[c2]]${}_{d2}$.
\end{coq}
The \coqe{inPar} function is relatively simple, but more involved than the corresponding \qwire definition because it requires relabeling the qubits in program $c_2$.

General composition in \sqir (including sequential composition) requires even more involved relabeling functions that are less straightforward to describe.
For example, consider the composition expressed in the following \qwire program:
\begin{coq}
box (ps, q) =>
  let (x, y, z) <- unbox c1 ps;
  let (q, z) <- unbox c2 (q, z);
  (x, y, z, q).
\end{coq}
This program connects the last output of program $c_1$ to the second input of program $c_2$. This operation is natural in \qwire, but describing this type of composition in \sqir requires some effort. In particular, the programmer must determine the required size of the new global register (in this case, 4) and explicitly provide a mapping from qubits in $c_1$ and $c_2$ to indices in the new register (for example, the first qubit in $c_2$ might be mapped to the fourth qubit in the new global register). 
When \sqir programs are written directly, this puts extra burden on the programmer.
When \sqir is used as an intermediate representation, however, these mapping functions should be produced automatically by the compiler.
The issue remains, though, that any proofs we write about the result of composing $c_1$ and $c_2$ will need to reason about the mapping function used (whether produced manually or automatically).

As an informal comparison of the impact of \qwire's and \sqir's representations on proof, we note that while proving the correctness of the \coqe{inPar} function in \sqir took a matter of hours, there is no correctness proof for the corresponding function in \qwire, despite many months of trying.
Of course, this comparison is not entirely fair: \qwire's \coqe{inPar} is more powerful than \sqir's equivalent. \sqir's \coqe{inPar} function does not require every qubit within the global register to be used -- any gaps will be filled by identity matrices. Also, \sqir does not allow introducing or discarding qubits, which we suspect will make ancilla management difficult to reason about.

\paragraph*{Quantum Data Structures}

\sqir also lacks some other useful features present in higher-level languages. For example, in QIO~\cite{Altenkirch2010} and Quipper~\cite{Green2013} one can construct circuits that compute on quantum data structures, like lists and trees of qubits. In \qwire, this concept is refined to use more precise dependent types to characterize the structures; e.g., the type for the \coqe{GHZ} program indicates it takes a list of $n$ qubits to a list of $n$ qubits. More interesting dependently-typed programs, like the quantum Fourier transform, use the parameter $n$ as an argument to rotation gates within the program.

Regrettably, these structures can make reasoning about programs difficult. For instance, as shown in \Cref{fig:pattern}, the \coqe{GHZ} program written in \qwire emits a list of qubits while the \coqe{fredkin_seq} circuit takes in a tree of qubits. Connecting the qubits from a \coqe{GHZ} to a \coqe{fredkin_seq} circuit with the same arity requires an intermediate gadget. And if we want to verify a property of this composition, we need to prove that this gadget is an identity.
In \sqir, which has neither quantum data structures nor typed circuits, this issue does not present itself.

\begin{figure}[t]
\centering
\includegraphics[scale=0.3]{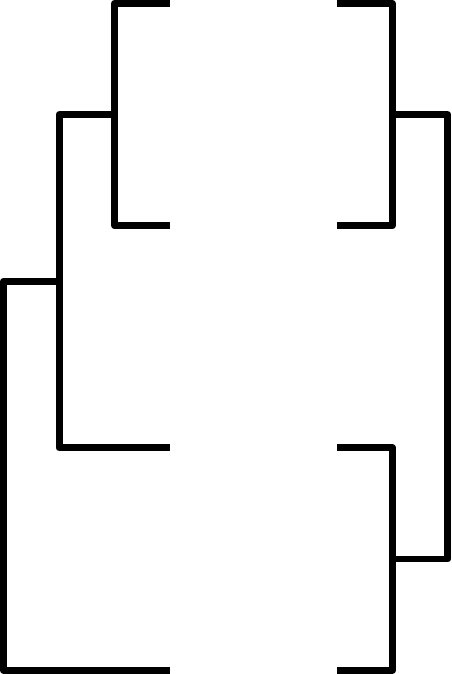}
\caption{The patterns for the output of \coqe{ghz} (left) and the input to \coqe{fredkin_seq} (right) over four qubits. \coqe{ghz} produces a list of qubits \coqe{(((q1,q2),q3),q4)} whereas \coqe{fredkin_seq} expects a tree \coqe{((q1,q2),(q3,q4))}. In composition, the mismatched patterns require an extra gadget to transform the former into the latter.}
\label{fig:pattern}
\end{figure}

\paragraph*{Dynamic Lifting}

\sqir also does not support \emph{dynamic lifting}, which refers to a language feature that permits measuring a qubit and using the result as a Boolean value in the host language to compute the remainder of a circuit \cite{Green2013}. Dynamic lifting is used extensively in Quipper and \qwire. Unfortunately, its presence complicates the denotational semantics, as the semantics of any Quipper or \qwire program depends on the semantics of Coq or Haskell, respectively. In giving a denotational semantics to \qwire, \citet{Paykin2017} assume an operational semantics for an arbitrary host language, and give a denotation for a lifted circuit only when both of its branches reduce to valid \qwire circuits. 

Although \sqir does not support dynamic lifting, its \coqe{meas} construct is a simpler alternative. Since the outcome of the measurement is not used to compute a new circuit, \sqir does not need a classical host language to do computation: It is an entirely self-contained, deeply embedded language. As a result, we can reason about \sqir circuits in isolation, and also easily reason about families of \sqir circuits described in Coq.

\paragraph*{Other Differences}

Another important difference between \qwire and \sqir is that \qwire circuits cannot be easily decomposed into subcircuits because output variables are bound in different places throughout the circuit. By contrast, a \sqir program is an arbitrary nesting of smaller programs. This means that for \sqir, the program \coqe{c1;((c2;(c3;c4));c5)} is equivalent to \coqe{c1;c2;c3;c4;c5} under all semantics, whereas every \qwire circuit (only) associates to the right. This allows us to arbitrarily flatten \sqir programs into a convenient list representation, as is done in \voqc (\Cref{sec:voqc}), and makes it easy to rewrite using \sqir identities. 

Also, unlike most quantum languages and as already discussed in the main body of the paper, \sqir features a distinct core language of unitary operators; the full language adds measurement to this core. The semantics of a unitary program is expressed directly as a matrix, which means that proofs of correctness of unitary optimizations (the bulk of \voqc) involve reasoning directly about matrices. Doing so is far simpler than reasoning about functions over density matrices, as is required for the full \sqir language or any program in \qwire.

\paragraph*{Concluding thoughts}

Upon reflection, we can see that the differences between \qwire and \sqir ultimately stem from their design goals. \qwire was developed as a general-purpose programming language for quantum computers~\cite{Paykin2017}, with ease of programmability as a key concern; only later was it extended as a tool for verification~\cite{Rand2017, Rand2018}. By contrast, \sqir was designed from the start with verification in mind, with by-hand programmability a secondary consideration; we expected \sqir would be compiled from another language such as Q\#~\cite{Svore2018}, Quipper~\cite{Green2013} or even \qwire itself. That said, for near-term quantum programs \sqir's lower-level abstractions have not proved difficult to use, even for source programming \cite{CPPsub}. As programs scale up, finding the right way to extend \sqir (or a language like it) with higher level abstractions without overly complicating verification will be an important goal.

}



\end{document}